\documentclass{aastex}

\usepackage[T1]{fontenc}
\usepackage{ae,aecompl}

\usepackage{graphicx}
\usepackage{epstopdf}
\usepackage{amsmath}

\usepackage{natbib}

\begin{document}

\title{G2 and Sgr $A^*$: A Cosmic Fizzle At The Galactic Center}


\author{Brian J. Morsony}
\affil{Dept. of Astronomy, University of Maryland, 1113 Physical Sciences Complex, College Park, MD, 20742-2421, USA}
\altaffiliation{morsony@astro.umd.edu}
\affil{Dept. of Astronomy, University of Wisconsin-Madison, 2535 Sterling Hall, 475 N. Charter Street, Madison, WI, 53706-1582, USA}
\affil{NSF Astronomy and Astrophysics Postdoctoral Fellow}

\author{Brandon T. Gracey}
\affil{Dept. of Physical and Environmental Sciences, Colorado Mesa University, Grand Junction, CO, 81501, USA}

\author{Jared C. Workman}
\affil{Dept. of Physical and Environmental Sciences, Colorado Mesa University, Grand Junction, CO, 81501, USA}

\author{DooSoo Yoon}
\affil{Dept. of Astronomy, University of Wisconsin-Madison, 2535 Sterling Hall, 475 N. Charter Street, Madison, WI, 53706-1582, USA}



\begin{abstract}

We carry out a series of simulations of G2-type clouds interacting with the black hole at the galactic center, to determine why no large changes in the luminosity of Sgr $A^*$ were seen, and to determine the nature of G2.
We measure the accretion rate from the gas cloud onto Sgr $A^*$ for a range of simulation parameters, such as cloud structure, background structure, background density, grid resolution, and accretion radius.
For a broad range of parameters, the amount of cloud material accreted is small relative to the amount of background material accreted.
The total accretion rate is not significantly effected for at least 30 years after periapsis.

We find that reproducing observations of G2 likely requires two components for the object: an extended, cold gas cloud responsible for the Br-$\gamma$ emission, and a compact core or dusty stellar object dominating the bolometric luminosity.
In simulations, the bolometric and X-ray luminosity have a peak lasting from about 1 year before to 1 year after periapsis, a feature not detected in observations.
Our simulated Br-$\gamma$ emission is largely consistent with observations leading up to periapsis, with a slight increase in luminosity and a large increase in the FWHM of the line velocity.
All emission from a gaseous component of G2 should fade rapidly after periapsis and be undetectable after 1 year, due to shock heating and expansion of the cloud.  
Any remaining emission should be from the compact component of G2.






\end{abstract}



\section{Introduction}
\label{sec:introduction}

%
%

We carry out a new set of simulations of G2-type gas clouds interacting with Sgr $A^*$.
Our goals are to 1) determine how the total accretion rate is affected by the cloud, and the relative amounts of cloud and background material accreted to see if changes in the luminosity of Sgr $A^*$ would be expected in the short or long term; and 2) model the emission from clouds with a range of internal structures and compare to observations, to constrain the nature of the G2 object.  
Rather than attempt to produce one exact model of G2 and its interaction, we vary several internal and external conditions to span a range of possibilities.

There has been a high level of interest in the object G2 since it was realized that it would pass very close to Sgr $A^*$, the supermassive black hole at the Milky Way's center \citep{gillessen12}.
Initially interpreted as a gas cloud, the tidal disruption of G2 could  potentially lead to a large increase in the rate of mass accretion onto the black hole, bringing about  a corresponding increase in the luminosity of Sgr $A^*$ \citep[e.g.][]{schartmann12, anninos12, saitoh14, decolle14}.
Extensive observations of G2 in Br-$\gamma$ hydrogen line emission and infrared L' band emission have found that G2 is closely following a highly eccentric orbit ($\epsilon$ from .9384 to .9814) and with a closed approach to Sgr $A^*$ of $1.92\times10^{15}$~cm to $3.96\times10^{15}$~cm (3000 to 6200 times the gravitational radius) \citep{gillessen12, gillessen13a, phifer13, gillessen13b}, reaching periapsis in March/April 2014.

The physical structure of G2 is still unclear.  
Br-$\gamma$ observations \citep[e.g.][]{gillessen12, gillessen13a, gillessen13b} have found that G2 is spatially extended, with a FWHM size of 42~mas (350~AU) in \citet{gillessen13b}.  
This  observed size and the brightness of Br-$\gamma$ emission could be explained if G2 is a gas cloud, with a mass of $\sim3$ Earth masses \citep{gillessen12}.
However, in L' band IR observations, G2 is unresolved, with an upper limit for its diameter of 32~mas (260~AU) \citep{witzel14}, and $>90\%$ of the emission from a FWHM of 20~mas (165~AU) \citep{eckart13}.
These observations have led to an alternative model of G2 as a dusty stellar object (DSO) \citep{eckart13, witzel14}, possibly the product of the recent creation of a 2 solar mass star through the merger of two smaller stars \citep{witzel14}.

A gas cloud would be expected to be tidally compressed as it heads towards periapsis, leading to an increase in IR emission.
No such increase was seen for G2 as it reached periapsis \citep{witzel14}, supporting the DSO model of G2.
An increase in X-ray emission, reaching $\sim10^{34}$~erg/s \citep{gillessen13b}, would also be expected for a gas cloud, but has not been seen in {\it Chandra} observations \citep{haggard14}.
If gas from G2 leads to an increase in mass accretion onto Sgr $A^*$, it would be detectable as an increase in luminosity.  
However, to date, no significant increase in the luminosity of Sgr $A^*$ has been seen in radio to millimeter \citep{tsuboi15, bower15}, IR \citep{witzel14}, or X-ray \citep{haggard14, degenaar15} observations.

Br-$\gamma$ observations \citep{gillessen12, gillessen13a, gillessen13b, phifer13, pfuhl15, valencia15} have shown a large and increasing velocity dispersion, up to $720$~km/s, as G2 approached periapsis, along with a possible increase in luminosity \citep{pfuhl15}.
After periapsis, there is an apparent drop in Br-$\gamma$ luminosity and a decrease of the velocity dispersion, to $220$~km/s \citep{valencia15}.
The Br-$\gamma$ observation may be better explained by a gaseous component rather than a DSO (see sect.~\ref{sec:cloud_density_profiles} for further discussion).

In its typical state, Sgr $A^*$ is believed to accrete $2\times10^{-9}$~M$_{\sun}$/year to $2\times10^{-7}$~M$_{\sun}$/year ($1.3\times10^{17}$~g/s to $1.3\times10^{19}$~g/s) from the background environment \citep{aitken00, bower03, marrone07}.
Sgr $A^*$ has a bolometric luminosity of $< 10^{37}$~erg/s \citep{narayan98} and an average X-ray luminosity of $3.4\times10^{33}$ erg/s \citep{nowak12, neilsen13}.
This translates to an overall radiative efficiency of $10^{-3}$ to $10^{-5}$ times the rest mass energy of the accreted material, making Sgr $A^*$ a very radiatively inefficient accretor.


Numerous previous simulations of G2 have been carried out to examine the evolution of the object and it's subsequent mass accretion onto Sgr $A^*$.
Simulations have been carried out with gas cloud \citep{schartmann12, burkert12, anninos12, sadowski13a, abarca14, saitoh14} and stellar wind \citep{ballone13, zajacek14, decolle14} origin models for G2.
Simulations measuring the accretion onto Sgr $A^*$ have generally predicted a high accretion rate.
For example, with an accreting inner boundary at $10^{15}$~cm, \citet{anninos12} found that the amount of cloud material accreted was $3$ -- $12\times10^{18}$~g/s ($5$ -- $19\times10^{-8}$~M$_{\sun}$/year).
With a stellar wind model for G2 and the same radius for the accreting inner boundary, $10^{15}$~cm, \citet{decolle14} found that the amount of cloud material accreted was about $2\times10^{18}$~g/s ($3\times10^{-8}$~M$_{\sun}$/year).
These simulations had background densities of $1.69\times10^{-21}$ and $9.5\times10^{-22} \times (R/10^{16}\textrm{cm})$~g~cm$^{-3}$.
These assumptions led, in both cases, to high background accretion rates, about $4\times10^{-6}$~M$_{\sun}$/year, a factor of 20 higher than the upper end of the allowed range for Sgr $A^*$ from \citet{marrone07}.
\citet{saitoh14} carried out SPH simulations of a cloud with an accretion radius at $4.5\times10^{14}$~cm and background number densities of 100~cm$^{-3}$ and 1000~cm$^{-3}$.
These simulations found cloud accretion rates from $1\times10^{-8}$~M$_{\sun}$/year to $2\times10^{-7}$~M$_{\sun}$/year and background accretion rates of $10^{-6}$~M$_{\sun}$/year and $10^{-5}$~M$_{\sun}$/year.
The cloud, therefore, accounts for only $1-5\%$ of the total accretion rate in any of these simulations.


In sec.~\ref{sec:methods} we lay out the numerical methods and setups of our simulations.
In sec.~\ref{sec:results} we examine how a range of numerical factors, including background velocity, accretion radius, background density, and radiative cooling change the accretion onto Sgr $A^*$ (sec.~\ref{sec:numerical}).
We then look at how cloud density structure affects accretion rate and the observational properties of the cloud, such as Br-$\gamma$, bolometric, and X-ray radiation (sec.~\ref{sec:cloud_density_profiles}).
We examine the long-term evolution of G2 in sec.~\ref{sec:long-term}, and summarize out conclusions in sec.~\ref{sec:conclusions}.

\section{Methods}
\label{sec:methods}

We carry out a series of 3-dimensional simulations using the FLASH \citep[v. 4.2.2][]{Fryxell00, Dubey08} grid-based adaptive-mesh hydrodynamics code.
.  
We use a directionally unsplit staggered mesh solver accurate to 2nd order in time and space \citep{lee13}.
Gas in our simulations is treated as quasi-adiabatic, with an adiabatic index of $5/3$.
For most simulations, gas is also allowed to cool following the cooling curve of \citet{sutherland93} assuming solar metallicity and a cooling floor of $10,000$~K.
A constant gravitational field from a point source (the black hole) is included.

Two tracer fluids are used to follow both the cloud and the background material separately.
This allows us to separate the motion and accretion of cloud material from the background flow.
Tracer particles are also included to follow the motion and orbital evolution of parcels of gas from G2.


For most simulations, gas is allowed to accrete onto the black hole by removing it from the simulation.
%

In these simulations, gas in cells within a defined radius, $R_{acc}$, of the black hole is allowed to accrete. 
Typically, the cells where accretion happens are only the central 8 grid points (2 by 2 by 2 pixel cube) surrounding the black hole. (The resolution of our simulations is $1.17 \times 10^{14}$~cm near the black hole.)

Not all material within the accretion radius is removed instantaneously.
Instead, a fraction of the gas is removed at each time step such that all gas would be removed in the free-fall time at that location.
For example, if the free-fall time at a given cell inside $R_{acc}$ is 10 time steps, $1/10$th of the gas at that location is removed in one time step.
Both density and pressure are reduced by the same amount, so that the temperature of the gas is not changed.
The amount of cloud and background mass removed from each cell at each time step is recorded.
Although the accretion radius is poorly resolved, this does not have a large impact on the accretion rate because of the slow removal of material.

This slow removal of gas is an appropriate prescription for gas accretion, given that we remain very far from the event horizon of the black hole, typically 184 times the gravitational radius.
The gravitational radius is defined as $R_{g} \equiv G M_{BH} / c^2$.
In reality, the accretion radius for a black hole will be at $1$ - $6$ $R_{g}$, much smaller than our simulations can resolve.  
The accretion radius we pick is, however, several times smaller than the accretion radius used by \citet{saitoh14} of 704 $R_{g}$ and \citet{anninos12} of 1565 $R_{g}$.


\subsection{Simulations Setups}
\label{sec:sim_setups}

The initial setup of our simulations consist of a spherically symmetric cloud in orbit about Sgr $A^*$, 5 years before periapsis.
We use a black hole mass of $8.62\times10^{39}$~g ($4.33\times10^{6}$~M$_{\sun}$) \citep{ghez08, gillessen09, genzel10} (Gravitational radius of $R_{g} = 6.39\times10^{11}$~cm).
We use orbital parameters from \citet{gillessen13b} with a semi-major axis of 1048 mas and an eccentricity of 0.9762.
We orient our simulations such that the cloud orbits in the x-y plane and the cloud velocity is initially in the $+$\^x direction.
The  initial position of the could center is  $[-3.4169332e16, -1.1264558e16, 0]$ (cm) and the initial velocity is $[1.6588336e8, 0, 0]$ (cm/s) with the a black hole at $[0, 0, 0]$ (cm) and stationary.


For our default backgrounds, the background gas is initially set to a constant density with a velocity of either $[1.659e8, 0, 0]$ (cm/s), the same as the initial cloud velocity (``wind'' models) or zero relative to the black hole (``no wind'' models).
For the ``wind'' model, this initial velocity is imposed everywhere, and there is an inflow boundary condition with constant velocity and density at the -x boundary.
The ``wind'' background model represents the case where the cloud is embedded in a large, low-density, co-moving stream of material.
A ``wind'' background minimizes the interaction of the cloud and background while the cloud is headed towards periapsis.
Note that after periapsis in the ``wind'' model, the cloud is traveling into the wind, which will enhance the dissipation of the cloud and, if anything, lead to an increase in the amount of cloud material accreted compared to a background flow that remains co-moving with the cloud.


Our computational domain is a cube $1.20 \times 10^{17}$~cm ($8038$~AU) on a side for the ``wind'' background simulations and 8 times larger for the ``no wind'' simulations (to ensure no inflow from the edges).
We allow a maximum resolution of $1.17 \times 10^{14}$~cm ($ = 0.95$~mas $ = 7.85$~AU $ = 184$ gravitational radii), equal to 8 levels of refinement for the ``wind'' simulations and 11 for the ``no wind'' simulations.
All locations on our grid with more than $90$\% cloud material are refined up to the maximum level.
All cells within $1.88\times10^{15}$~cm (16 grid points) of the black hole are also refined to the maximum level.


The initial background density in the default cases is $1.67 \times 10^{-23}$~g~cm$^{-3}$, with a temperature of $12,000$K.
The exceptions are the ``hot'' models, with a temperature of $1.2\times10^{8}$K, the 100bg models, which have 100 times the initial density, and the ``bg1'' and ``bg2'' with background profiles form \citet{anninos12} (see table \ref{table:models} and sec.~\ref{sec:background}).
Although our default background has a fairly low density \citep[see][]{anninos12, yuan2003}, the relative accretion rate of cloud and background material is largely independent of background density (see Sect.~\ref{sec:background}).
The initial temperature of the cloud material is $12,000$K in all cases.

\begin{figure*}
\includegraphics[scale=.9]{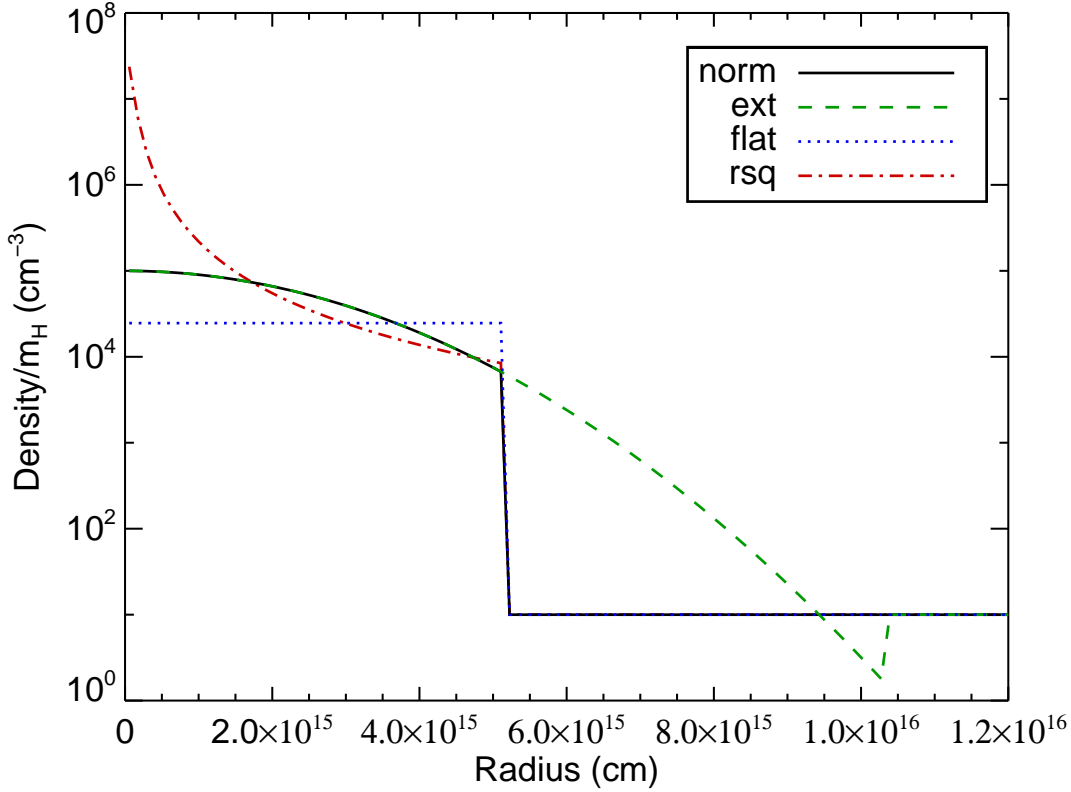}
\caption{
Initial density profiles of our 4 cloud models.
The total cloud mass in each model is 4 Earth masses, except for the ``ext'' model which has $15\%$ extra mass.
}
\label{fig:cloud_density_profiles}
\end{figure*}

To probe the effects of the physical structure of G2 on accretion luminosity  we model the cloud with one of 4 initial density profiles.
Our ``normal'' model is Gaussian density profile with a sigma of $2.20\times10^{15}$~cm (FWHM of 42 mas as seen from Earth, consistent with \citet{gillessen13b}).
The Gaussian model extends to a radius of $5.17\times10^{15}$~cm from the center (twice the half-max radius) after which it is cut off.
The central density is $1.67\times10^{-19}$~g~cm$^{-3}$, giving a total cloud mass of $2.41\times10^{28}$~g, about 4 Earth masses.
Our ``extende'' model is the same as the normal model, but the Gaussian profile extends to $1.03\times10^{16}$~cm, 4 times the half-max radius.
The total cloud mass is $15.7$\% higher in this case.
The ``flat'' model consists of a cloud with a constant density and a radius of $5.17\times10^{15}$~cm, the same as the normal model.  
The density used is $4.12\times10^{-20}$~g~cm$^{-3}$, giving the cloud the same mass at the normal model.
Finally, the ``rsq'' model has a density profile of $\rho = A r^{-2}$~g~cm$^{-3}$ from the center, modeling a stellar wind.  
The density is again cut off at $5.17\times10^{15}$~cm, the same as the normal and flat models.  
The constant is set to $A = 3.67\times10^{11}$~g~cm$^{-1}$, giving the same total cloud mass as the normal and flat models.
The initial density profiles are plotted in fig.~\ref{fig:cloud_density_profiles}.

Our standard accretion radius is $R_{acc} = 1.47\times10^{14}$~cm, which means only the central 8 grid points (2 by 2 by 2 pixel cube) are inside the accretion radius.
For some models (labeled 2r, 4r, etc.) a larger accretion radius (of 2 times larger, 4 times, etc.) is used.
Initial parameters of all simulations are summarized in table~\ref{table:models}.

\section{Results}
\label{sec:results}

\begin{figure*}
\includegraphics[scale=.3]{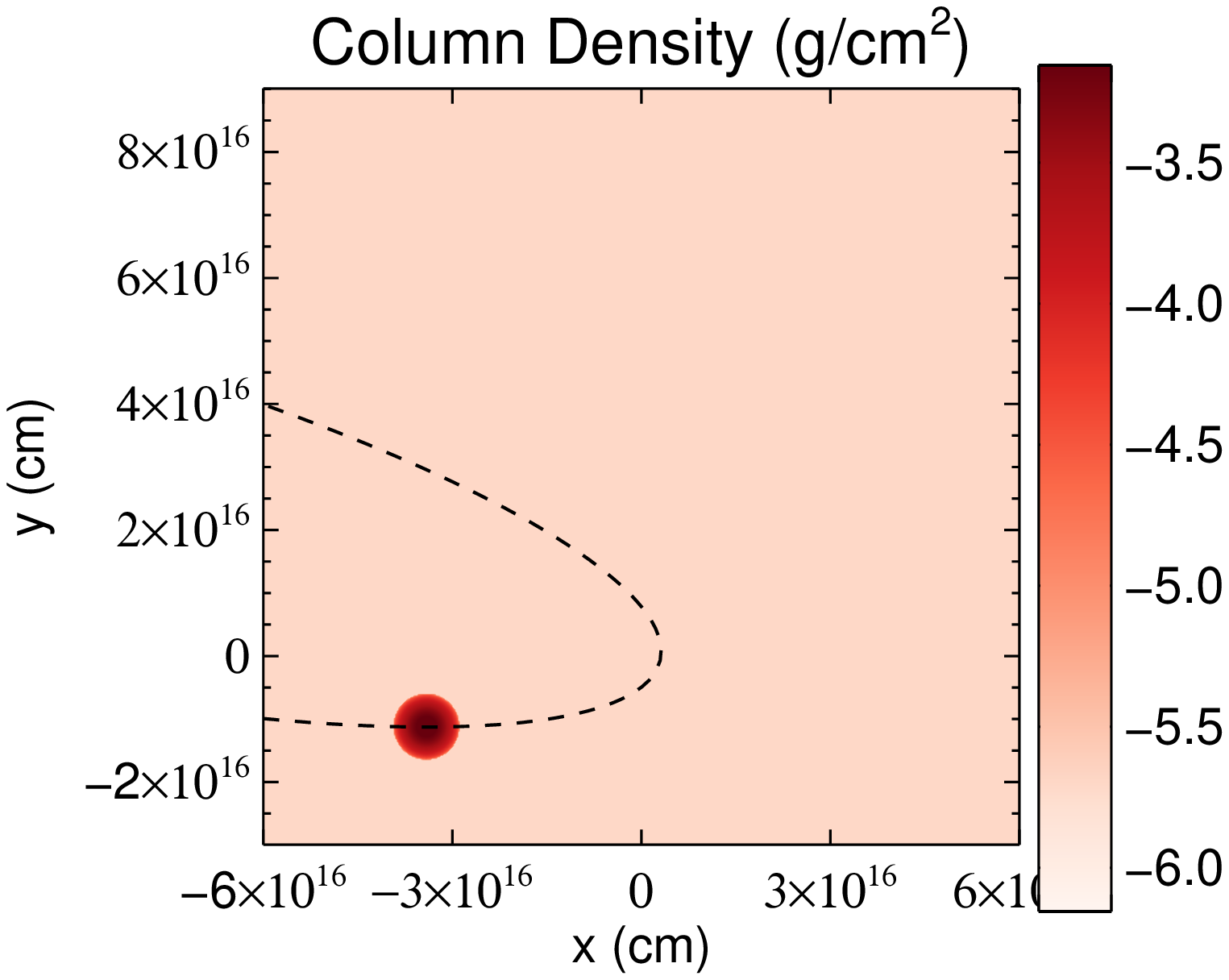}
\includegraphics[scale=.3]{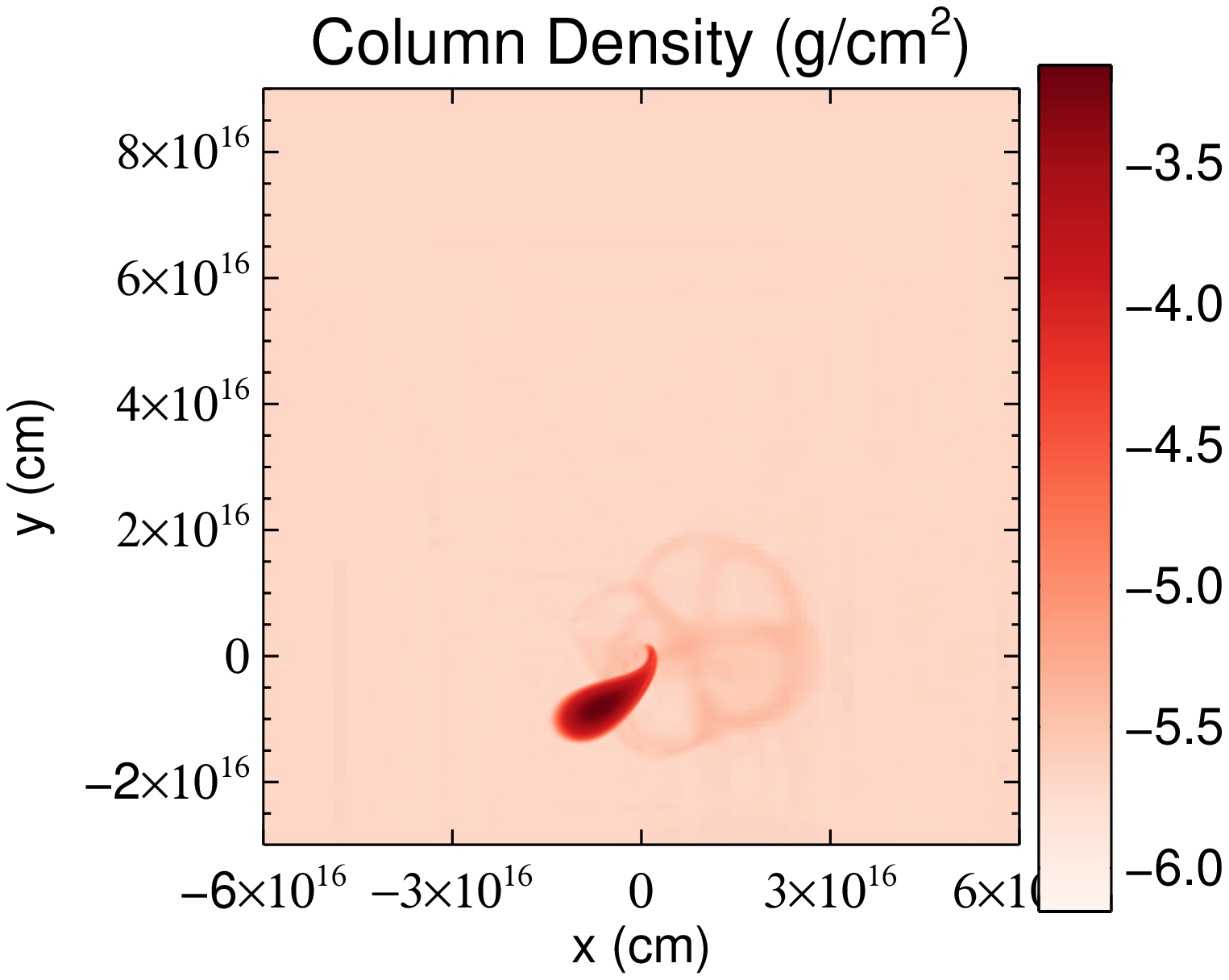}
\includegraphics[scale=.3]{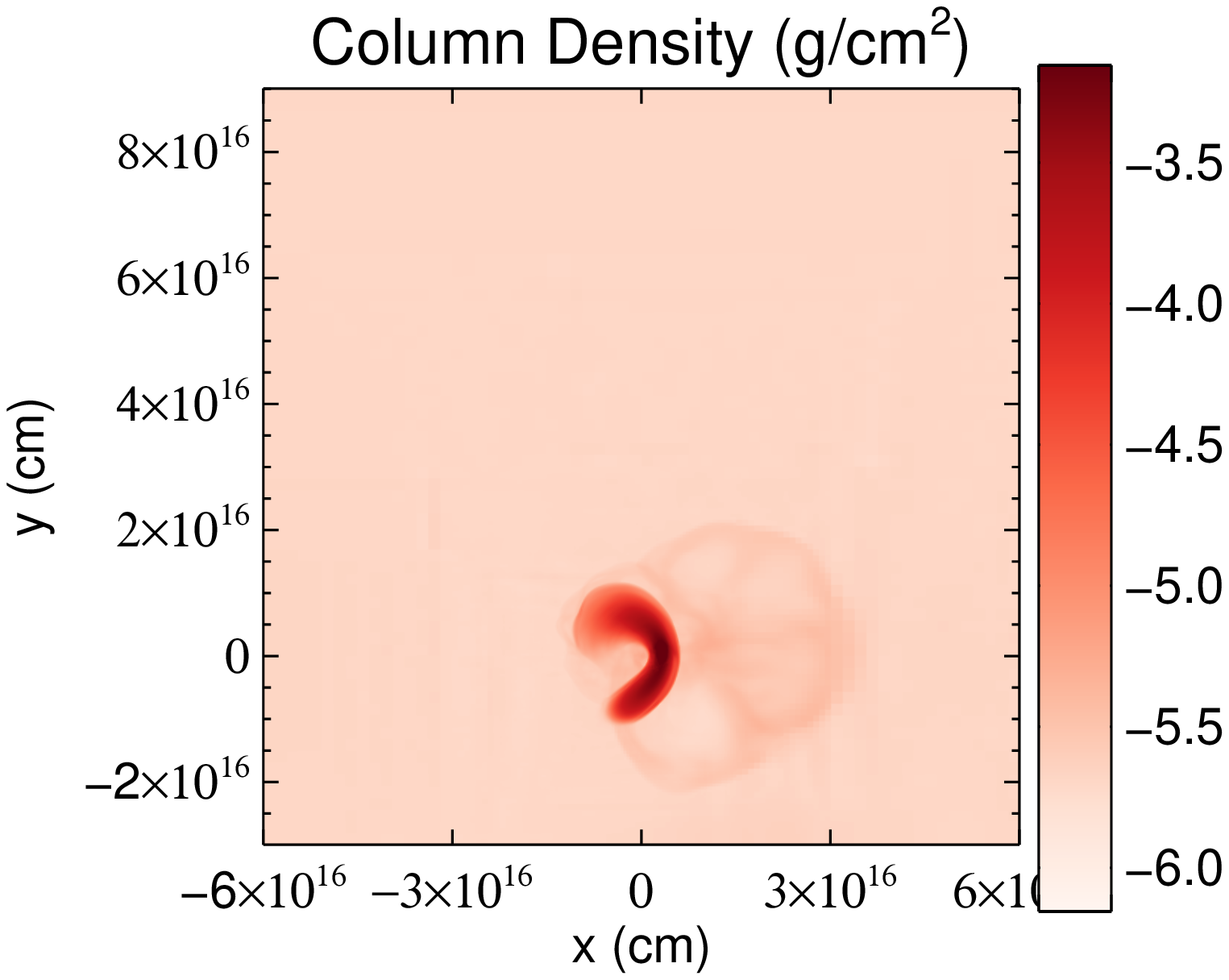}

\includegraphics[scale=.3]{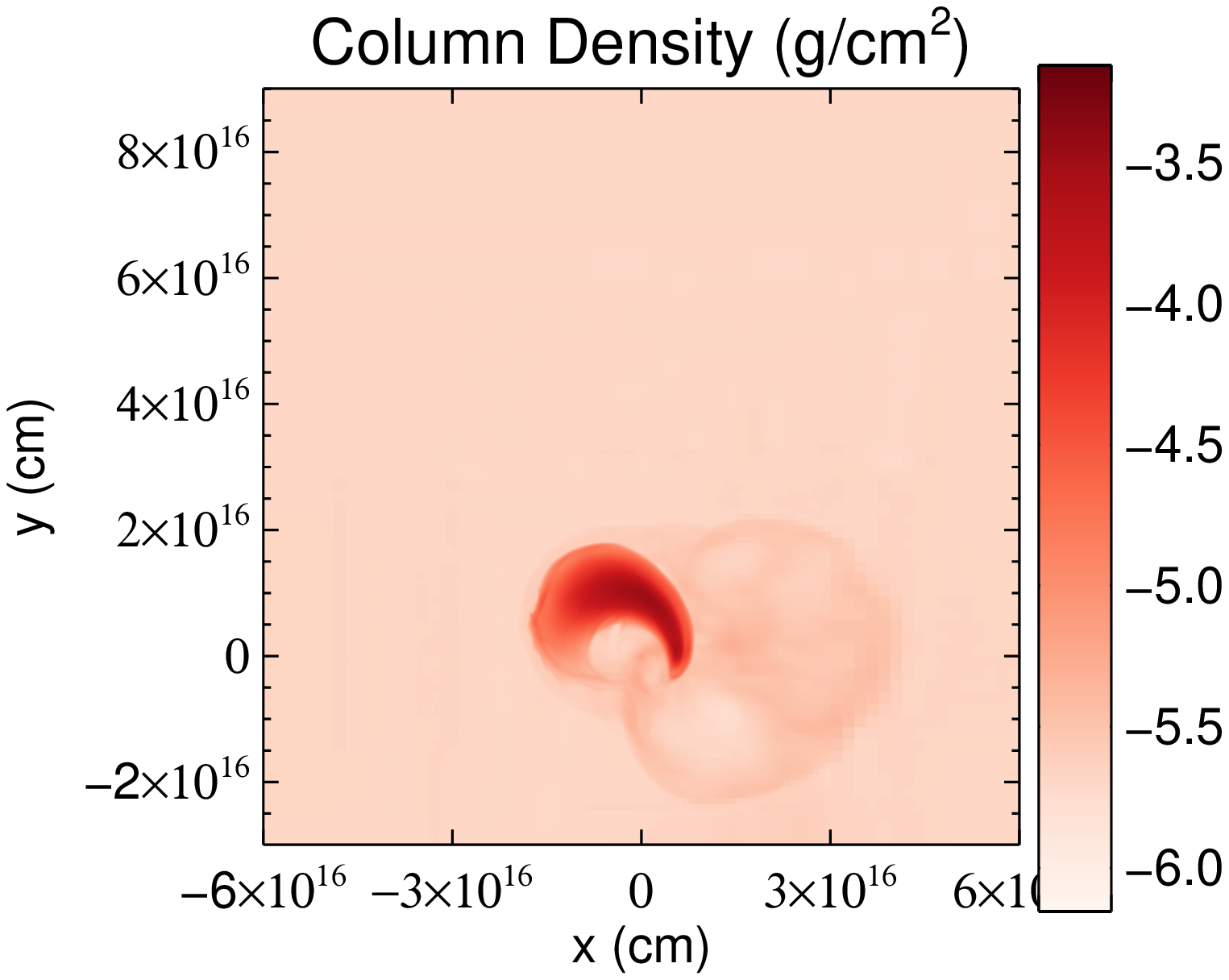}
\includegraphics[scale=.3]{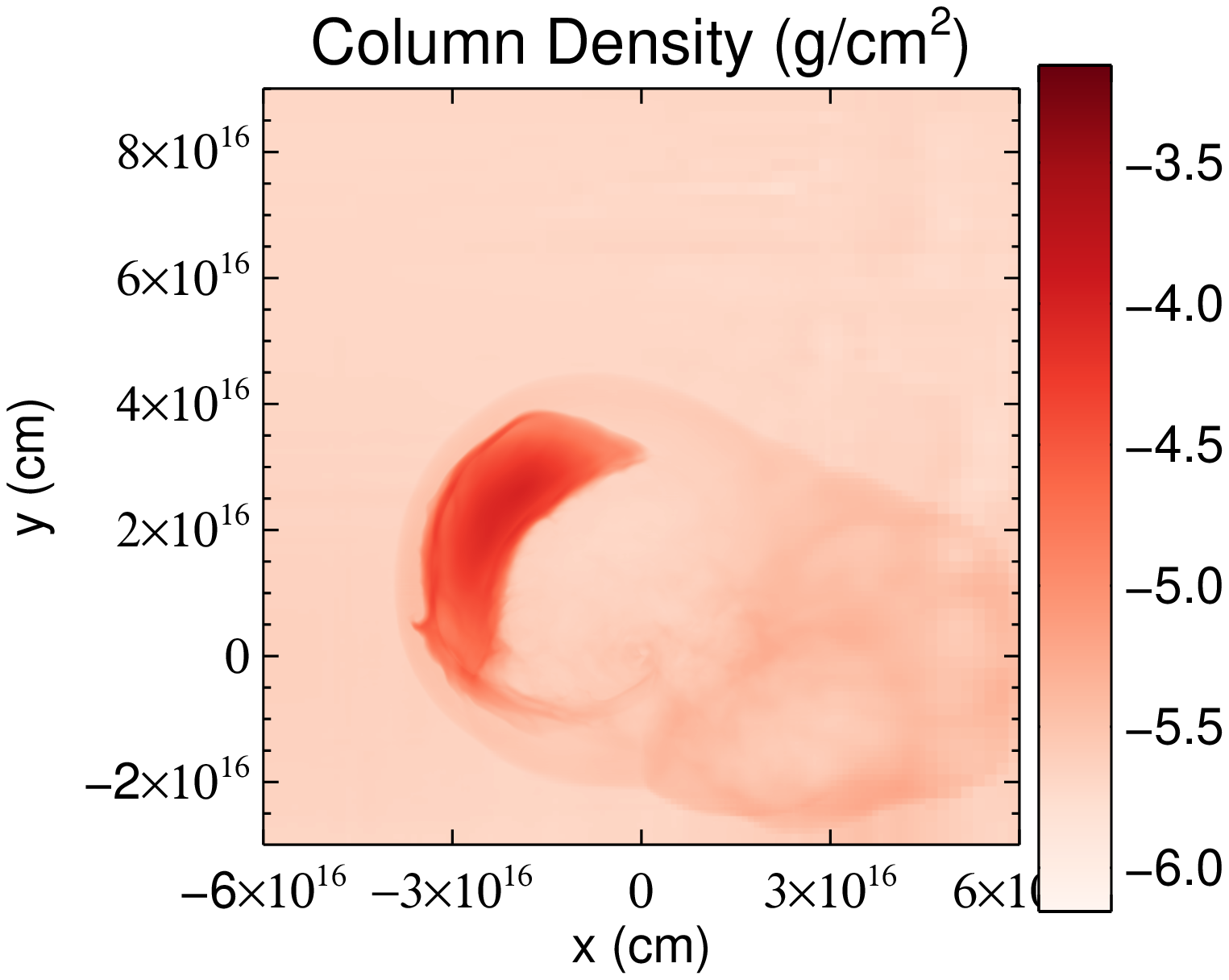}
\includegraphics[scale=.3]{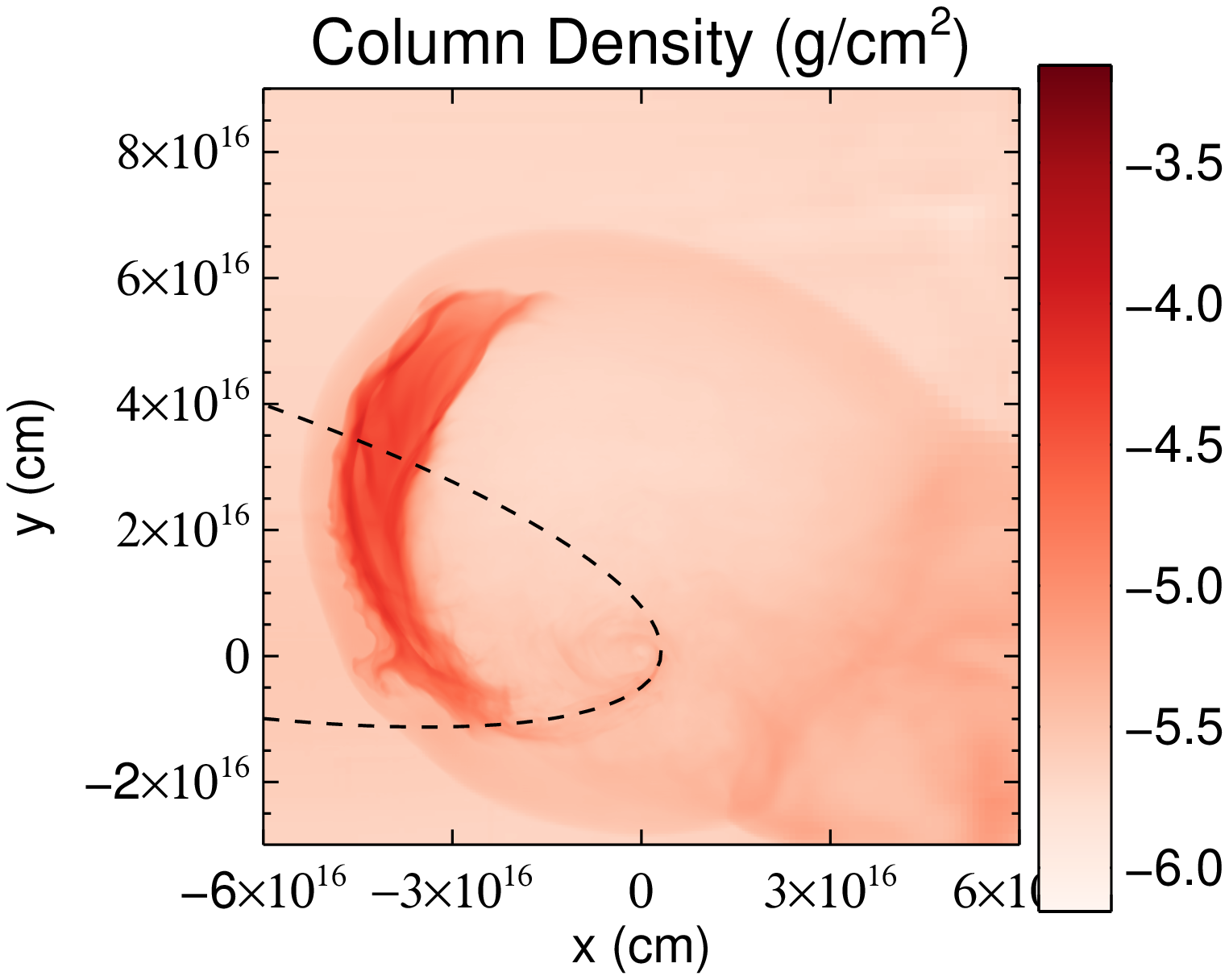}

\caption{
Column density (log scale) plots showing the cloud at time relative to periapsis of -5, -1, 0, 1, 5 and 10 years (left to right, top to bottom) for the ``norm'' model.
View is from above the orbital plane, with the black hole at (0,0).  
The path of the orbit is overlaid on the first and last snapshot (black dashed line).
As the cloud approaches periapsis, no part of the cloud passes within our accretion radius.
After closest approach, the cloud expands into a fan of material, with a very weak stream of material leading back towards the black hole.
At 10 years after periapsis, a tenuous elliptical ring of material can be seen.
}
\label{fig:dens_sequence}
\end{figure*}

For all of our simulations, the cloud is initially 5 years from periapsis.  
As the cloud approaches Sgr $A^*$, it is tidally stretched along the orbital path and compressed perpendicular to the orbital plane.
When cloud material passes periapsis, the converging gas flow creates a ``nozzle shock'' \citep{evans89, guillochon14} which heats the gas to about $10^7$~K.
After periapsis, the cloud expands in a fan of outgoing material, with a small accretion stream leading back to the black hole.
Fig.~\ref{fig:dens_sequence} shows the column density for a sequence of frames from our ``normal'' simulation, from above the orbital plane.
Interaction with the background material does not significantly effect the cloud, although a shock forms around the cloud, particularly visible in front of the outgoing fan after periapsis.
At 10 years after periapsis (last frame), the accretion stream from the fan connects to a tenuous, elliptical ring of material that can be seen around the black hole, formed by cloud material that has fallen back and begun to circularize.
The size of the ring is large compared to the accretion radius ($5\times10^{15}$~cm) and contains only $0.2\%$ of the cloud mass.

\begin{figure*}
\includegraphics[scale=.3]{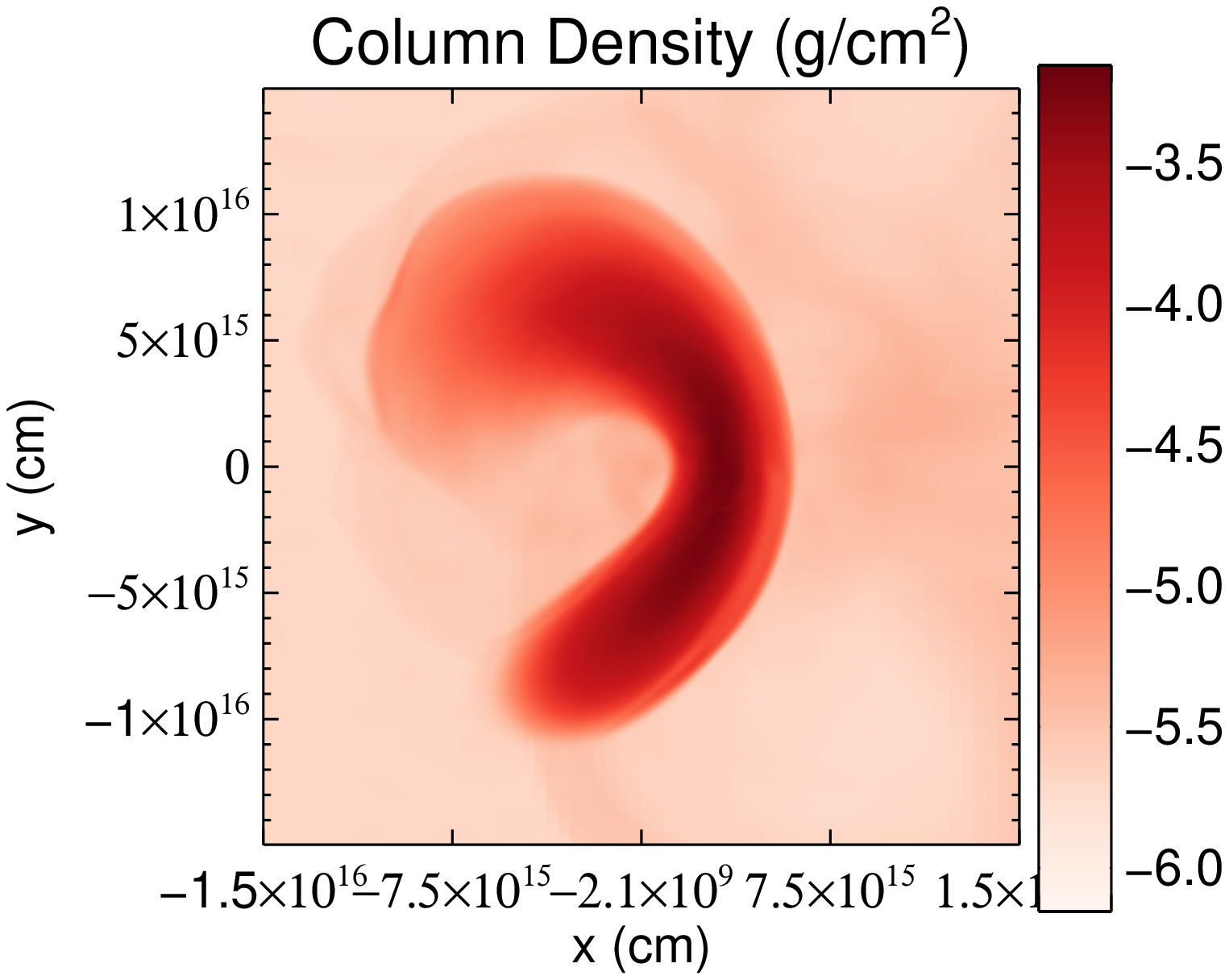}
\includegraphics[scale=.3]{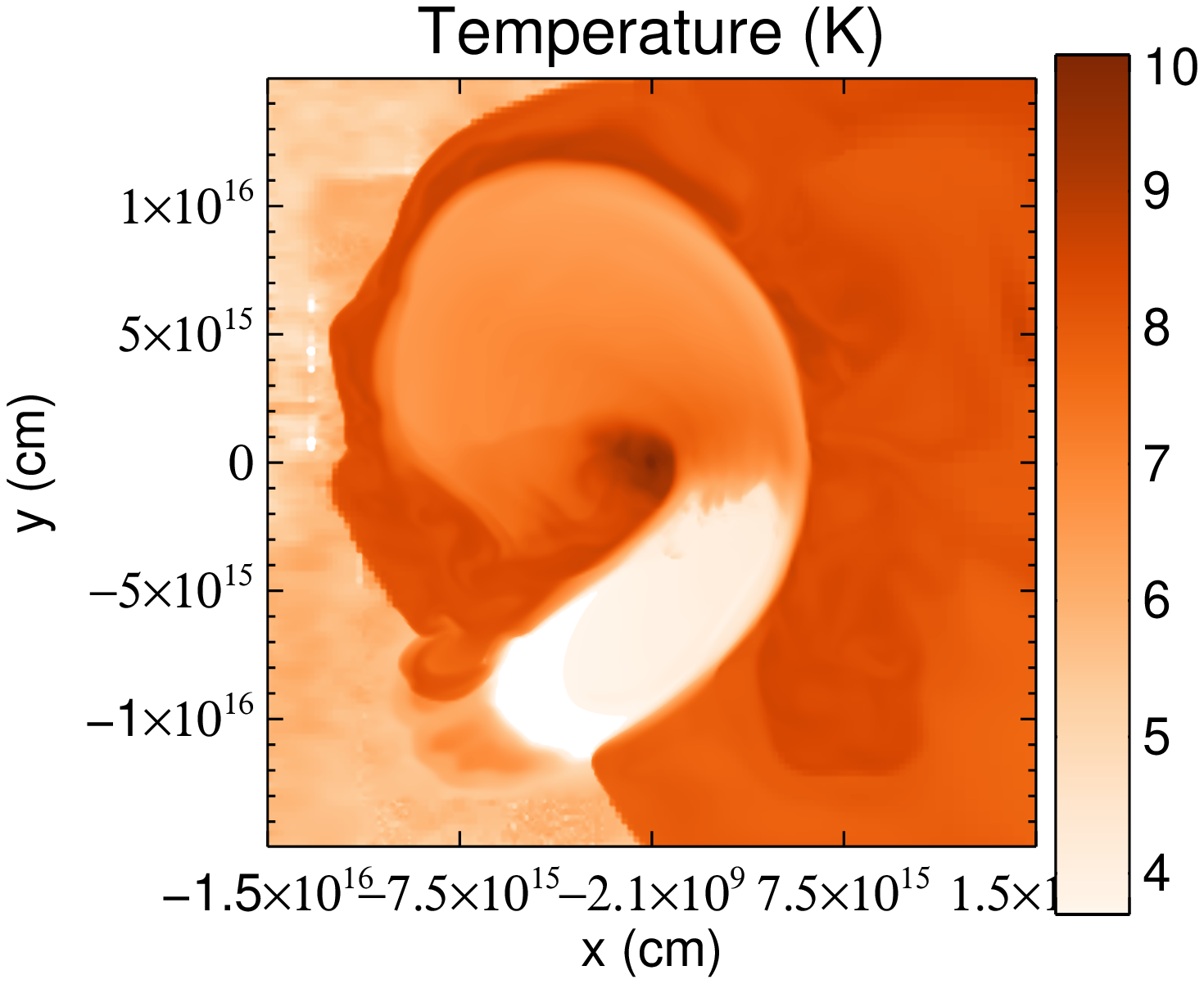}
\includegraphics[scale=.3]{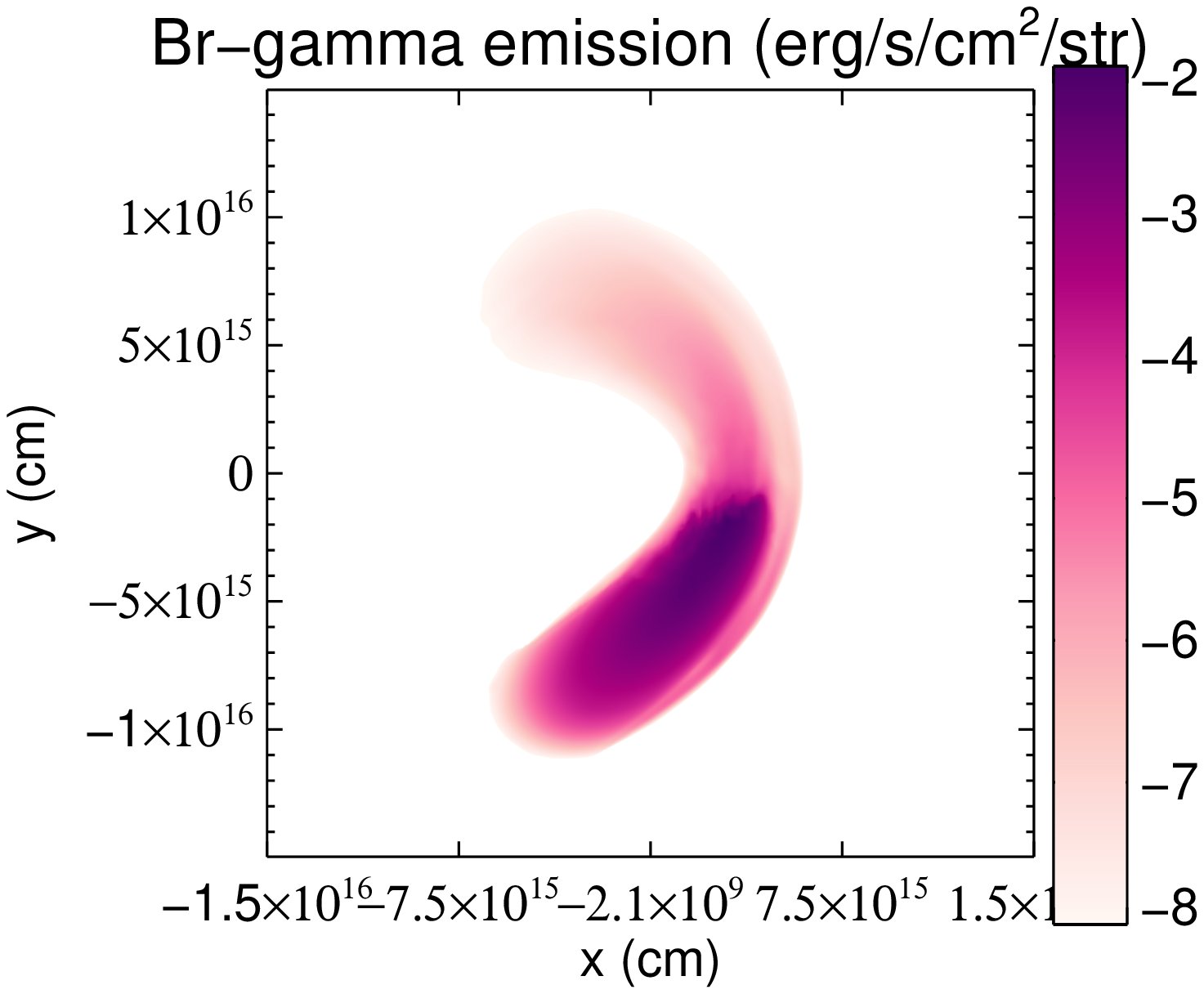}
\caption{
Zoom in images of the cloud at periapsis for the ``norm'' model.  
Column density (left) shows the tidal deformation of the cloud and expansion into a fan as material passes periapsis.
Temperature in a slice through the orbital plane (center) shows that the bulk of the cloud is still cold ($\sim 1.2\times10^4$~K) before periapsis, surrounded by a thin layer shocked by interaction with the ambient gas.
Near periapsis, the gas is heated by a nozzle shock up to $\sim 10^7$~K.
This hot gas then expands into a fan moving away from the black hole.
Br-$\gamma$ surface brightness (right) is primarily from cold gas, and decreases sharply as gas is heated at the nozzle shock.
All images are on log scales.
}
\label{fig:periapsis_zoom}
\end{figure*}

Fig.~\ref{fig:periapsis_zoom} shows a zoomed-in view of the cloud at periapsis.  
Column density (left) shows the tidal deformation of the cloud and its expansion into a fan as material passes periapsis.
A temperature slice through the orbital plane (center) shows that the bulk of the cloud is still cold ($\sim 1.2\times10^4$~K) before periapsis, surrounded by a thin layer shocked by interaction with the ambient gas.
Near periapsis, the gas is heated by the nozzle shock up to $\sim 10^7$~K.
%
The extended envelope of low density, hot gas ($T > 10^{8}$~K) in fig.~\ref{fig:periapsis_zoom} is due to shocked background material.

Br-$\gamma$ emissivity is modeled as 
\begin{equation}
3.44\times10^{-27} \times (T/10^4)^{-1.09} \times n_{p} \times n_{e}\rm{~erg~s}^{-1}\rm{~cm}^{-3}
\end{equation}
\citep{osterbrock06, ballone13}.
%
We assume the entire cloud is fully photoionized at all times.
The right panel of fig.~\ref{fig:periapsis_zoom} shows the modeled Br-$\gamma$ surface brightness at periapsis.
Due to the strong temperature dependence, Br-$\gamma$ emission is primarily from cold gas, and decreases by 3 orders of magnitude as gas is heated at the nozzle shock.

\subsection{Mass accretion - numerical considerations}
\label{sec:numerical}

In this section, we examine how various parameters of our numerical setup can effect the total amount of material accreted onto the black hole, and the relative amount of cloud and background material accreted.


In our simulations, material within the accretion radius is removed in a free-fall time, with the accretion rate being the rate at which material is removed.
Cloud and background material accretion rates are recorded separately.
The accretion radius for simulations is given in table~\ref{table:models}, with 1x corresponding to a radius of $1.47\times10^{14}$~cm.

We examine how background structure, accretion radius, background density, and cooling influence accretion.
Simulations are carried out with either our normal (Gaussian) cloud profile, or with no cloud present.
In the next section, we present simulations with different density profiles and compare them to observations.
For each simulation, the total mass and cloud-only mass accreted from 5 years before to 5 years after periapsis, the ratio of cloud to total mass accreted after periapsis, and the ratio of total mass accreted after periapsis to an equivalent simulation with no cloud present are listed in table~\ref{table:models}.

\subsubsection{Background velocity}

\begin{figure*}
\includegraphics[scale=.9]{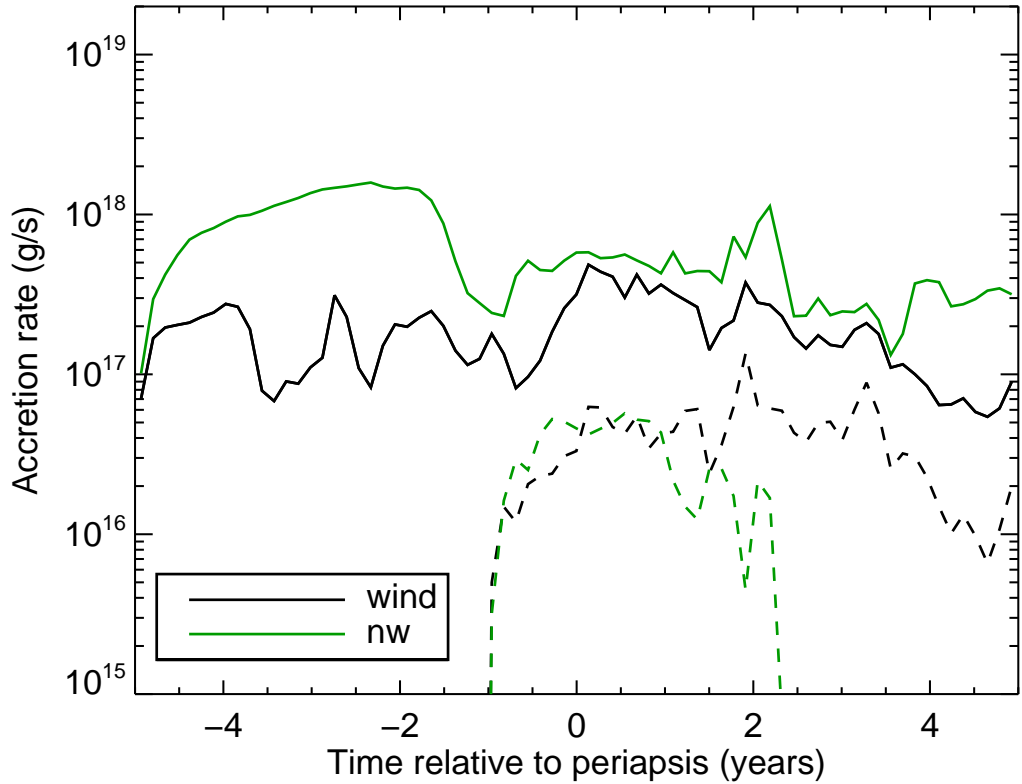}
\caption{
Accretion rate vs. time for our normal model with a wind background (black) and no wind background (green). 
In each case, the solid line is the total mass accretion and the dashed line is accretion of cloud material only. 
Each point is a 50-day average of accretion rate. 
The total accretion rate is lower with a wind background. 
Accretion of cloud material begins 1 year before periapsis, and continues at a rate at around $3\times10^{16}$~g/s. 
Although highly variable, there is no significant increase in the total accretion rate as the cloud reaches periapsis. 
Cloud material is not the dominant component of accreted material in either case, accounting for about $21\%$ and $4\%$ of the material accreted after periapsis, respectively. 
}
\label{fig:acc_norm_tot_v_cloud}
\end{figure*}

For our simulations, we use one of two background structures -- a background moving uniformly with the initial velocity of the cloud (``wind'') or a stationary background with respect to the black hole (``no wind'').
Fig.~\ref{fig:acc_norm_tot_v_cloud} plots the total and cloud-only accretion rates vs. time for a wind  and no-wind  background.

The total accretion rate is lower with a wind background.
The total accretion over the 10 years of the simulation is $6.01\times10^{25}$~g for the wind background and $2.05\times10^{26}$~g for the no wind background.
These accretion rates ($3\times10^{-9}$~M$_{\sun}$/year and $1\times10^{-8}$~M$_{\sun}$/year) are consistent with the low end of the real accretion rate onto Sgr $A^*$ ($2\times10^{-9}$~M$_{\sun}$/year to $2\times10^{-7}$~M$_{\sun}$/year) \citep{marrone07}.
In both cases, the accretion rate varies by an order of magnitude around its average value because the accretion is convectively unstable, producing large inhomogeneities in the temperature, density, and velocity of the inflowing material.

The background velocity does not have a large effect on the amount of cloud material accreted.
Cloud material begins to accrete 1 year before periapsis and continues at a rate of around $3\times10^{16}$~g/s ($5\times10^{-10}$~M$_{\sun}$/year) in both simulations, but for the no wind model the accretion rate drops by two orders of magnitude at 2 years after periapsis.  
This drop is due to the continued (unstable) accretion of background material pushing cloud-enriched gas away from the black hole in that case.   
The total amount of cloud material accreted by 5 years after periapsis is $7.50\times10^{24}$~g in the wind simulation and $3.33\times10^{24}$~g in the no wind simulation, about $0.03\%$ and $0.014\%$ of the total cloud mass.
In both models, the dominant component of the accreted mass is background material at all times.
From periapsis to 5 years after periapsis, only $21\%$ and $3\%$ of the accreted mass is from cloud material in the wind and no wind simulations, respectively.

\subsubsection{Accretion radius}
\label{sec:Accretion_radius}

\begin{figure*}
\includegraphics[scale=.45]{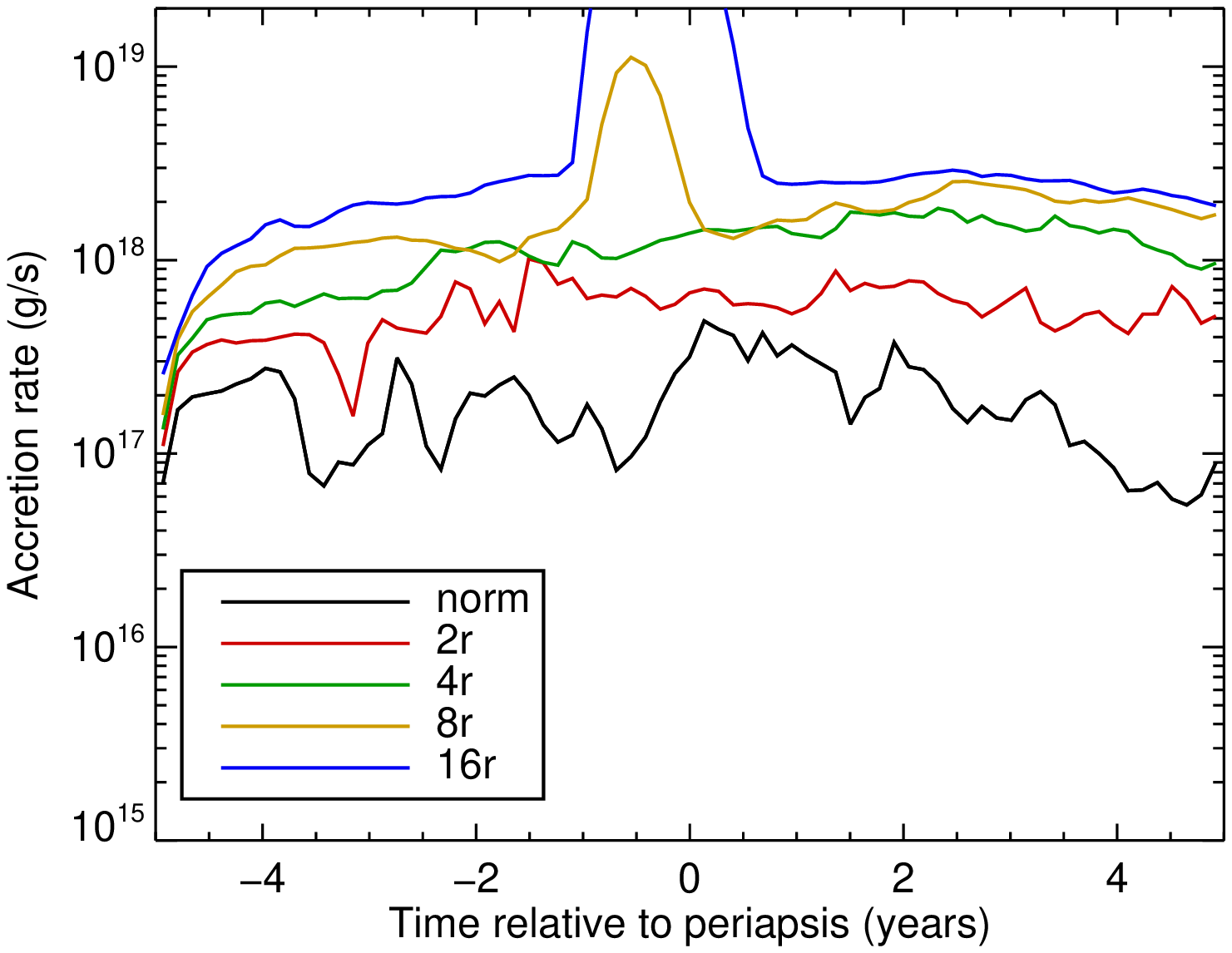}
\includegraphics[scale=.45]{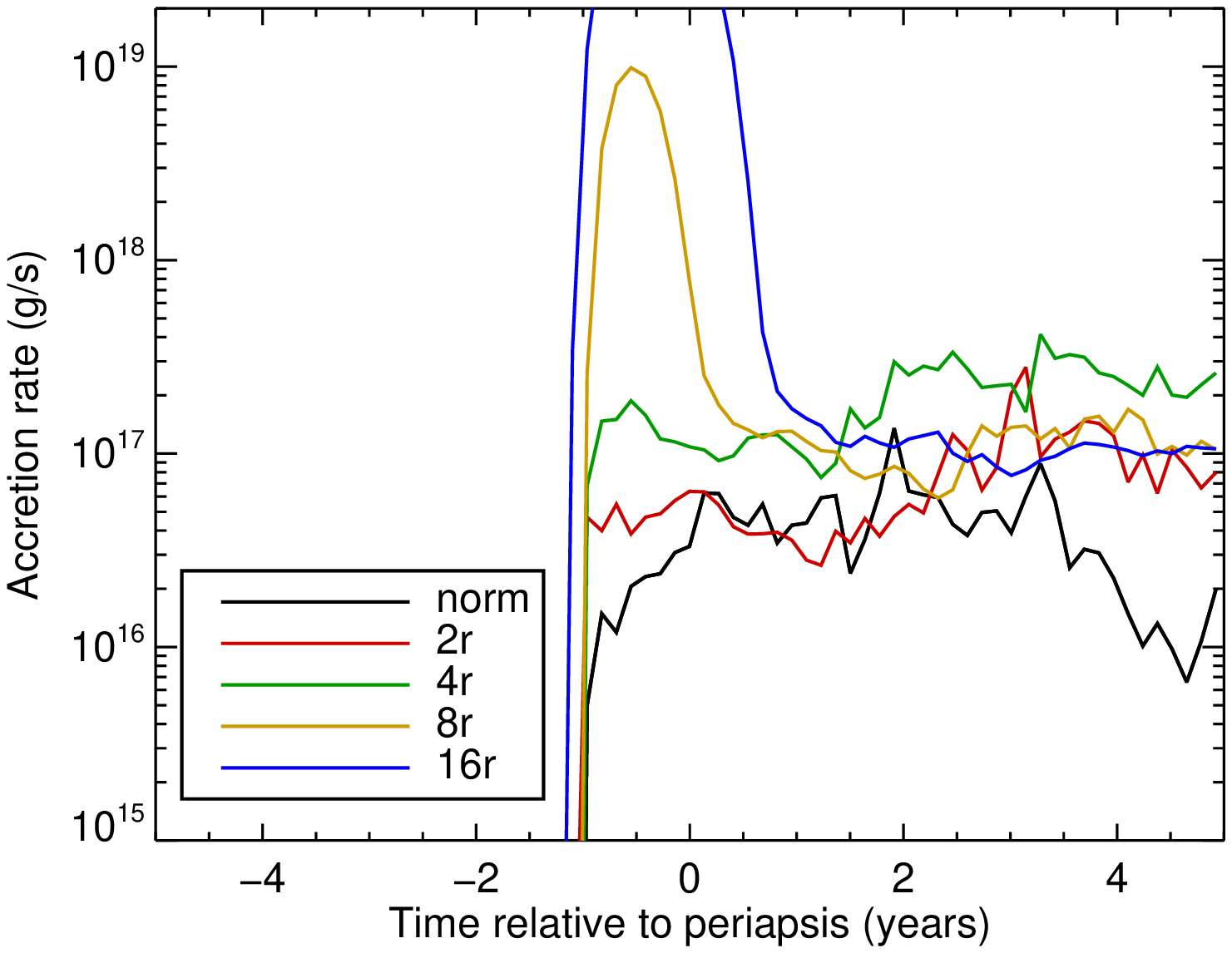}
\caption{
Total (left) and cloud only (right) accretion rates models with ``normal'' cloud profile and a wind background.  
The accretion radius ranges from 1 to 16 times our normal accretion radius of $R_{acc} = 1.47\times10^{14}$~cm.
Accretion rate increases as $R_{acc}$ increases, both for all material and the cloud only.
In all cases, the cloud is a sub-dominant part of total material accreted for most of the simulation.
The exception is for the 8r and 16r models close to periapsis.
During this time, part of the could passes inside the accretion radius and is removed from the simulation.  
However, this material is on highly elliptical orbits and should pass back out beyond the accretion radius again.
These accretion radii ($1.17\times10^{15}$~cm and $2.34\times10^{15}$~cm) result in an artificially high amount of cloud material being accreted.
}
\label{fig:acc_wind_racc_size}
\end{figure*}

Changing the accretion radius has a large effect on the accretion rate.
Fig.~\ref{fig:acc_wind_racc_size} shows accretion rates for wind background models with accretion radii of 1, 2, 4, 8, and 16 times the normal radius.
Total accretion rate (left panel) and cloud-only accretion (right panel) are plotted.


The accretion rate, both overall and for the cloud only, increases with increasing accretion radius.
Gas near the black hole is partially supported by the pressure of gas closer in, up to the accretion radius where the interior pressure drops rapidly.
By moving the accretion radius outward, the pressure support is lowered and more gas is able to flow in.
There is only pressure support in an unstable and/or shocked atmosphere, and would not be present if a Bondi inflow were able to develop.

There is a marked difference between the 8r and 16r  models and the other three.
For the two largest accretion radii ($R_{acc} > 10^{15}$~cm), part of the cloud passes inside the accretion radius and is directly accreted, producing an artificially high accretion rate.
This produces the large bump in the accretion rate in the year before periapsis in these two simulations, and a relatively low cloud accretion rate at late times.
The amount of cloud material accreted increases by a factor of 500 from the normal to 16r simulations, with almost $1/7$th of the cloud accreted in the 16r simulation.

The cloud material is never the dominate component of the total accretion in any simulation, except when part of the cloud is passing inside the accretion radius.
Otherwise, the cloud only accounts for $10 - 25\%$ of the total accretion.
Excluding the cloud material, the amount of background material increases by a factor of 13 between the normal and 16r simulations.

In a simulation where mass is accreted, this outward motion cannot happen for material inside the accretion radius.  
For the simulations with an accretion radius larger than $10^{15}$~cm, in particular, most of the mass accreted from the cloud should actually continues back out to a much larger distance rather than remaining close to the black hole and being quickly accreted.


\subsubsection{No cloud comparison}

\begin{figure*}
\includegraphics[scale=.45]{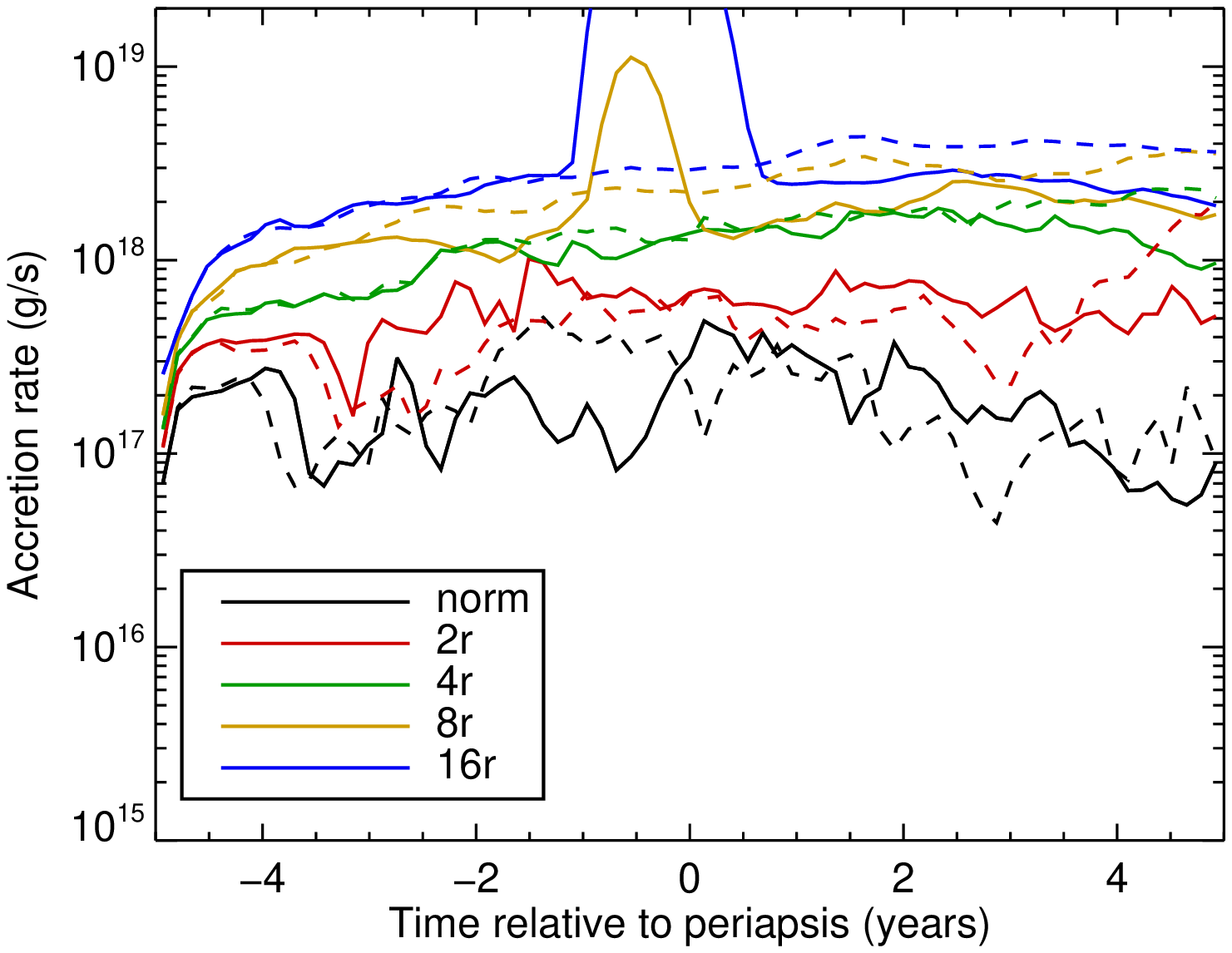}
\includegraphics[scale=.45]{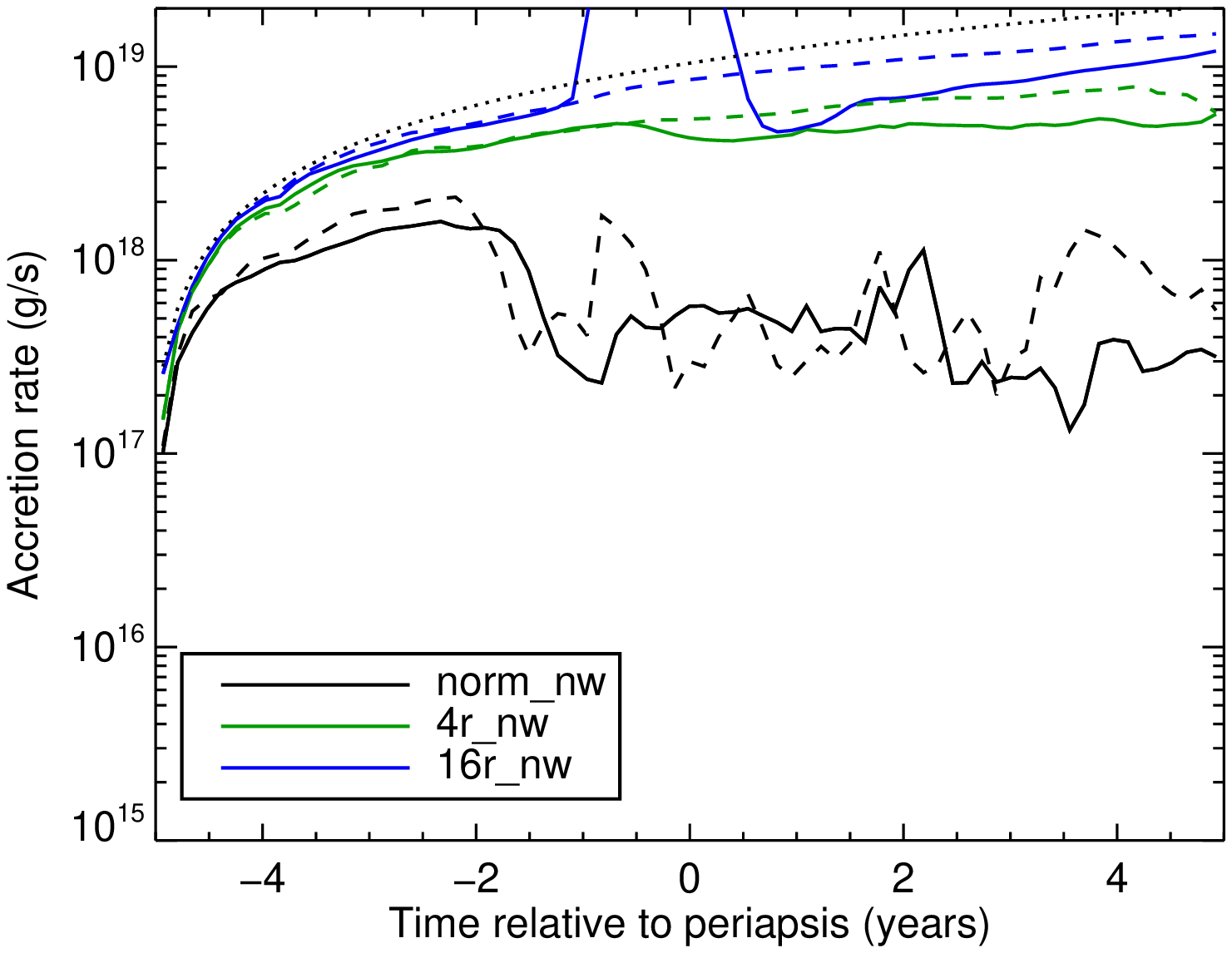}
\caption{
Comparison of total accretion rate in models with a cloud (solid lines) and without a cloud (dashed lines) with different values of $R_{acc}$. 
The accretion radius ranges from 1 to 16 times our normal accretion radius of $R_{acc} = 1.47\times10^{14}$~cm. 
Models with a wind background are on the left, those with a no wind background are on the right. 
In all cases, the overall accretion rate onto the black hole is similar whether or not a cloud is present in the simulation, except where a large $R_{acc}$ leads to an artificially high accretion rate near periapsis. 
For models with a large accretion radius, the accretion rate is generally higher at late times for the models without a cloud due to the cloud's passage disrupting the accretion stream from the wind material. 
For the no wind background models (right panel), the black dotted line represents the accretion rate for free-fall of background gas onto the black hole with no pressure support.   
The 16r\_nw\_nc accretion rate is only about $40\%$ below this free-fall accretion rate. 
}
\label{fig:cloud_v_nocloud_racc_size}
\end{figure*}

For each of the models in section~\ref{sec:Accretion_radius} with a different accretion radius, we also ran a model with the same setup but no cloud present, just a constant initial density.
Fig.~\ref{fig:cloud_v_nocloud_racc_size} compares simulations with and without a cloud for wind background (left panel) and no wind (right panel) simulations.
Solid lines are with a cloud, dashed lines are without a cloud.
For free-falling gas, starting from an initial constant density and zero velocity, and assuming the gas does not shock as it falls in, we would get a mass accretion rate of $\dot{M} = \frac{64}{3} \pi \rho_0 GMt$~g/s.  
This accretion rate is plotted as the black dotted line in fig.~\ref{fig:hot_comp}.

The total accretion rate is very similar between models with and without a cloud.  
The only exceptions are the 8r and 16r models, when part of the cloud passes inside the accretion radius.
Without a cloud present, the 16r\_nw (no wind) model comes close to the free-fall mass inflow rate.
The total accretion rate tends to be slightly higher for the models without a cloud, because the passage of the cloud partially disrupts the large scale accretion inflow of the background material.
This strongly indicates that, regardless of the absolute accretion rate onto Sgr $A^*$, G2 would not be expected to lead to a noticeable change in the accretion rate.

\subsubsection{Background structure \label{sec:background} }


To test how interactions between the cloud and background material effects accretion rate, we ran simulations with several different background structures: a hot background at $\sim10^8$~K (for the wind and no wind case), a background density 100 time normal (for wind and no wind cases), and for 2 background density profiles used in the simulations of \citet{anninos12}.

\begin{figure*}
\includegraphics[scale=.45]{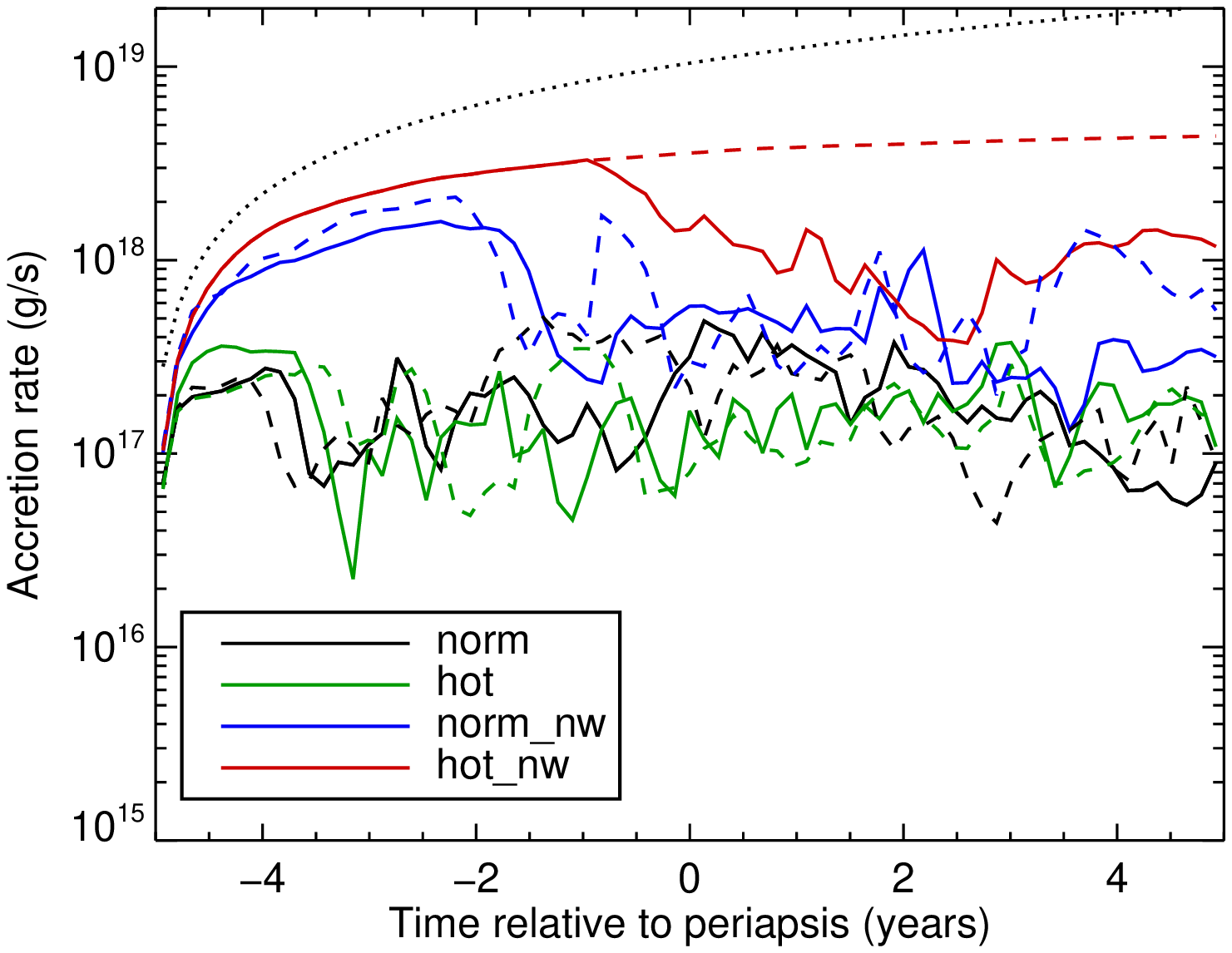}
\includegraphics[scale=.45]{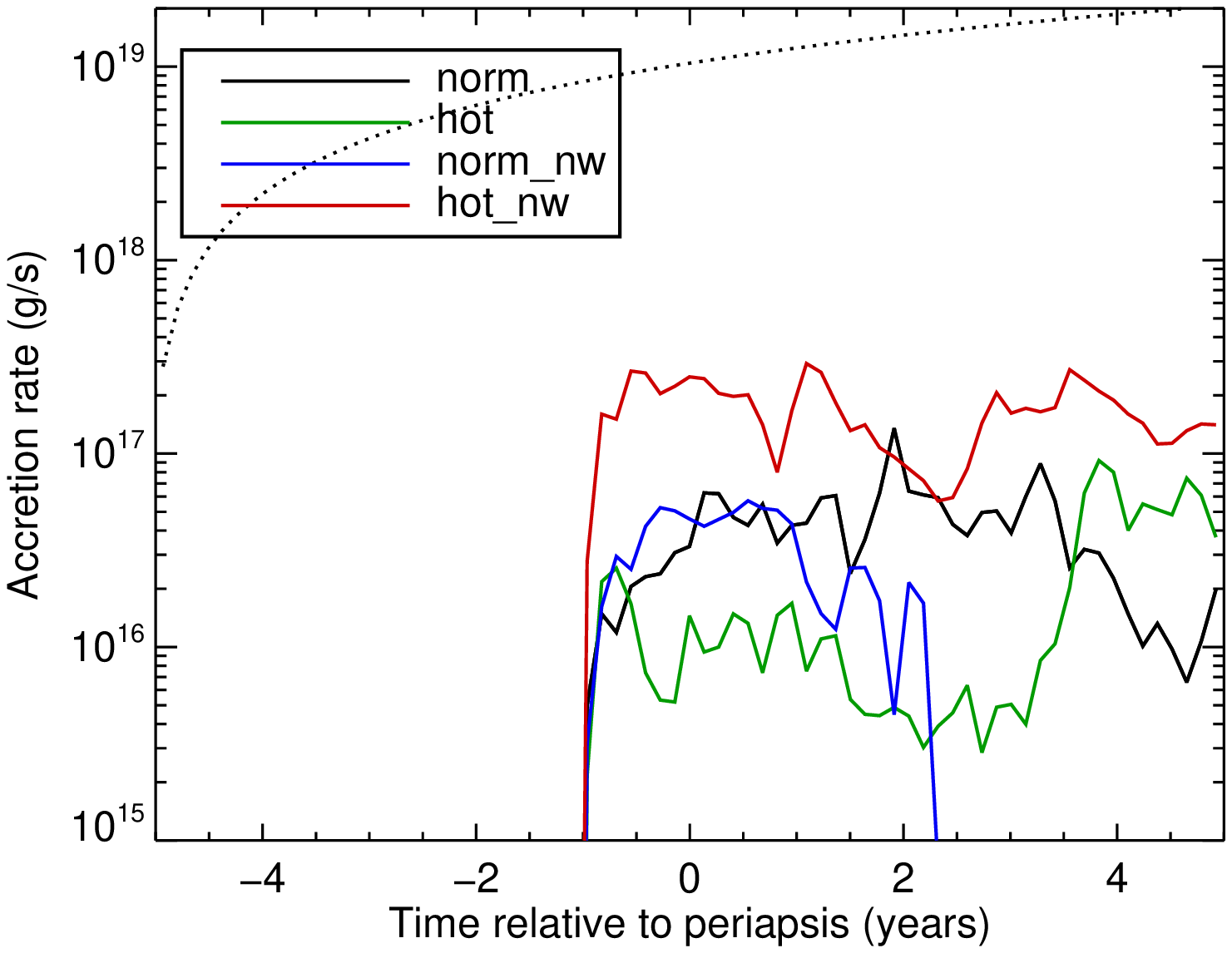}
\caption{
Comparison of accretion rate in models with standard background temperature of $10^{4}$~K (black and blue lines) and hot background temperature of $10^{8}$~K (green and red lines). 
Models with a wind background are black and green, those with a no wind background are blue and red. 
Solid lines are models with a cloud, dashed lines are for corresponding models with no cloud present. 
The black dotted lines represent the accretion rate for free-fall of the no wind background gas onto the black hole.   
For the wind background, the gas temperature makes very little difference to either the total accretion rate (left) or the cloud accretion rate (right). 
For the no wind background, the hot background (red lines) makes a substantial difference. 
In this case, the accretion inflow is stable, leading to a high, and increasing, accretion rate throughout the simulation if no cloud is present (dashed red). 
For the model with a cloud (solid red), the inflow is destabilized by the cloud 1 year before periapsis, and the total accretion rate drops to a level similar to the cold background model (blue lines). 
}
\label{fig:hot_comp}
\end{figure*}

We ran a set of simulations with ``hot'' backgrounds, with the same density as our normal backgrounds but a temperature of $1.2\times10^8$~K rather than $1.2\times10^4$~K, increasing the sound speed by a factor of 100.
Simulations were run for both a wind and no wind (initially static) background, with and without a cloud present.
The accretion rate vs time is plotted in fig.~\ref{fig:hot_comp}.

For a wind background velocity, the normal and hot background (black and green lines) show no significant differences in either the total or cloud only accretion rate.  
In both cases, the speed of the background is higher than the sound speed and dominates the accretion structure.
There is also very little difference between the cloud and no cloud cases (solid and dashed lines).
The Bondi-Hoyle-Lyttleton \citep{bondi1952} accretion rate, from the background only, would be about $1.5\times10^{19}$~g/s for the normal case and $8.8\times10^{18}$~g/s for the ``hot'' case, far above the actual values seen in our simulation, due to the instability of the accretion flow.

For the no wind backgrounds, with zero initial velocity of the background material, we see a different behavior between the normal and ``hot'' models (blue and red lines).
For a motionless, spherically symmetrical background, the most straightforward expectation is for a spherical Bondi flow to develop.
Our simulations do not run long enough to reach a stable Bondi rate, but if the system were evolving towards a Bondi flow, the accretion rate should follow the free-fall inflow rate.
For an adiabatic gas with an adiabatic index of $\Gamma = 5/3$, a Bondi flow is characterized by the inflowing gas never being shocked and a radial density profile of $\rho \propto r^{-3/2}$ \citep{bondi1952}.
For free-falling gas, starting from an initial constant density and assuming the gas does not shock as it falls in, we would get a density profile of $\rho \propto r^{-3/2} t$, where $t$ is the simulation time, and a mass accretion rate of $\dot{M} = \frac{64}{3}\pi \rho_0 GMt$~g/s.  
This accretion rate is plotted as the black dotted line in fig.~\ref{fig:hot_comp}.

However, in none of our simulations do we get a situation resembling a Bondi or free-fall inflow.
For our normal (cold) background simulations (blue lines), we instead have an convectively unstable, shocked inflow with a greatly reduced and approximately constant, but variable, accretion rate.
The onset of the inflow becoming unstable can be seen at about 2 years before periapsis, when the accretion rate changes from being slowly rising to above $10^{18}$~g/s to a highly variable rate around $5\times10^{17}$~g/s, even with no cloud present (blue dashed line).

The no wind case, however, has a stable inflow in the absence of a cloud (red dashed line).
This inflow is not at the free-fall inflow rate.
Instead, the inflow is shocked but does not become convectively unstable, resulting in a stable shocked inflow that is spherically symmetric.
The difference between these stable and unstable states is the zero angular momentum analogue to the difference between a convectively stable and unstable ADAF disk inflow.
See, for example, \citet{narayan_yi1994}, \citet{blandford_begelman1999}, and \citet{quataert_gruzinov2000} for analytic discussion, and \citet{stone1999}, \citet{igumenshchev1999}, and \citet{pang2011} for simulation results.

In the stable shocked state, the inflow has a density profile at a give time of $\rho \propto r^{-1}$, and the accretion rate increases as about $\dot{M} \propto t^{1/3}$.
In either the stable or convective state, the sound speed of the gas is close to $c_{s} = \sqrt{\frac{2}{3} \frac{GM}{r}}$~cm/s, the expected value for a hydrostatic atmosphere.

When the cloud is present in the hot, no wind simulation (solid red line), the cloud passage destabilizes this shocked stable inflow at 1 year before periapsis, and pushes into the convectively unstable state.
The accretion rate decreases to a value to that seen in the normal background cases (blue lines).
The change of accretion state results in much less overall material being accreted when the cloud is present ($5.87\times10^{26}$~g rather than $1.00\times10^{27}$~g).
This large difference is not seen in the normal background simulations, but it is due only to the change from a stable to unstable accretion inflow, not to a fundamental difference in the behavior of the cloud between the normal and hot background simulations.

The initial shocked stable state of the background in the hot, no wind simulations results in a higher density surrounding the black hole when the cloud passes periapsis than in the normal, no wind simulation.  This results in more dissipation and a higher rate of cloud material accretion (red line vs. blue line in fig.~\ref{fig:hot_comp}, right panel).  The relative amount of cloud material is still fairly small, only $16\%$ of the total material accreted after periapsis (see Table~\ref{table:models}).

The stable shocked inflow will only develop if the background is perfectly spherically symmetric and has no bulk velocity or angular momentum relative to the black hole.  
This not a realistic expectation for the gas surrounding Sgr $A^*$.
Thus, we do not expect an order of magnitude drop in the accretion rate onto Sgr $A^*$ due to G2.

\begin{figure*}
\includegraphics[scale=.45]{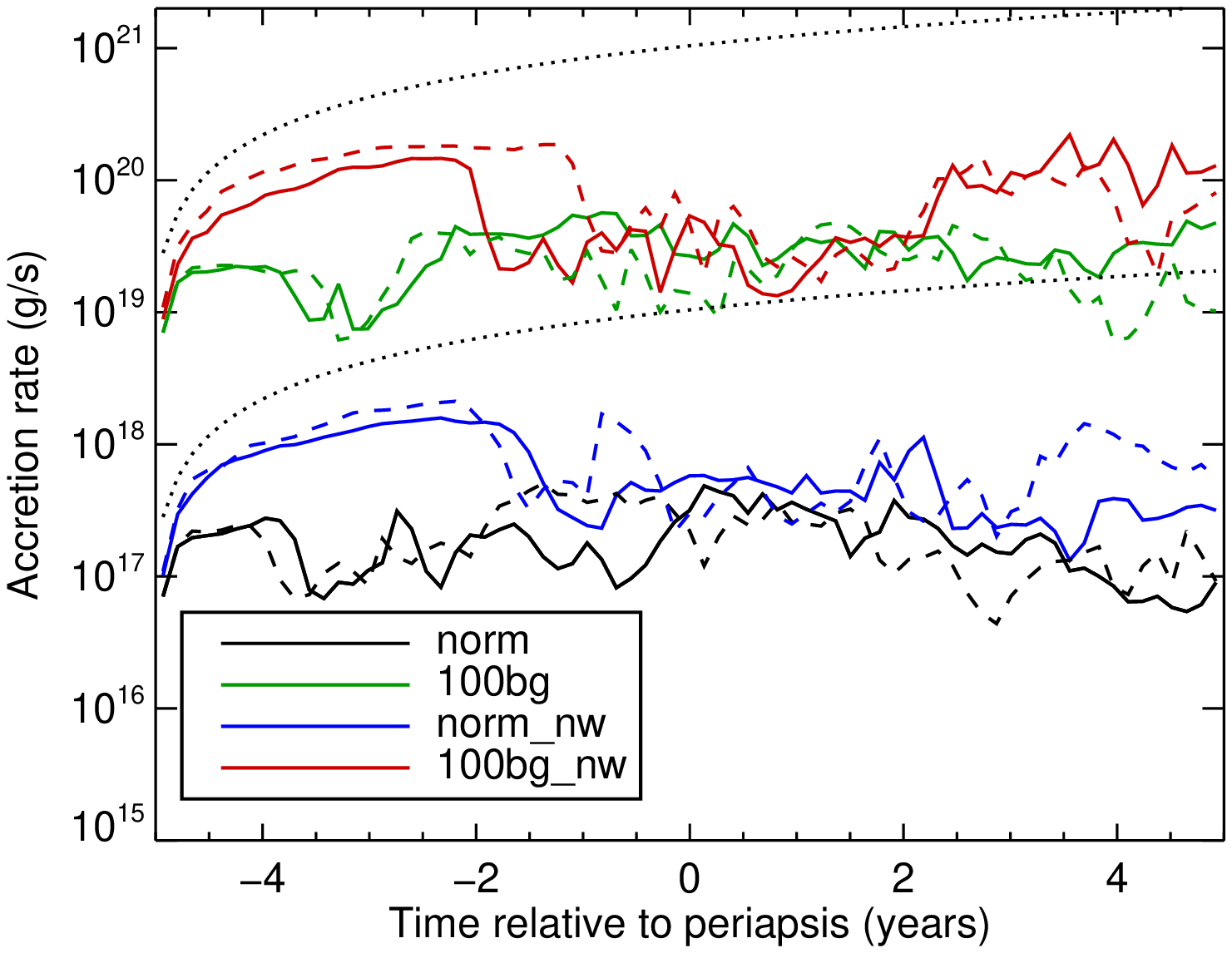}
\includegraphics[scale=.45]{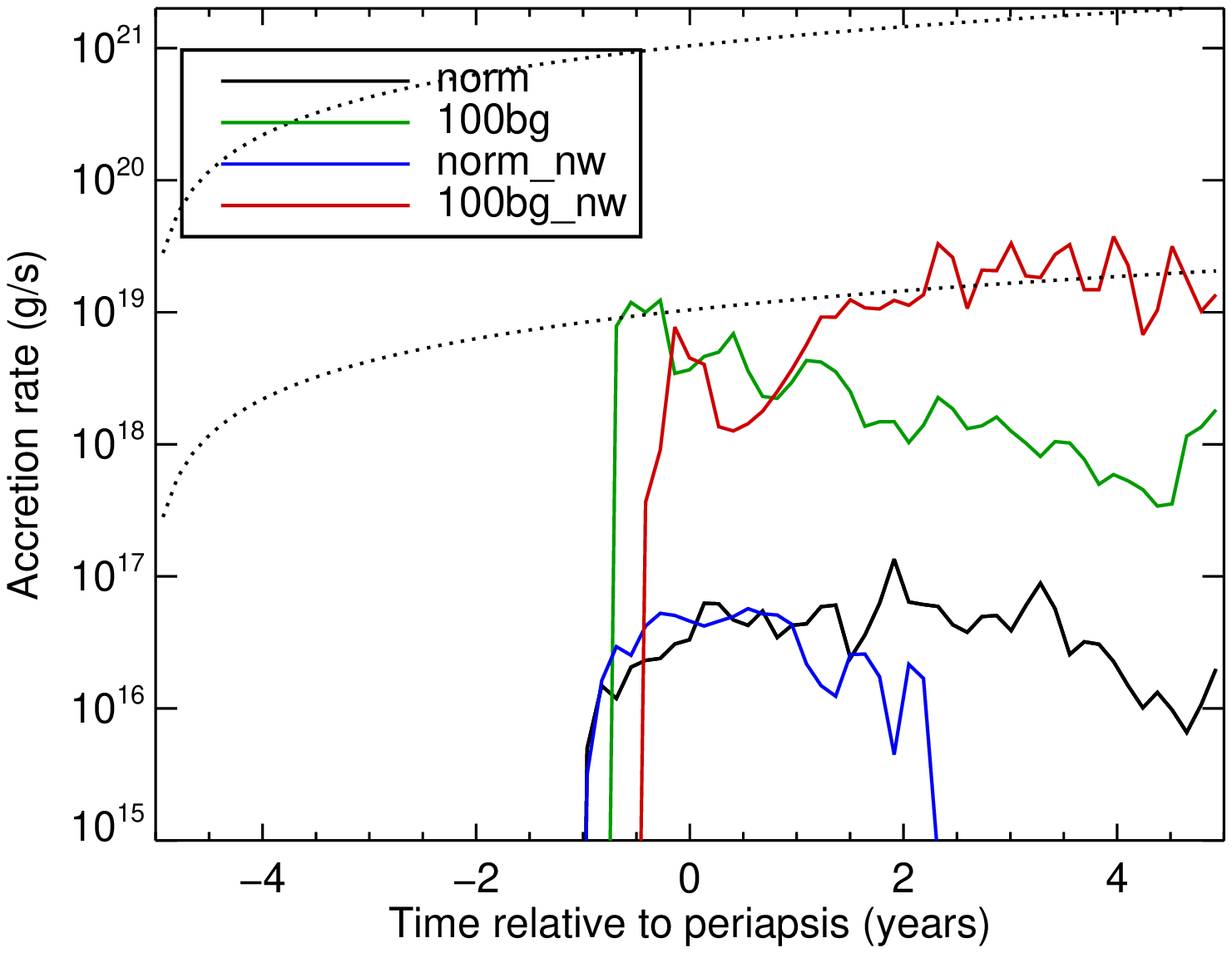}
\caption{
Comparison of accretion rate in models with standard background density (black and blue lines) and 100 times the background density (green and red lines). 
Models with a wind background are black and green, those with a no wind background are in blue and red. 
Solid lines are simulations with a cloud, dashed lines are with no cloud present.
The black dotted lines represent the accretion rate for free-fall of the no wind background gas onto the black hole for standard (lower) and 100 times (upper) density.   
The total accretion rate (left) increases by a factor of 155 for the wind and 115 for the no wind models. 
The amount of cloud material accreted (right) increases by slightly smaller factors of 70 times for the wind background and a larger factor of 700 times for the no wind background (due to the higher late-time accretion rate). 
Even with the increased dissipation of the cloud due to interaction with a denser background, cloud accretion remain a sub-dominant component.   
Overall, less than $10\%$ of the total cloud mass is accreted in the high background cases by 5 years after periapsis. 
}
\label{fig:100bg_comp}
\end{figure*}

Simulations with a background density 100 times the normal density of $1.67\times10^{-23}$~g~cm$^{-3}$, for wind and no wind background models, are shown in fig.~\ref{fig:100bg_comp}.
The total and cloud-only accretion rate both increase dramatically for the high background density simulations.
The total accretion rate (left) increases by a factor of 155 for the wind and 115 for the no wind models with clouds (solid lines). 

In all these cases, the the accretion flow is convectively unstable whether or not a cloud is present, leading to very similar total accretion rates between the models with a cloud (solid lines) and without a cloud (dashed lines).
Note that the typical late-time accretion rate of around $3\times10^{19}$~g/s ($5\times10^{-7}$~M$_{\sun}$/year) exceeds the upper end of expected mass accretion rate for Sgr $A^*$ of $1.3\times10^{19}$~g/s ($2\times10^{-7}$~M$_{\sun}$/year) \citep{marrone07}.
This high accretion rate could indicates that this background density is too high, or that the accretion rate in the simulations is too high due to numerical effects, e.g. a lack of resolution and overly large accretion radius.
As \citet{anninos12} point out, the accretion rate of material reaching the black hole is not necessarily the same as the accretion rate measured at 100's of gravitational radii.

The amount of cloud material accreted (right) increases by slightly smaller factors of 70 times (vs. 155) for the wind background and a larger factor of 700 times (vs. 115) for the no wind background, due to the higher late-time accretion rate. 
Although there is increased dissipation of the cloud due to interaction with a denser background, cloud accretion remain a sub-dominant component.   
Even in the 100bg\_nw case, only $10\%$ of the total cloud mass is accreted by 5 years after periapsis, and only 18\% of the total mass accreted after periapsis is from the cloud.

We also carried out simulations with the structured, initially static background densities used by \citet{anninos12}.
The first, ``bg1'', has a structure of $\rho = 1.3\times10^{-21} (r/r_0)^{-1}$~g~cm$^{-3}$ and $T = 1.2\times10^{8} (r/r_0)^{-1}$~K, where $r_0 = 10^4 R_s = 1.3\times10^{16}$~cm.
This background structure was used by \citet{burkert12}, \citet{schartmann12}, and \citet{anninos12}.
The second background, ``bg2'' has a structure of $\rho = 1.3\times10^{-21} (r/r_0)^{-1.125}$~g~cm$^{-3}$ and $T = 10^{8} (r/r_0)^{-0.75}$~K.
This background profile was designed by \citet{yuan2003} to match {\it Chandra} X-ray observations, and also used in \citet{anninos12}.

\begin{figure*}
\includegraphics[scale=.45]{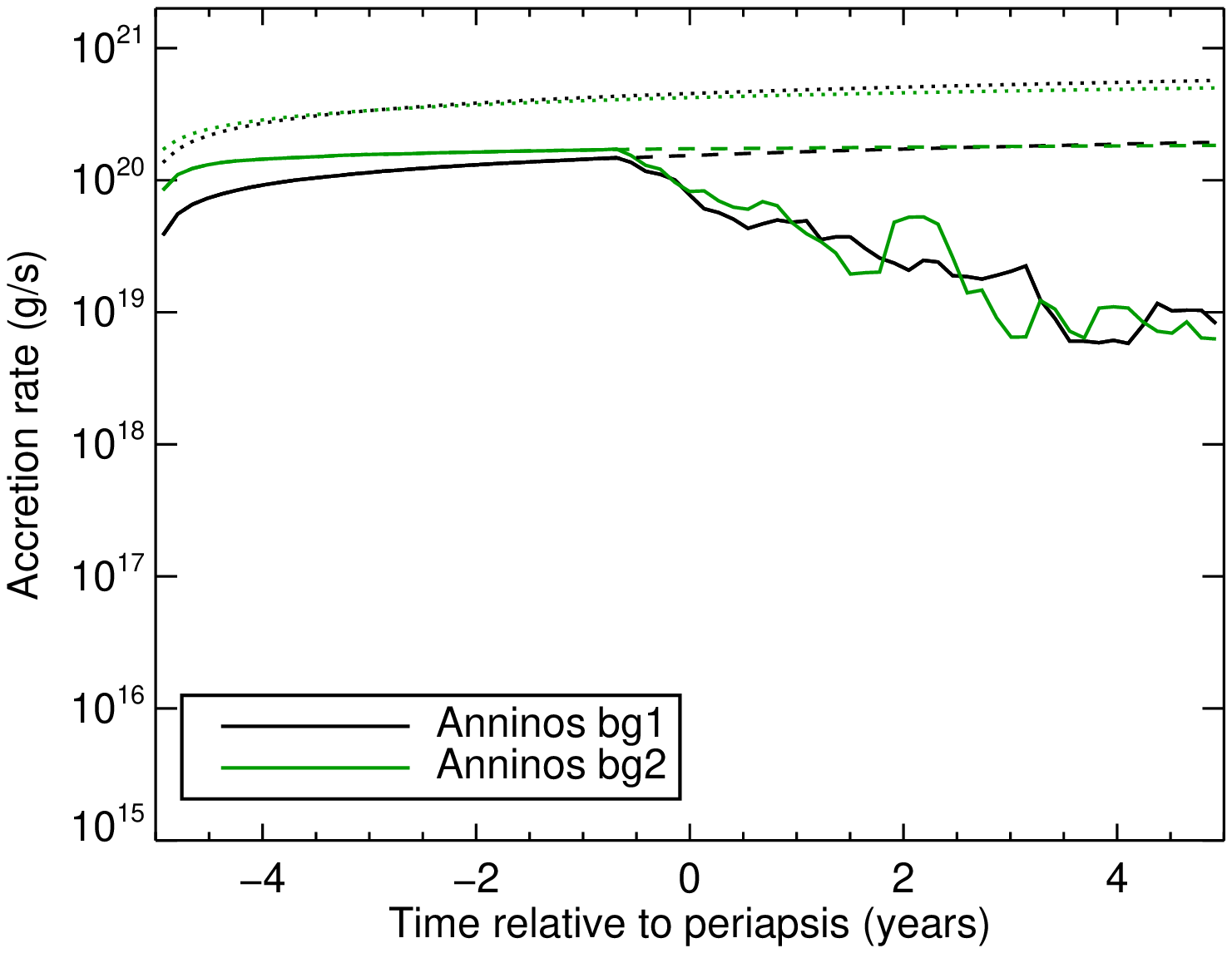}
\includegraphics[scale=.45]{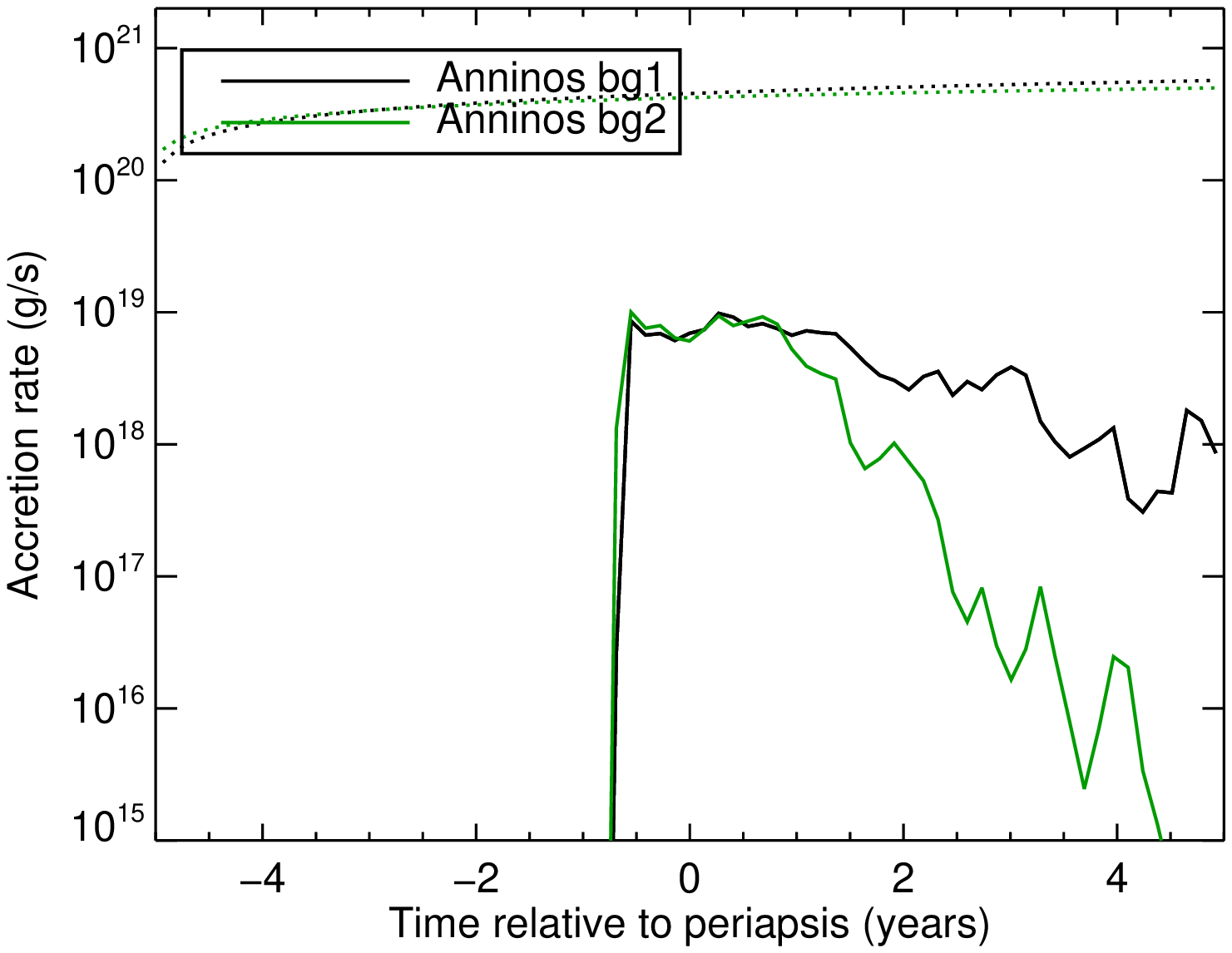}
\caption{
Comparison of accretion rate in models with bg1 (black) and bg2 (green) from \citet{anninos12}. 
Models with a cloud are solid lines, those with no cloud are dashed lines. 
The dotted lines represent the accretion rate for free-fall of the gas onto the black hole for bg1 (black) and bg2 (green).   
In the absence of a cloud, both these background create a stable accretion flow with a slowly increasing accretion rate (left). 
The passage of the cloud destabilizes the accretion flow about 1 year before periapsis, pushing it into a convective regime. 
This leads to a reduction of the total accretion rate for the cloud models by a factor of $>10$ by 5 years after periapsis. 
As a result, the total amount of material accreted over the 10 years of the simulation being reduced by a factor of 2 when a cloud is present. 
}
\label{fig:bg1bg2_comp}
\end{figure*}

The accretion rates for our simulations using these backgrounds are shown in fig.~\ref{fig:bg1bg2_comp}.
Solid lines are simulations with a cloud, dashed lines are simulations with no cloud present.
The dotted lines are the free-fall inflow rate for ``bg1'' (black) and ``bg2'' (green).

With these two backgrounds, the initial accretion flow is shocked such that the sound speed has close to the hydrostatic profile of $c_{s} = \sqrt{\frac{2}{3} \frac{GM}{r}}$~cm/s.
The inflow remains in a spherically symmetric shocked stable state, similar to the behavior of the ``hot'' no wind background (above).
When a cloud is present, the inflow is destabilized about 1 year before periapsis and pushed into the convectively unstable state, leading to a reduction in the overall accretion rate by a factor of $>10$.
This results in the total amount of material accreted over the 10 years of the simulation being reduced by a factor of 2 when the cloud is present.
Similar to all our other simulations, the accretion rate of cloud material (right panel of fig.~\ref{fig:bg1bg2_comp}) is much less than the total accretion rate, accounting for $15\%$ and $7\%$ of the mass accreted after periapsis.
These results are consistent with \citet{anninos12}, using these same background models, both in terms of the cloud accretion rate and their finding the cloud accretion would equal $1$\% - $5$\% of the background accretion rate.

Overall, altering the temperature, density, and structure of the background {\bf does not} change the relative amount of background and cloud material accreted.
The accretion rate scales roughly with the background density.
The only large differences are seen when a background results in a shocked stable inflow, as in the case of the ``hot'' no wind background, ``bg1'', and ``bg2''.
In these cases, the passage of the cloud destabilizes the spherically symmetric inflow and changes it to a convectively unstable inflow state with a much lower accretion rate.
However, there is no reason to believe that the actual accretion inflow around Sgr~$A^*$ is stable or perfectly spherically symmetric, so this large state change is unlikely to be reflective of reality.
Note that the absolute accretion rates in our simulations are highly dependent on the specific numerical parameters of the setups used, and should not be taken as a guide to determining the real density profile around Sgr $A^*$.


\subsubsection{Radiative cooling}

All of our simulations include optically thin radiative cooling, with a cooling curve following \citet{sutherland93}.
Although the shortest cooling times are short compared to the orbital period (400 years), they are still long compared to the simulation time step (hours) and at most comparable to the total length of the simulations (10 years).
Therefore, the gas everywhere behave nearly adiabatically.

To verify that cooling is not significant contributor to accretion rates, we run one simulation with cooling disabled.  
This simulation is otherwise identical to our norm simulation with a wind background.


The total and cloud-only accretion rates are similar to an identical setup with cooling.
The amount of cloud material accreted is higher by $20\%$ with cooling turned on, suggesting that cooling aids accretion.
The total amount of material accreted, however, decreases slightly with cooling turned on in the cloud model, and increases slightly in the no cloud model.
Given the stochastic nature of the accretion seen in our simulations, these changes in cloud and total accretion rates are likely not significant.

\subsection{Long-term evolution}
\label{sec:long-term}

To examine the long-term evolution of accretion onto the black hole, we carry out two long duration simulations.
One is our normal simulation, carried out to 11 years after periapsis.
The other is the same setup with 1/2 the resolution, expect within $1.87/times10^{15}$~cm of the black hole, where we use our normal resolution and standard accretion radius ($1.47\times10^{14}$~cm).
This simulations is run to 30 years after periapsis.

\begin{figure*}
\includegraphics[scale=.45]{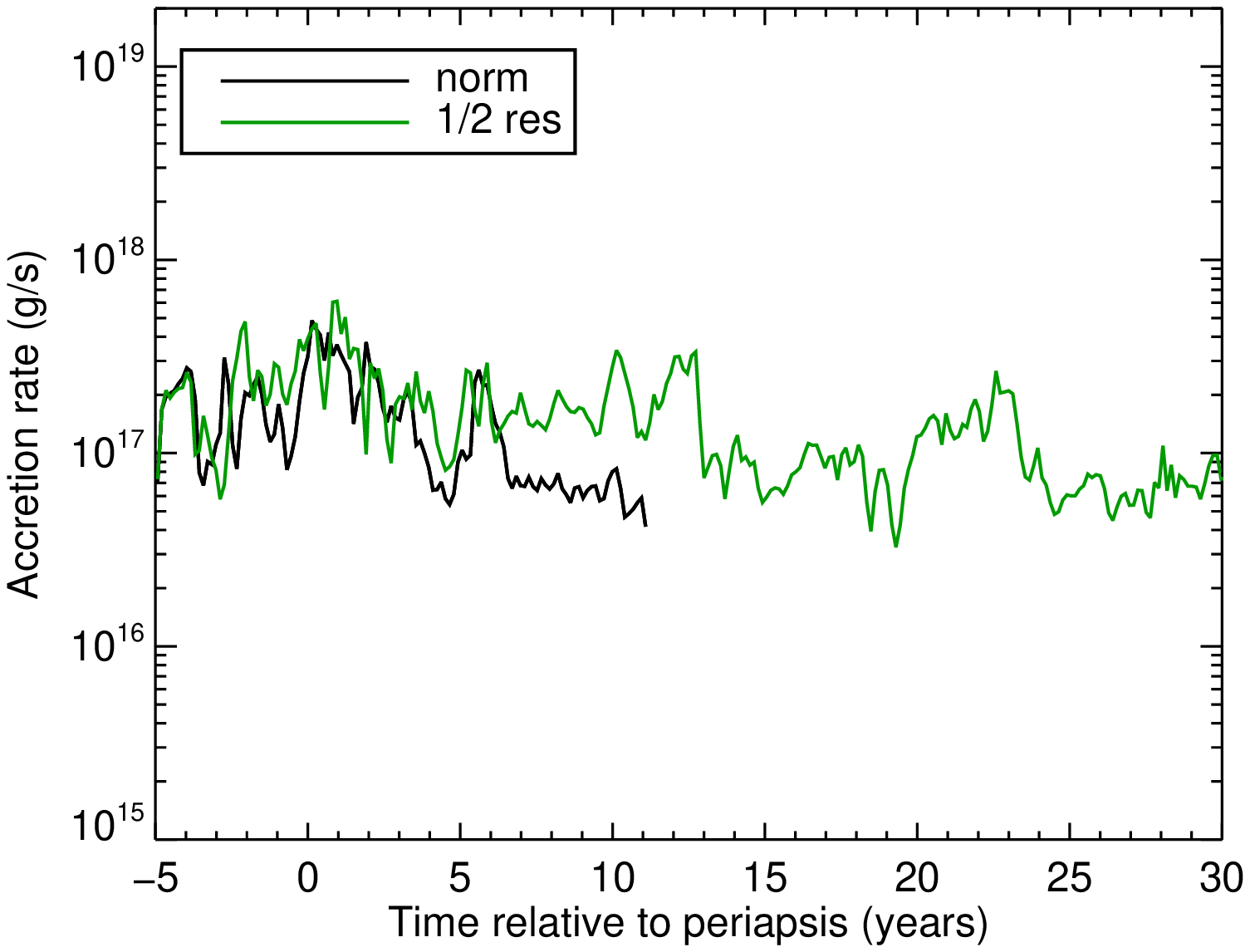}
\includegraphics[scale=.45]{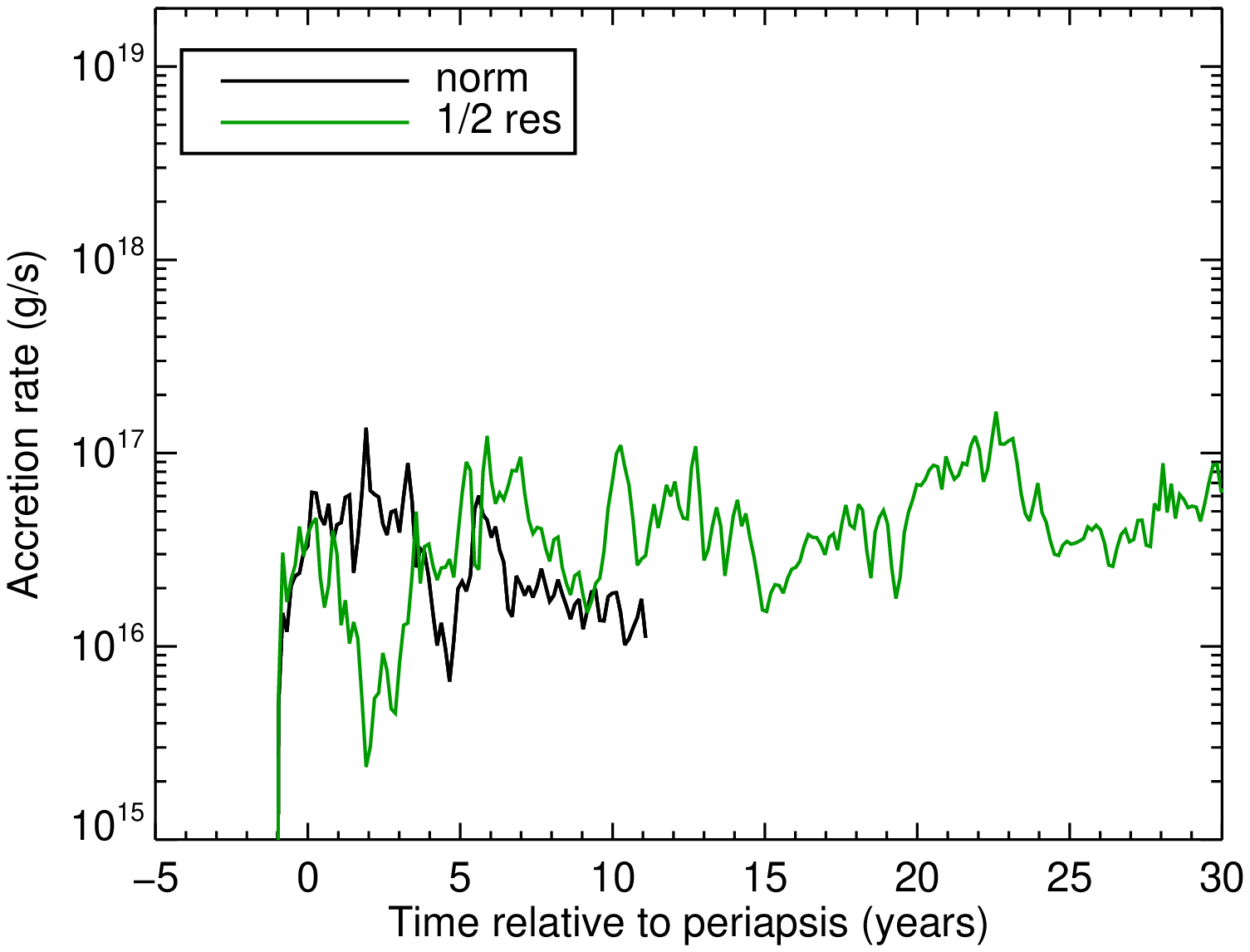}
\caption{
Long term accretion rate in models with normal cloud density profiles for regular resolution (black) and 1/2 resolution in the cloud (green).
Both simulations are carried out with a wind background model.
The total accretion rates (left) and cloud accretion rates (right) are similar for the two models out to 11 years after periapsis, when the regular resolution simulated was terminated.  
A brief increase in accretion rate, by about a factor of two, is seen near periapsis, but this likely a stochastic variation in the (unstable) background accretion, not directly attributable to the cloud.
However, there is no long-term increase in accretion rate, even out to 30 years after periapsis in the 1/2 resolution model.
At 30 years after periapsis, only $0.2\%$ of the cloud has been accreted.
}
\label{fig:long_comp}
\end{figure*}

Fig.~\ref{fig:long_comp} shows the total and cloud only mass accretion rate for the two simulations.  
The accretion rates are similar overall out to the end of the full resolution simulation at 11 years after periapsis.
Even out to 30 years, the overall accretion rate remains low; peaking at several $\times 10^{17}$~g/s and falling to about $10^{17}$~g/s by 10 to 15 years after periapsis.

At late times, the cloud accretion (right panel) can dominate the total accretion.
The cloud accretion rate remains consistently between $2\times10^{16}$~g/s and $10^{17}$~g/s for the entire 30 years after periapsis, with no significant long-term increase.
By 30 years after periapsis, the total amount of cloud material accreted is only $4.8\times10^{25}$~g, $0.2\%$ of the cloud mass.
Based on this simulation, we find it unlikely that G2 will lead to a noticeable increase in accretion onto Sgr $A^*$ over the next several decades.

For the long duration normal resolution simulation we also included tracer particles in the cloud and are able to follow the evolution of the orbits of individual particles.
4500 particles are initially placed in the cloud, distributed such that each particles traces and equal amount of mass.
%
At each time step, we use the position and velocity information of each tracer particle to calculate the orbital parameters of that particle as if it were on an independent Keplerian orbit around the black hole.  
These orbital parameters (e.g. orbital period, eccentricity, semi-major axis, periapsis distance) will change with time due to hydrodynamic interactions.
We can then examine how the cumulative distribution and overall average of these orbital parameters change with time to gain insight into how cloud material become available for accretion.

\begin{figure*}
\includegraphics[scale=.45]{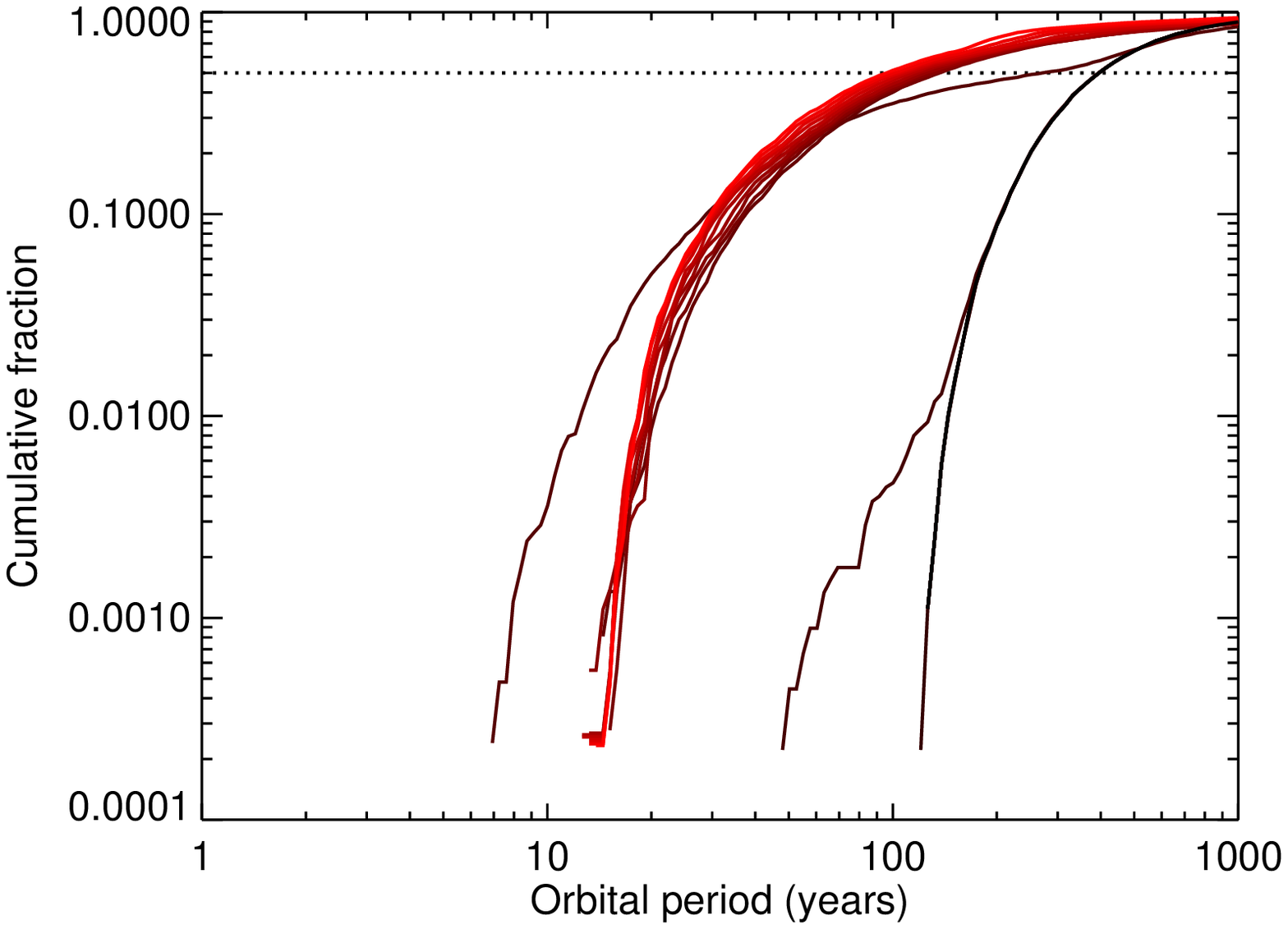}
\includegraphics[scale=.45]{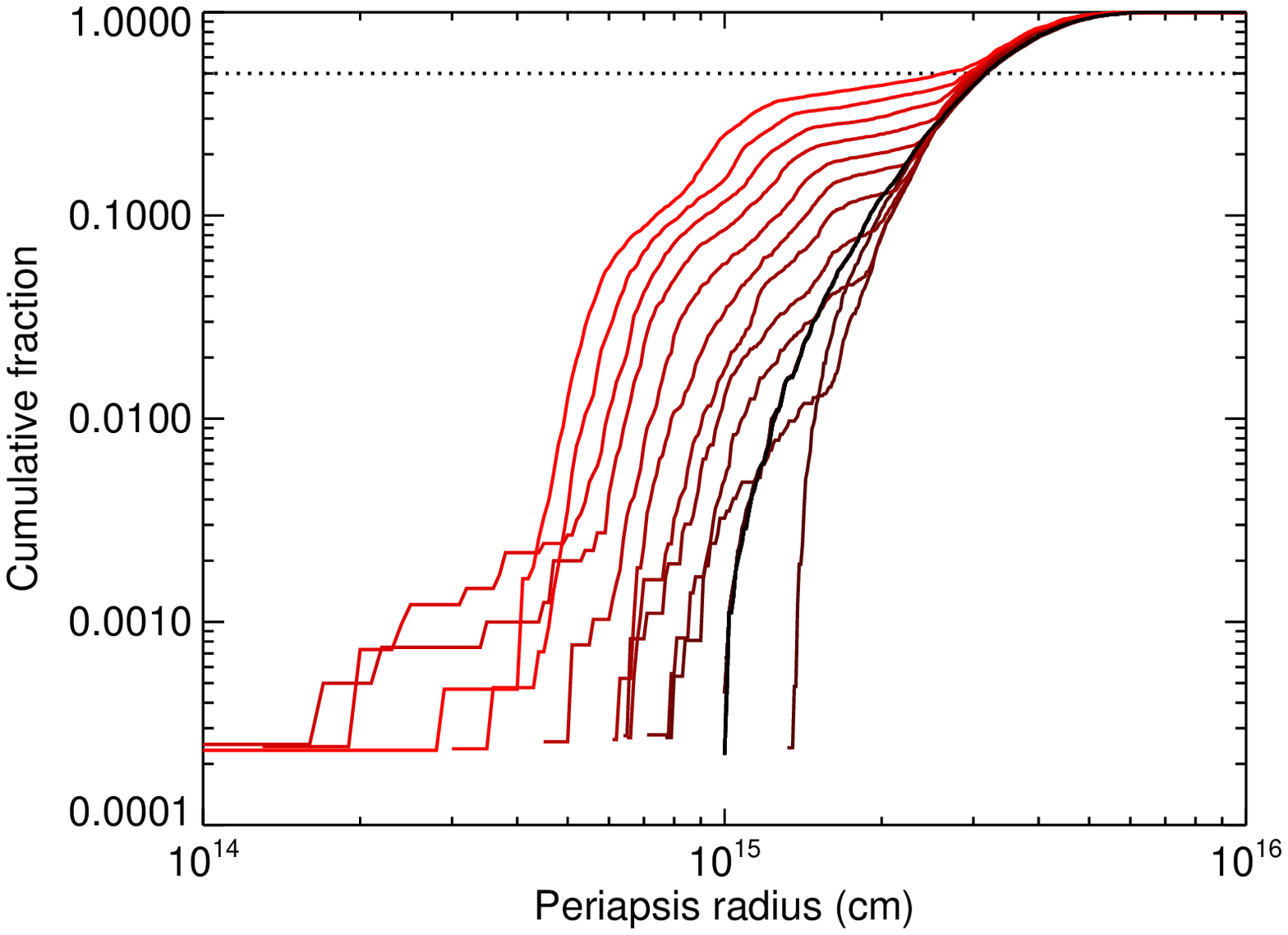}
\caption{
Cumulative distribution of orbital period (left) and periapsis radius (right) of cloud material from 5 years before (black) to 10 years after (bright red) cloud periapsis.
Distributions are plotted at one year intervals proceeding from black to red with time.
Dotted line is at 0.5 (median value).
The orbital period undergoes a rapid change as the cloud passes periapsis, jumping from a median period of 400 years and minimum value of 120 years to a median of 140 years and minimum of 14 years.
There is a corresponding decrease in the eccentricity of the orbits at this time.
However, the periapsis distance (right) does not change rapidly.  
It is stable until periapsis, after which there is a slow evolution inward due to interaction with the background material.
By 10 years after the initial periapsis only $1\%$ of cloud material has a periapsis radius of less than $5\times10^{14}$~cm.
Even at this time, only 1 tracer particle, representing $1/4500$th of the cloud, is within our normal accretion radius.
}
\label{fig:tracer_cumulative}
\end{figure*}

Fig.~\ref{fig:tracer_cumulative} shows the cumulative distribution of orbital period and periapsis radius of the tracer particles for each year of the simulation.
The orbital period undergoes a rapid change as the cloud passes periapsis, jumping from a median period of 400 years and minimum value of 120 years to a median of 100 years and minimum of 14 years.
There is a corresponding decrease in the eccentricity of the orbits at this time.
However, the periapsis distance (right) does not change rapidly.  
It is stable until periapsis, after which there is a slow evolution inward due to interaction with the background material.
By 10 years after the initial periapsis, $1\%$ of cloud material has a periapsis radius of less than $5\times10^{14}$~cm.
Even at this time, only 1 tracer particle, representing $1/4500$th of the cloud, is within our normal accretion radius.

\begin{figure*}
\includegraphics[scale=.45]{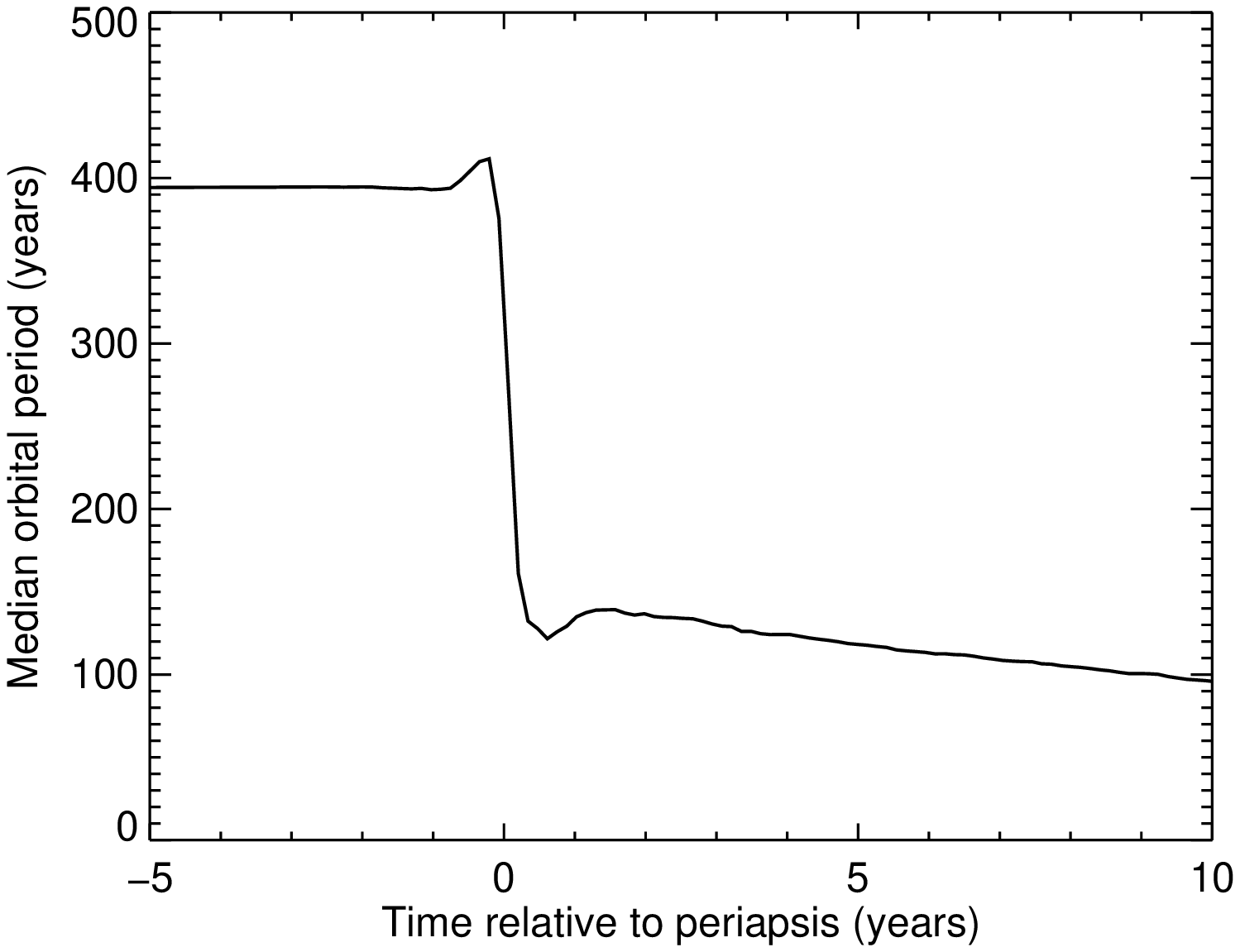}
\includegraphics[scale=.45]{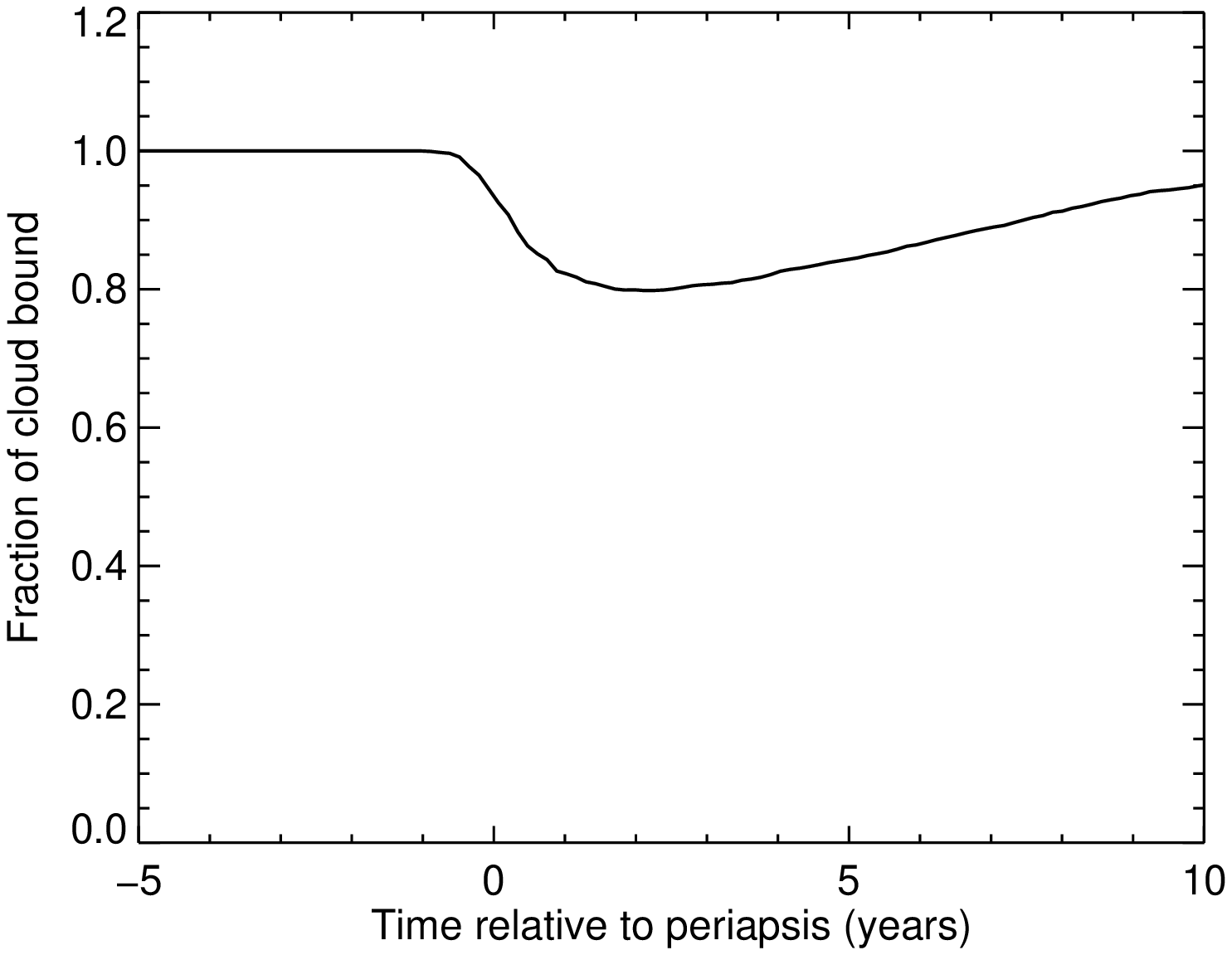}
\caption{
Median orbital period of cloud material vs. time (left) and fraction of cloud mass bound to black hole (right).
Near periapsis, the median period drops from 400 years to 140 years.
Also at this time, about $20\%$ of the cloud becomes unbound.
Most of the unbound cloud material again becomes bound due to drag from the background material, resulting in $95\%$ of cloud material being bound at 10 years after periapsis.
}
\label{fig:tracer_median}
\end{figure*}

Fig.~\ref{fig:tracer_median} shows the median orbital period (left panel) and fraction of cloud mass bound to the black hole (right panel).
This shows clearly the drop in median orbital period from 400 years to 140 years near periapsis.

At periapsis, about $20\%$ of the cloud becomes unbound, accounting for the reduction in orbital period.
Most of the unbound cloud material again becomes bound due to drag from the background material, resulting in $95\%$ of cloud material being bound at 10 years after periapsis.
Drag from background material is also responsible for the gradual decrease in median orbital period after periapsis.

The tracer particle evolution indicates that most of the cloud remains on long, highly elliptical orbits.
Only about $10\%$ of the cloud material has an orbital period less than $30$~years, and even on the second pass through periapsis very little material will pass close enough to be accreted.
This is consistent with the low accretion rate seen 30 years after periapsis in the very long duration simulation.

\subsection{Cloud density profiles}
\label{sec:cloud_density_profiles}

To examine the effect of the initial cloud density profile, we compare models with 4 different density profiles (``norm'', ``extended'', ``flat'', ``rsq'') and one with no cloud present (``no cloud''), which are otherwise identical.
With these models, we can determine how cloud structure impacts accretion rate and the observable properties of cloud emission, such as Br-$\gamma$ line properties and bolometric and X-ray luminosity.
The models can then be compared to real observations to constrain the nature of the G2 object.

\begin{figure*}
\includegraphics[scale=.45]{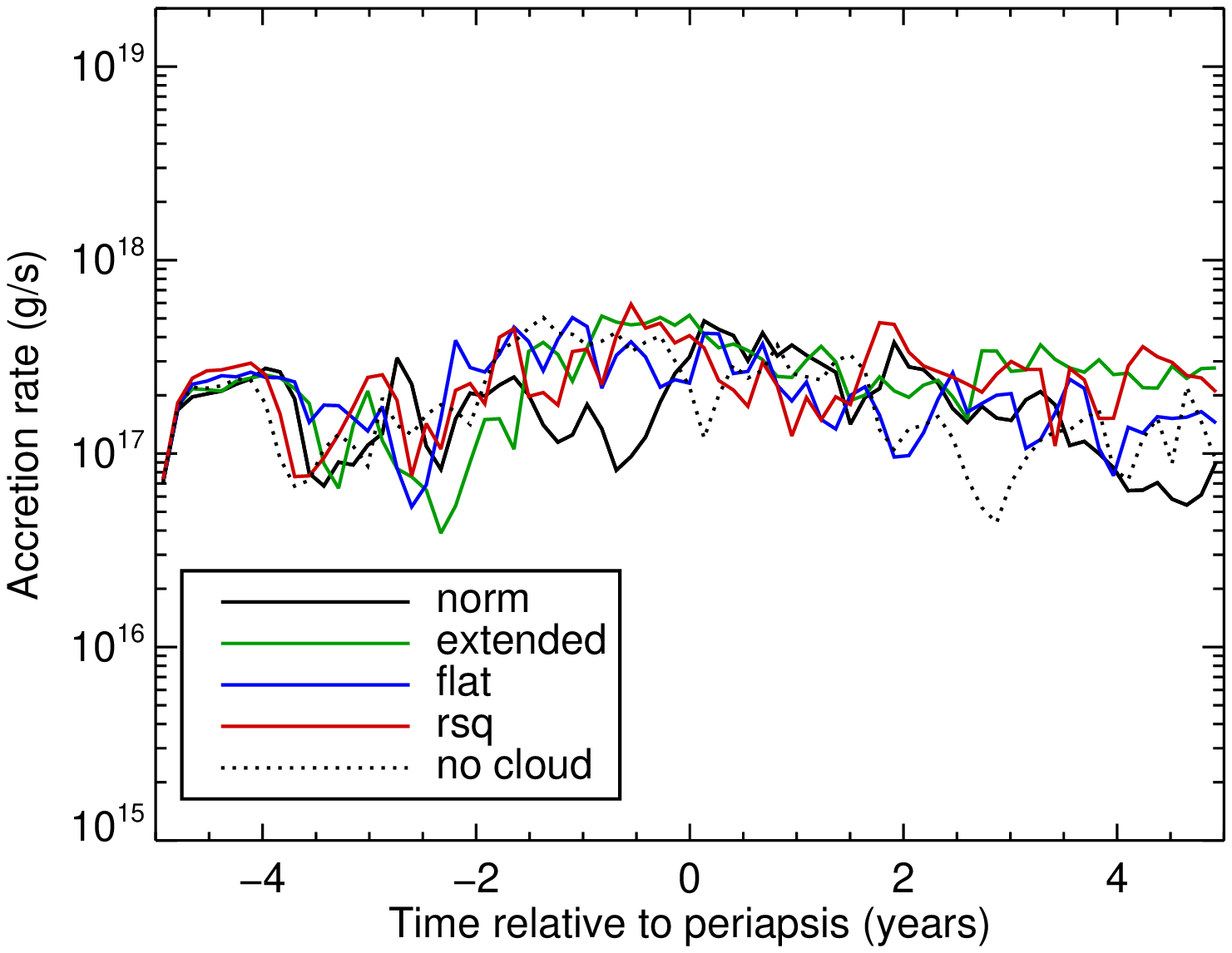}
\includegraphics[scale=.45]{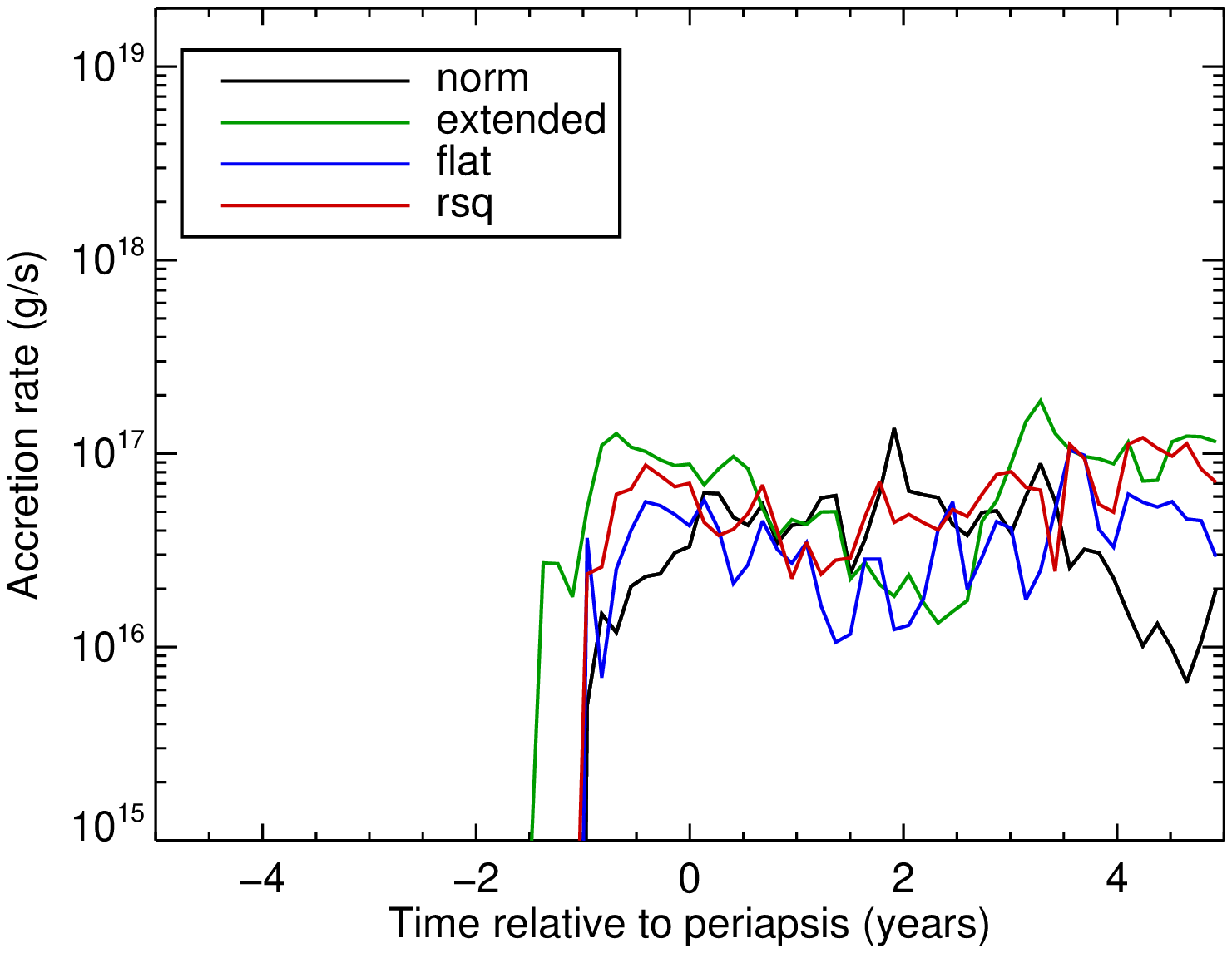}
\caption{
Comparison of total accretion rate in models with normal Gaussian (black), extended Gaussian (green), flat (blue), and $r^{-2}$ (red) cloud density profiles, and a simulation with no cloud present (dashed black).
All simulations are carried out with a wind background model.
The total accretion rates (left) are similar for all models, with a somewhat higher rate at late times for the ``extended'' and ``rsq'' models.
A brief increase in total accretion rate, by about a factor of two, is seen near periapsis for the ``norm'' model.
However, there are increases similar in magnitude seen in other models both before and after periapsis, including the simulation with no cloud present.
Cloud material begins to accrete earlier for the ``extended'' cloud models (right), at 1.5 years before periapsis rather than 1 year before.
In all cases, the amount of cloud material accreted is small, making up less than $50\%$ of the material being accreted at any time.
The overall amount of material cloud material accreted is small.
In the most extreme case, the ``extended'' model, only $0.06\%$ of the total cloud mass has been accreted by 5 years after periapsis.
}
\label{fig:profiles_comp}
\end{figure*}

The total and cloud-only accretion rate for the 5 models are plotted in fig~\ref{fig:profiles_comp}.
The accretion rate are variable but not significantly different overall for all models.
The total accretion rates (left panel) are somewhat higher at late times for the ``extended'' and ``rsq'' models.
A brief increase in total accretion rate, by about a factor of two, is seen near periapsis for the ``norm'' model.
However, there are increases similar in magnitude seen in other models both before and after periapsis, including the simulation with no cloud present.
Looking at the cloud material accreted (right panel), the ``extended'' cloud begins to accrete earlier, at 1.5 years before periapsis rather than 1 year before, due to the larger physical size of the cloud.
In all cases, the amount of cloud material accreted is small, making up less than $50\%$ of the material being accreted at any time.

\begin{figure*}
\includegraphics[scale=.45]{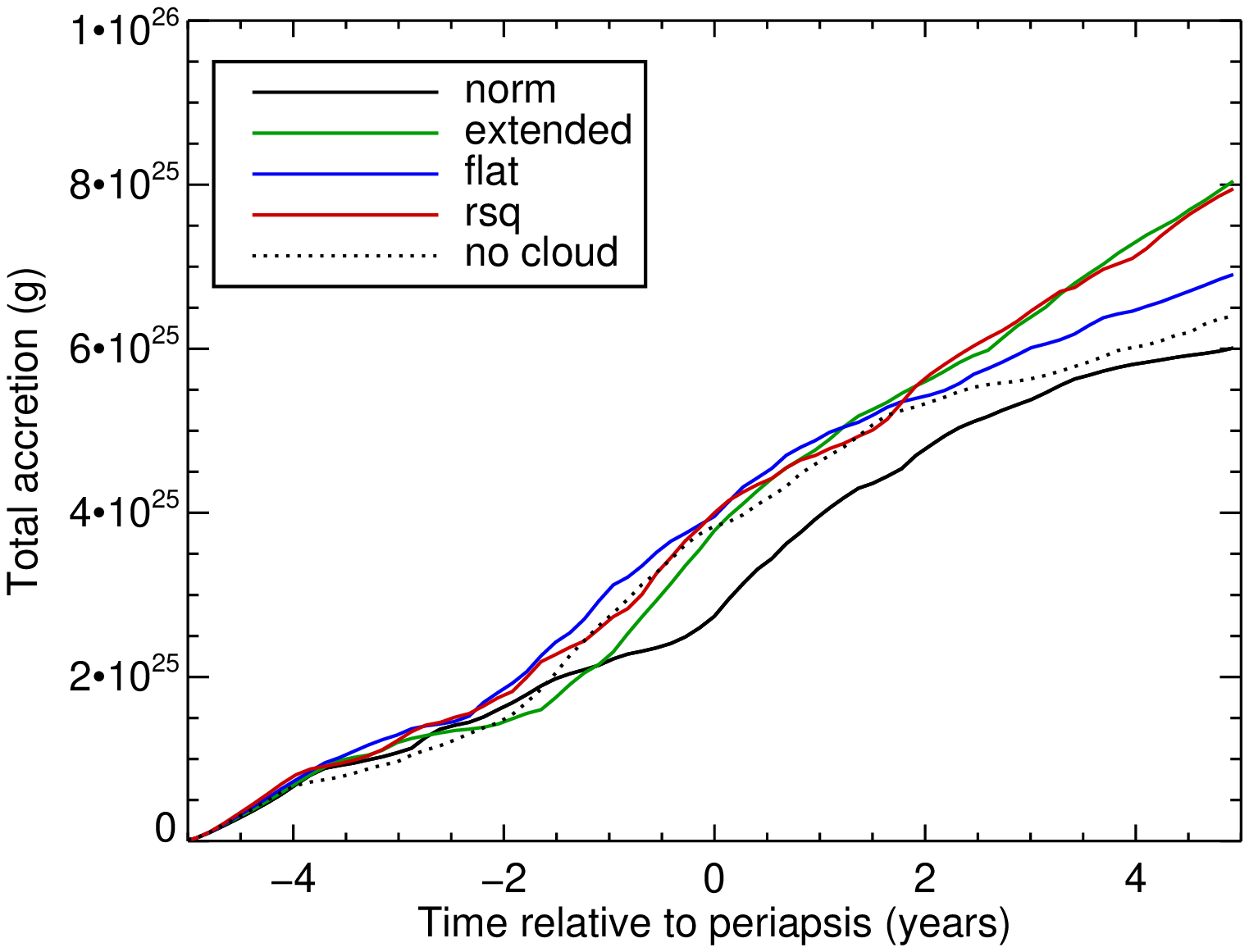}
\includegraphics[scale=.45]{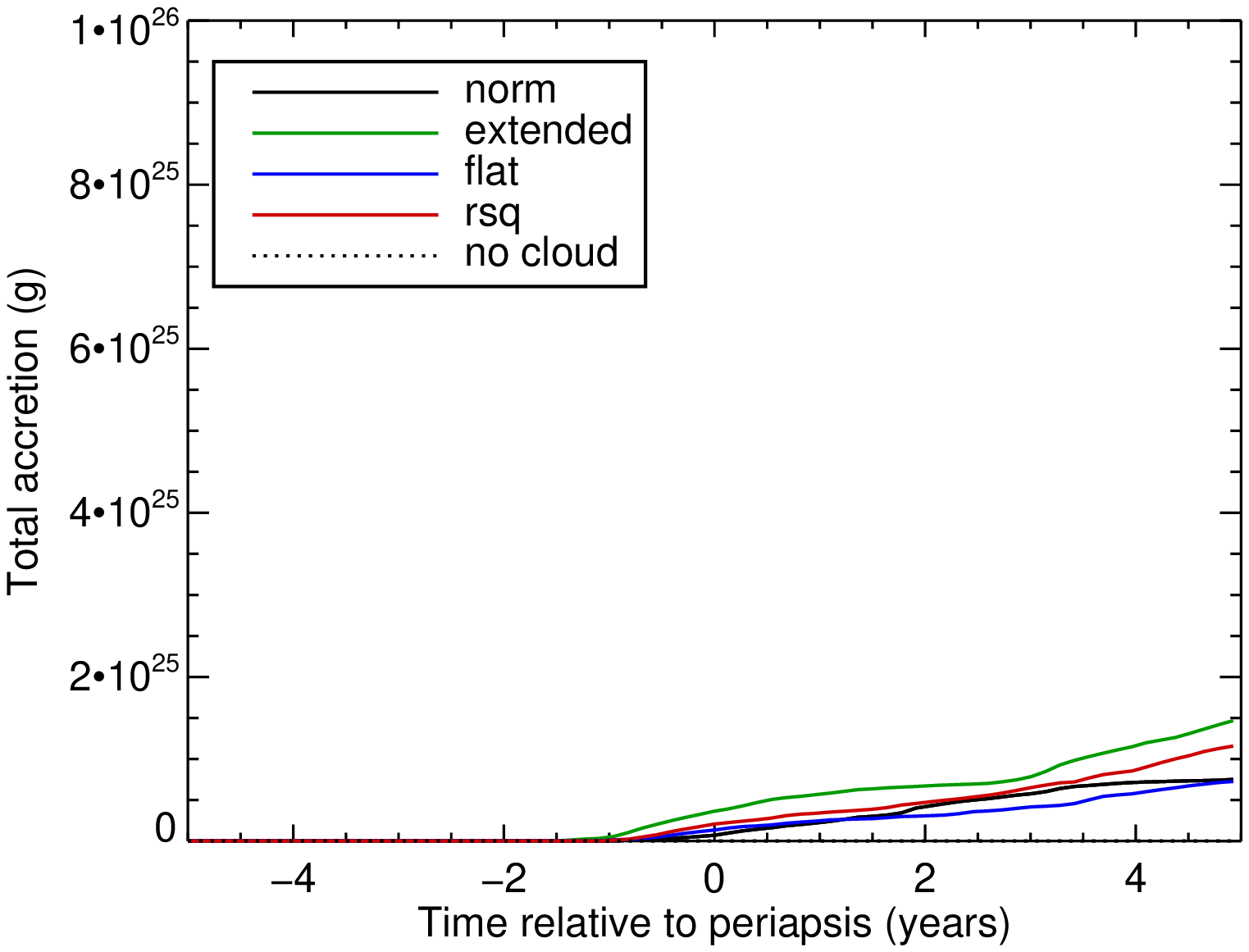}
\caption{
Cumulative mass accreted in models with normal Gaussian (black), extended Gaussian (green), flat (blue), and $r^{-2}$ (red) cloud density profiles, and a simulation with no cloud present (dashed black).
All simulations are carried out with a wind background model.
The total mass accreted (left) is similar for all models, with the most mass accreted in the ``extended'' and ``rsq'' models.
In the most extreme case, the ``extended'' model, only $25\%$ more mass has been accreted than in the ``no cloud'' model.
In the ``norm'' model, less mass is accreted than the model with no cloud present.
The cumulative amount of cloud material accreted (right) is higher by a factor of two for the ``extended'' model compared to the ``norm'' model, but even in this case accounts for less than $20\%$ of the total mass accreted.
The overall amount of material cloud material accreted is small.
In the most extreme case, the ``extended'' model, only $0.06\%$ of the total cloud mass has been accreted by 5 years after periapsis.
}
\label{fig:profiles_comp_cumulative}
\end{figure*}

The cumulative amount of mass accreted is plotted in fig.~\ref{fig:profiles_comp_cumulative}.
The total mass accreted (left panel) is similar for all models, with the most mass accreted in the ``extended'' and ``rsq'' models.
In the most extreme case, the ``extended'' model, only $25\%$ more mass has been accreted than in the ``no cloud'' model.
In the ``norm'' model, less mass is accreted than the model with no cloud present.
The cumulative amount of cloud material accreted (right panel) varies by a factor or two between models, but by 5 years after periapsis is less than $20\%$ of the total mass accreted in all cases, and at most $26\%$ of the mass accreted after periapsis.
Even in the most extreme case, the ``extended'' model, only $0.06\%$ of the total cloud mass has been accreted by 5 years after periapsis.


Modeling of Br-$\gamma$, bolometric, and X-ray luminosity of the clouds in our simulations provides a direct comparison with observational data.  
Because this emission comes from the bulk of the cloud material, it is not effected by the small amount of material accreted onto the black hole and removed from the simulation.

\subsubsection{Br-$\gamma$ luminosity}

\begin{figure*}
\includegraphics[scale=.9]{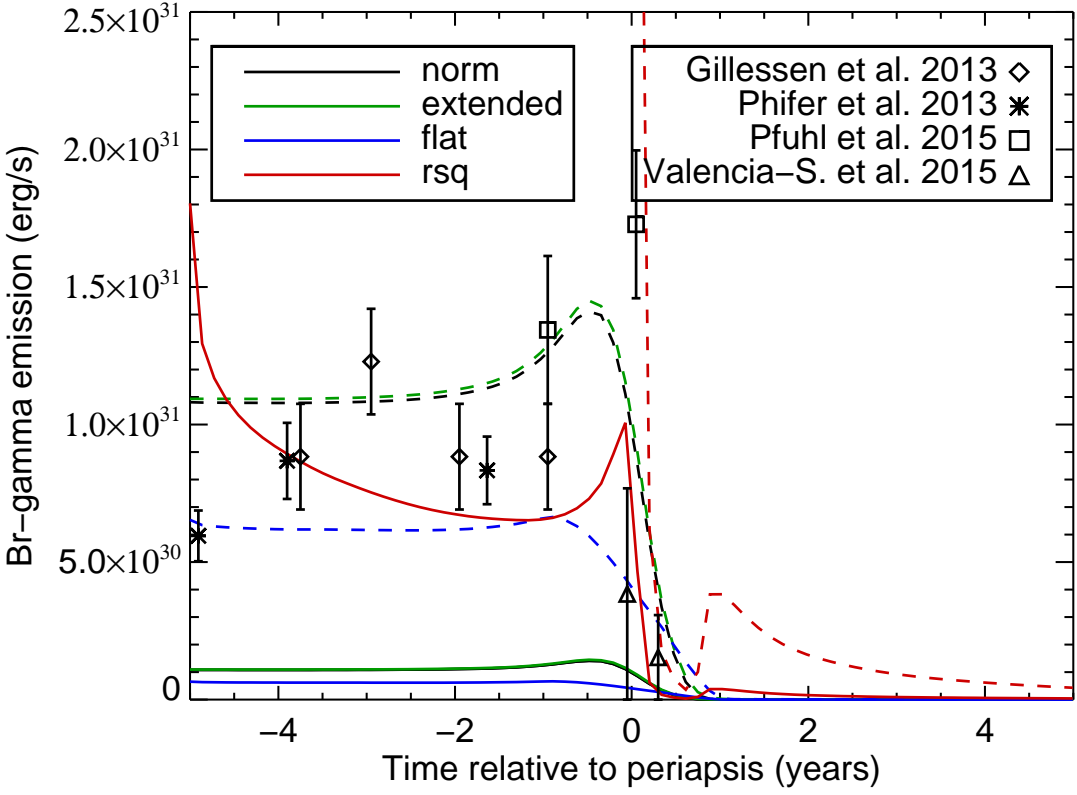}
\caption{
Comparison of Br-$\gamma$ emission in models with normal Gaussian (black), extended Gaussian (green), flat (blue), and $r^{-2}$ (red) cloud density profiles.
All simulations are carried out with a wind background model.
Solid lines are directly models from simulations, dashed lines are 10 times brighter, equivalent to an intial temperature of $1,500$K rather than $12,000$K.
Also plotted are observed Br-$\gamma$ luminosity of G2 from \citet{gillessen13b} (diamonds), \citet{phifer13} (stars), \citet{pfuhl15} (squares), and \citet{valencia15} (triangles).
Periapsis of G2 is assumed to be at year 2014.25.
}
\label{fig:profiles_br3_data}
\end{figure*}

Unlike the accretion rate, different cloud profiles produce significantly different emissions.
The total Br-$\gamma$ luminosity from our 4 simulations of different cloud profiles are shown in fig.~\ref{fig:profiles_br3_data} on a log scale.
At the beginning of the simulation, the spatial FWHM of Br-$\gamma$ emission is 30~mas for the ``norm'' and ``extended'' clouds, 50~mas for the ``flat'' profile cloud, and just 2~mas for the ``rsq'' cloud.
The ``rsq'' model is initially much brighter, by a factor of 16 compared to the ``norm'' model, because its luminosity is dominated by the very dense center of the cloud.

The ``norm'' and ``extended'' profiles produce a nearly identical evolution, as the Br-$\gamma$ emission is dominated by the dense center of the cloud, not the extended wings.
For these two models the Br-$\gamma$ luminosity is initially flat
and then rises as the cloud tidally compresses, peaking about $30\%$ above the initial value 6 months before periapsis.
At this point, heating of the cloud at the nozzle shock begins to reduce emissivity enough to counteract the increase due to compression, leading to a rapid drop in luminosity.  
By 1 year after cloud-center periapsis, nearly the entire cloud has been shocked, and Br-$\gamma$ luminosity has dropped a factor of 1000.

For the ``flat'' model, the initial luminosity is lower by about $45\%$, because the cloud is less compact.
The luminosity has a small peak of about $8\%$ 1 year before periapsis, followed by an extended decline ending 1 year after periapsis.
The more extended density of this cloud, compared to the ``norm'' Gaussian profile, leads to a decline in luminosity as soon as the cloud begins to shock at periapsis.
By 1 year after cloud-center passes periapsis, Br-$\gamma$ luminosity has again dropped a factor of 1000.



The ``rsq'' profile initially has a very large initial pressure and density gradient in the center.  
The pixels closest to the cloud center have 9 times the density and pressure as the next-closest pixels.  
This large gradient leads to a rapid drop in the central density by a factor of 9 over three years, the sound crossing time of one pixel.
Because the luminosity is dominated by the very dense central region, this leads to an initial drop in luminosity.
A minimum luminosity of about 6 times the initial ``norm'' model luminosity is reached 1 year before periapsis.  
After this, there is a sharp increase of $50\%$, peaking 1 to 2 months before periapsis.
This is followed by a rapid decline, reaching a minimum of $1/300$th the peak luminosity by 6 months after periapsis.
There is a re-brightening after this because the core of the cloud is dense enough to cool somewhat.
This re-brightening reaches $6\%$ the peak brightness by 1 year after periapsis, after which the Br-$\gamma$ luminosity declines again due to expansion of the cloud.

Regardless of the cloud profile, we find a sharp drop in Br-$\gamma$ emission following periapsis, which should make G2 undetectable by a few months to 1 year after periapsis.

Comparing with observational data from \citet{gillessen13b}, \citet{phifer13}, \citet{pfuhl15}, and \citet{valencia15} in fig.~\ref{fig:profiles_br3_data}, we find that our ``norm'', ``extended'' and ``flat'' (solid black, green and blue line) models do not produce enough emission to be compatible with observations of G2.

However, the Br-$\gamma$ luminosity is very sensitive to initial temperature and density in the cloud. We compare the results of simulations with varying mass and temperature to see what effects these changes would have.
The dashed lines represent $10\times$ the Br-$\gamma$ luminosity for each model, equivalent to lowering the initial cloud temperature from $12,000$K to $1,500$K or a cloud mass increase of $\sqrt{10}$.

An order of magnitude increase in Br-$\gamma$ luminosity makes the ``norm'', ``extended'', and ``flat'' models roughly consistent with the pre-periapsis observations of G2, and with the observation of weak post-periapsis emission from G2 in August 2014 \citep{valencia15}.
The ``rsq'' model without any changes (solid red line) is also roughly consistent with this data.

None of our models are consistent with the factor of $\sim2$ increase very close periapsis (2014.3) reported by \citet{pfuhl15}.
The largest increase, about $50\%$, is seen for the ``rsq'' model.
A density profile even more compact than our ``rsq'' model may be able to give an increase as large as a factor of 2, as the emission would be dominated by gas that is not shock heated until just before periapsis.

The \citet{pfuhl15} observation also appears inconsistent with the observation of \citet{valencia15} just before periapsis, which reported a flux 4.5 times lower than \citet{pfuhl15}.
\citet{valencia15} also demonstrated that the aperture (the physical size and shape of the area assigned to the cloud and background) used to select the Br-$\gamma$ emission, made a difference of a factor of two in the flux detected.  
This error is incorporated into the large error bars on the \citet{valencia15} data in fig.~\ref{fig:profiles_br3_data}.
The incompatibility between the \citet{valencia15} and \citet{pfuhl15} observational data may be due to \citet{pfuhl15} not taking the effect of different aperture selections into account.

\subsubsection{Bolometric luminosity}

\begin{figure*}
\includegraphics[scale=.9]{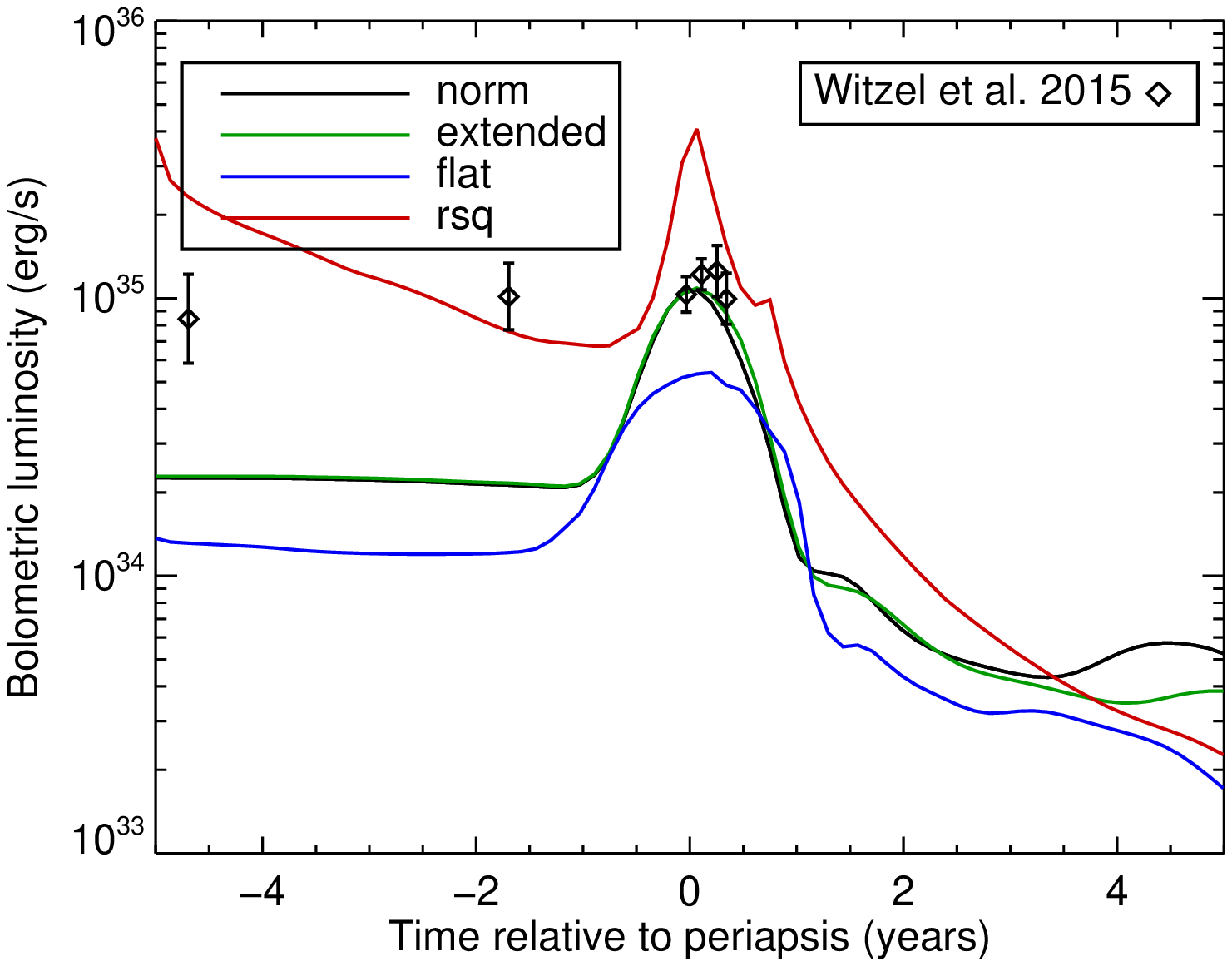}
\caption{
Bolometric cloud emission in models with normal Gaussian (black), extended Gaussian (green), flat (blue), and $r^{-2}$ (red) cloud density profiles.
Emission is calculated as the cooling function from \citet{sutherland93}.
Also plotted are values of G2 bolometric luminosity from \citet{witzel14}, derived from L' observations and assuming a black-body dust temperature of $550$K.
In our simulations, emission is initially flat, then rises steadily as the clouds are compressed, peaking at periapsis and then declining.  
The bulk of the emission occurs from 1 year before to 1 year after periapsis.
The ``norm'' and ``extended'' models have a peak luminosity of about $1.1\times10^{35}$~erg/s.
The ``flat'' model peaks at $40\%$ of this luminosity, and the ``rsq'' model at 5 times as much.
In all models, the luminosity at late times falls to a few~$\times10^{33}$~erg/s.
Note that the ``rsq'' model has a very narrow peak and rapid decline as the dense center of the cloud is shocked at periapsis.
A significant peak is not seen in broadband observations of G2.
}
\label{fig:profiles_bolo_data}
\end{figure*}

The bolometric luminosity of the clouds in our simulations is shown in fig.~\ref{fig:profiles_bolo_data}.
Emission is calculated using the cooling function of \citet{sutherland93} and assuming the cloud is fully photoionized.
Also plotted are observed luminosities from L' data from \citet{witzel14}, assuming a black-body temperature of $550$K.
To simplify our radiation modeling, we are assuming that all emission from the cloud is absorbed and re-emitted by dust between us and the galactic center, with a dust temperature of $550$K.


In our simulations, bolometric emission is initially flat or declining (particularly in the ``rsq'' case).  
It then rises steadily as the clouds are compressed, peaking at periapsis, and then declines.  
The rise and fall occurs from 1 year before to 1 year after periapsis.
The ``norm'' and ``extended'' models have a peak luminosity of about $1.1\times10^{35}$~erg/s.
The ``flat'' model peaks $40\%$ this luminosity, and the ``rsq'' model at 5 times as much.
In all models, the luminosity at late times falls to a few~$\times10^{33}$~erg/s
The width of the peak is narrower for the ``rsq'' model and wider for the ``flat'' model.

The bolometric luminosity in the initial (flat) portion of the light curve should not be trusted, as it is very sensitive to the low-temperature portion of the cooling function used.
The peak luminosity, however, is dominated by higher temperature gas and is not sensitive to the cooling function or initial cloud temperature.

No significant peak at periapsis is seen in the observational data \citep{witzel14}.
There is an increase in the combined flux of Sgr $A^*$ and G2 in L' band observations near periapsis, but \citet{witzel14} attribute this increase to Sgr $A^*$, not G2, due to a corresponding increase in K' band emission assumed to be solely due to Sgr $A^*$.
The lack of an observed peak in G2 emission is difficult to reconcile with our simulations, which all have a peak with a luminosity at least $40\%$ as bright as the observed flux attributed to G2 at periapsis (for the ``flat'' model).
%
%
%
%
In order to have a constant flux at early times and no peak at periapsis, another source, such as a dusty stellar object \citep{eckart13, witzel14}, is likely required to dominate the broadband emission of G2.


If the L' luminosity is dominated by cloud emission, it should fade rapidly in the year following periapsis due mainly to the expansion and cooling of the cloud material.

\begin{figure*}
\includegraphics[scale=.9]{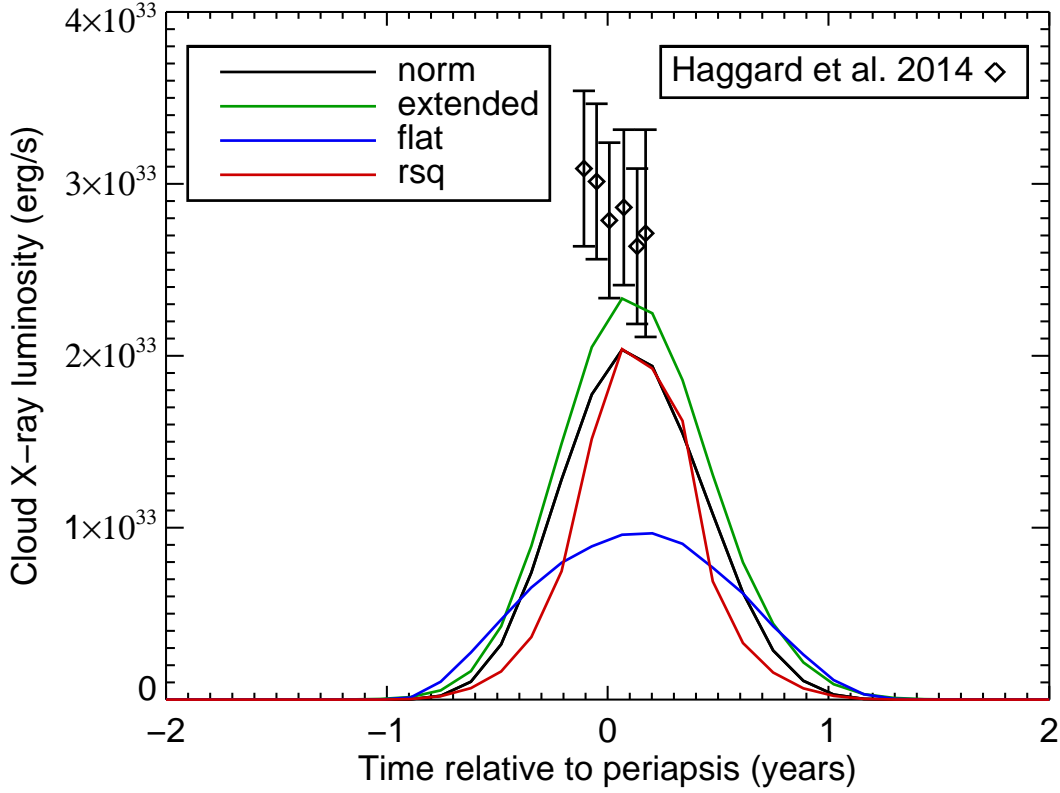}
\caption{
X-ray emission from 2 - 8 keV in models with normal Gaussian (black), extended Gaussian (green), flat (blue), and $r^{-2}$ (red) cloud density profiles.
Emission is calculated using XIM \citep{heinz09}.
Also plotted are X-ray luminosity observations of Sgr $A^*$ from \citet{haggard14}, derived from {\it Chandra} observations in 2014.
In our simulations, X-ray emission is only significant from about 6 months before to 6 months after periapsis.
The peak luminosity ranges from $9.7\times 10^{32}$~erg/s to $2.3\times10^{33}$~erg/s.
The long term average for the luminosity of Sgr $A^*$ is $3.4\times10^{33}$ erg/s in X-rays \citep{nowak12, neilsen13}, comparable to the 2014 observations.
Even our least compact simulation, the ``flat'' cloud, would have produced a detectable increase in X-ray luminosity, indicating that the gaseous component of G2 is either less that a few Earth masses or extremely extended.
}
\label{fig:profiles_xray_data}
\end{figure*}

X-ray emission from the shocked cloud, modeled using the XIM software package \citet{heinz09}, is shown in fig.~\ref{fig:profiles_xray_data}.
X-ray emission is only significant from about 6 months before to 6 months after periapsis.
The peak luminosity ranges from $9.7\times 10^{32}$~erg/s to $2.3\times10^{33}$~erg/s.
Also plotted are X-ray luminosity observations of Sgr $A^*$ from \citet{haggard14}, derived from {\it Chandra} observations in 2014.
The long term average for the luminosity of Sgr $A^*$ is $3.4\times10^{33}$ erg/s in X-rays \citep{nowak12, neilsen13}, comparable to the 2014 observations.
Even our least compact simulation, the ``flat'' cloud, would have produced a detectable increase in X-ray luminosity.
The lack of an increase in either bolometric or X-ray emission near periapsis indicates that the gaseous component of G2 is either less massive than assumed in our simulations (4 Earth masses) or extremely extended if we favor a single component model for G2.

\subsubsection{Br-$\gamma$ line velocity and line width}

In addition to flux, we can also model the apparent line-of-sight velocity and FWHM of the velocity dispersion of the Br-$\gamma$ line.
We rotate our Br-$\gamma$ emission data to an Earth-based frame using the orbital parameters from \citet{gillessen13b}, the same used for our cloud orbit.
%
%
Br-$\gamma$ luminosity from each grid point in our simulation is binned by line-of-sight velocity (bins width $20$~km/s).
We then fit a Gaussian to the resulting velocity vs. luminosity data.
This Gaussian fit provides the line center velocity and FWHM velocity of the line.

\begin{figure*}
\includegraphics[scale=.9]{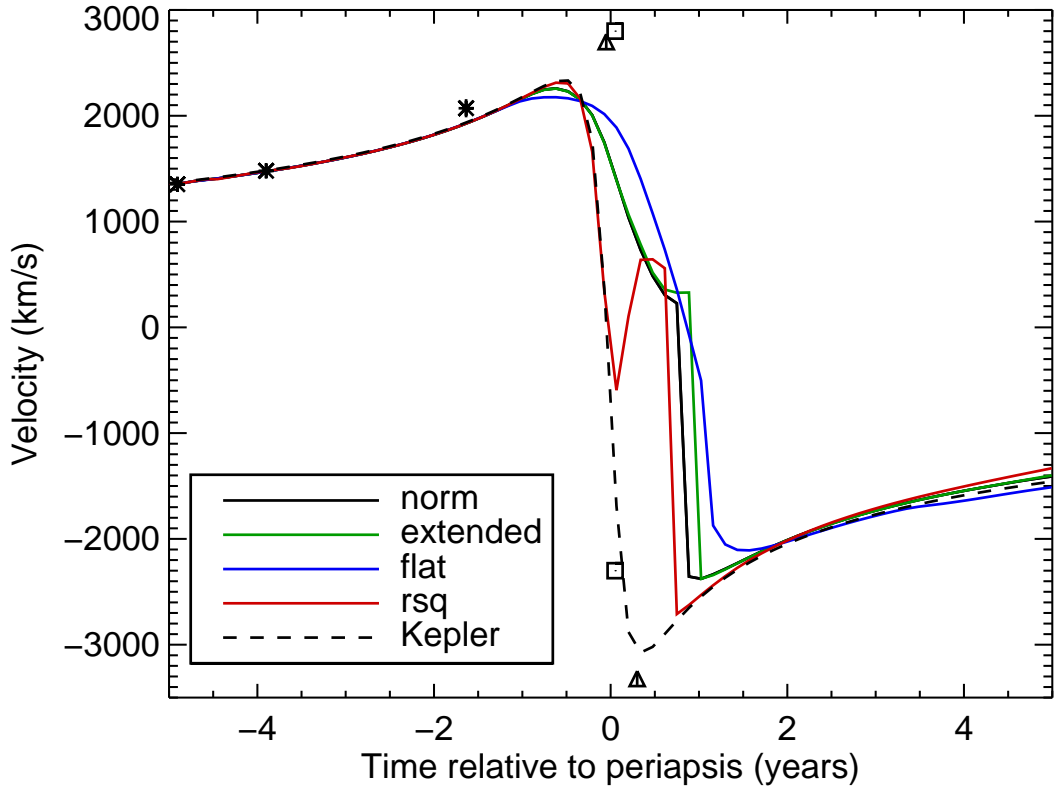}
\caption{
Comparison of Br-$\gamma$ line velocity in models with normal Gaussian (black), extended Gaussian (green), flat (blue), and $r^{-2}$ (red) cloud density profiles.
All simulations are carried out with a wind background model.
Dashed black lines represent the velocity expected for the cloud center, based on the Keplerian orbit from \citet{gillessen13b} used in our simulations.
Also plotted are observed Br-$\gamma$ velocities of G2 from \citet{phifer13} (stars), \citet{pfuhl15} (squares), and \citet{valencia15} (triangles).
Periapsis of G2 is assumed to be at year 2014.25.
The two squares are the red- and blue-shifted line components observed by \citet{pfuhl15} in April 2014.
The Br-$\gamma$ velocity in our simulations closely follow the Keplerian orbit, except from a few months before until 8 to 14 months after periapsis.
During this time, when the cloud is passing through the nozzle shock, emission is dominated by the unshocked, pre-periapsis portion of the cloud.
}
\label{fig:profiles_velocity_data}
\end{figure*}

Fig.~\ref{fig:profiles_velocity_data} shows the the fitted Br-$\gamma$ velocity for our four cloud density profiles, along with the Keplerian velocity expected for the orbital data from \citet{gillessen13b} (dashed line).
Also plotted are observational data from \citet{phifer13}, \citet{pfuhl15}, and \citet{valencia15}.
Our simulated velocity closely follows the Keplerian orbit of the cloud center up until shortly before the maximum (red-shifted) velocity is reached.  
The maximum velocity reaches between $90\%$ and $100\%$ of the expected value for all models, with the most compact (``rsq'') following most closely.

At periapsis, however, the expected rapid change to a negative (blue-shifted) velocity is not seen in our simulations.
Because the luminosity of the gas declines rapidly as it goes through the nozzle shock, the Br-$\gamma$ emission continues to be dominated by the unshocked, pre-periapsis material.
Therefore the velocity of the dominant emission component continues to be positive for 9 to 14 months after periapsis.

In the simulation with the most compact emission, ``rsq'', the velocity becomes briefly negative as the dense core of the cloud center reaches periapsis. 
The velocity then becomes positive again once the core is shocked, and remains positive until about 9 months after periapsis.
After this, the velocity rejoins the Keplerian orbit with a strongly negative (blue-shifted) value.
For the less compact clouds, this sharp decrease in velocity happens progressively later because the emission remains dominated by unshocked material that has not passed periapsis for a longer time.

None of our models reach the highest observed red-shifted velocities, which is not surprising given that our cloud orbit does not reach velocities this high.
Our simulation do reach very close to the largest velocities predicted for the orbital parameters used.

More worryingly for the cloud model of G2, none of our simulations are able to reproduce the very large negative velocity observed in August 2014 from \citet{valencia15}.
%
All of our clouds still have a dominant red-shifted component to the velocity at this time because, although most of the cloud mass has passed periapsis, it has been shock heated, suppressing its Br-$\gamma$ emission.
Detecting this blue-shifted velocity would required either a very compact source where nearly all emitting material has passed periapsis, or a source that does not decrease significantly in line emission at the nozzle shock (i.e. something other than a gas cloud).

\begin{figure*}
\includegraphics[scale=.9]{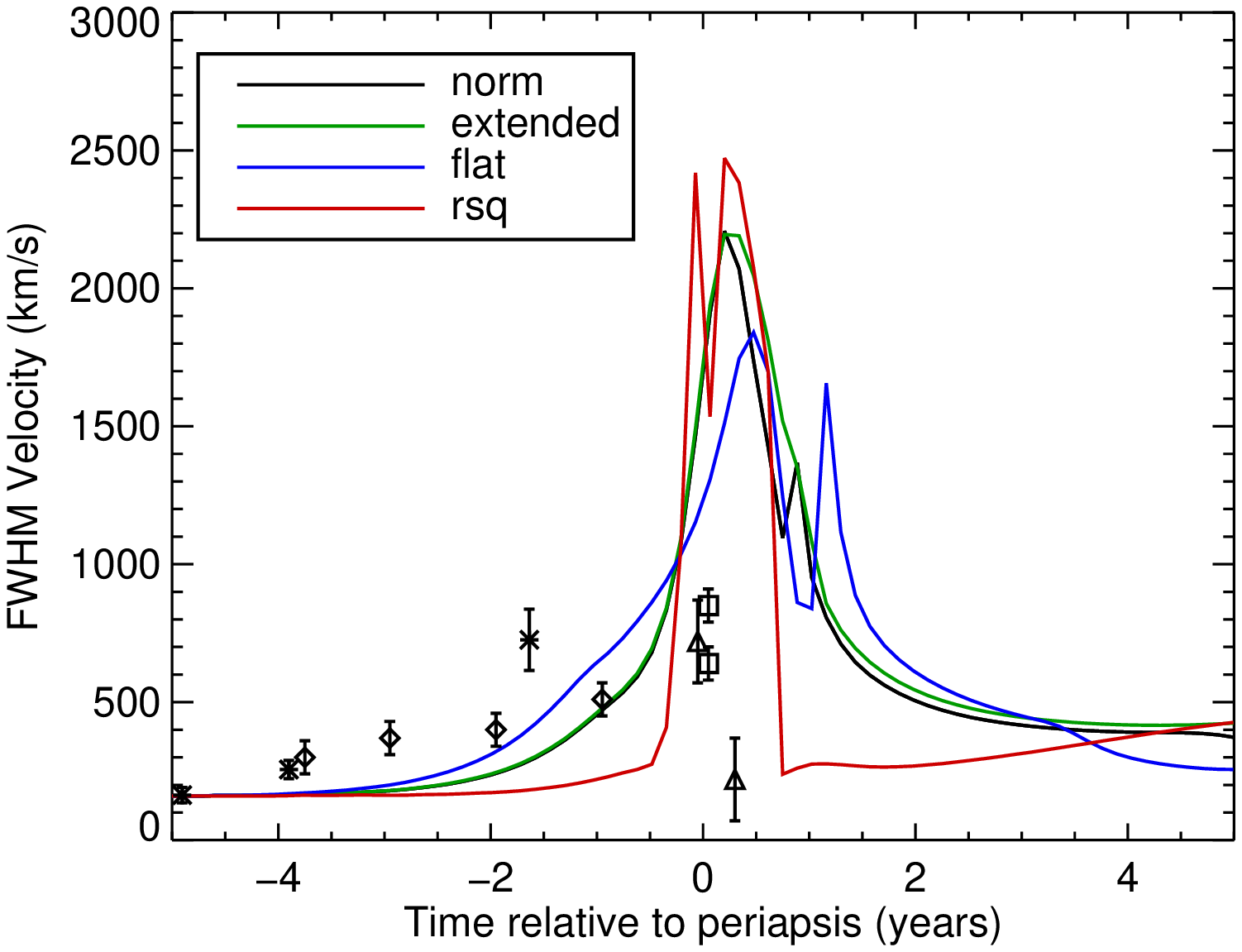}
\caption{
Comparison of FWHM of Br-$\gamma$ line velocity in models with normal Gaussian (black), extended Gaussian (green), flat (blue), and $r^{-2}$ (red) cloud density profiles. 
All simulations are carried out with a wind background model. 
Also plotted are observed FWHM of Br-$\gamma$ velocities of G2 from \citet{gillessen13b} (diamonds), \citet{phifer13} (stars), \citet{pfuhl15} (squares), and \citet{valencia15} (triangles). 
Periapsis of G2 is assumed to be at year 2014.25. 
The two squares are the red- and blue-shifted line components observed by \citet{pfuhl15} in April 2014. 
The ``norm'', ``extended'' and ``flat'' models show a gradual increase and decrease of the velocity FWHM before and after periapsis. 
The ``rsq'' model has a very small FWHM, except for a few months before to about 8 months after periapsis, when the cloud is passing the nozzle shock. 
}
\label{fig:profiles_FWHM_data}
\end{figure*}

From fitting a Gaussian to the velocity profile of Br-$\gamma$ emission, we obtain the FWHM of the fit, which corresponds to how much the emitting material is spread out along the orbit of G2.
The FWHM of the velocity of the Br-$\gamma$ line is shown in fig.~\ref{fig:profiles_FWHM_data}, along with observation data.

For the ``norm'', ``extended'' and ``flat'' models, we find that the FWHM begins to increase 3 to 4 years before periapsis, reaches very large values ($>1000$~km/s) from 6 months before to about 1 year after periapsis, and then slowly narrows over the next few years.
The very compact ``rsq'' model, by contrast, has a small FWHM, $<200$~km/s, until a few months before periapsis.
It is then very wide ($1500 - 2000$~km/s) until about 8 months after periapsis when it rapidly narrows back to $\sim300$~km/s.
This behavior is consistent with expectations: for the less centrally concentrated clouds emitting material is more spread out along the cloud's orbit, while in the centrally concentrated cloud most of the emitting material is moving at the same velocity. 
Near periapsis, the line is very broad due to the large change in ($\sim5000$~km/s) occurring in the orbital velocity.

Observations of the Br-$\gamma$ line FWHM velocity show a gradual increase from $\sim200$~km/s 5 years before periapsis up to $600 - 900$~km/s at periapsis.
There is then an observation of a very narrow, blue-shifted line in August 2015 ($\sim4$ months after periapsis) from \citet{valencia15}, with a FWHM of $220\pm150$~km/s.
The gradual increase in FWHM observed is consistent with an extended gas cloud being tidally stretched, rather than a very compact source like our ``rsq'' model.
The post-periapsis observation, however, are difficult to reconcile.  
%
All of our models have very broad lines at this time, which would make the line difficult to detect at all because the flux would be so spread out in wavelength.
The very narrow , highly blue-shifted line seen at this time is more consistent with very compact emission from, for example, a DSO.


\subsubsection{Br-$\gamma$ spatial extent}

To determine how the spatial extent of the cloud changes with time, we again rotate our Br-$\gamma$ emission data to an Earth-based frame using the orbital parameters from \citet{gillessen13b}.
We then construct an image of the total Br-$\gamma$ emission at all velocities, and apply a Gaussian smoothing with a FWHM of 37.5~mas, comparable to the spatial resolution of SINFONI \citep{gillessen13b}.
We fit a 2D elliptical Gaussian to the resulting image, and deconvolve the result with the smoothing Gaussian.
The result is the apparent ``true'' extent of the cloud along both a long and short axis of the ellipse.

\begin{figure*}
\includegraphics[scale=.9]{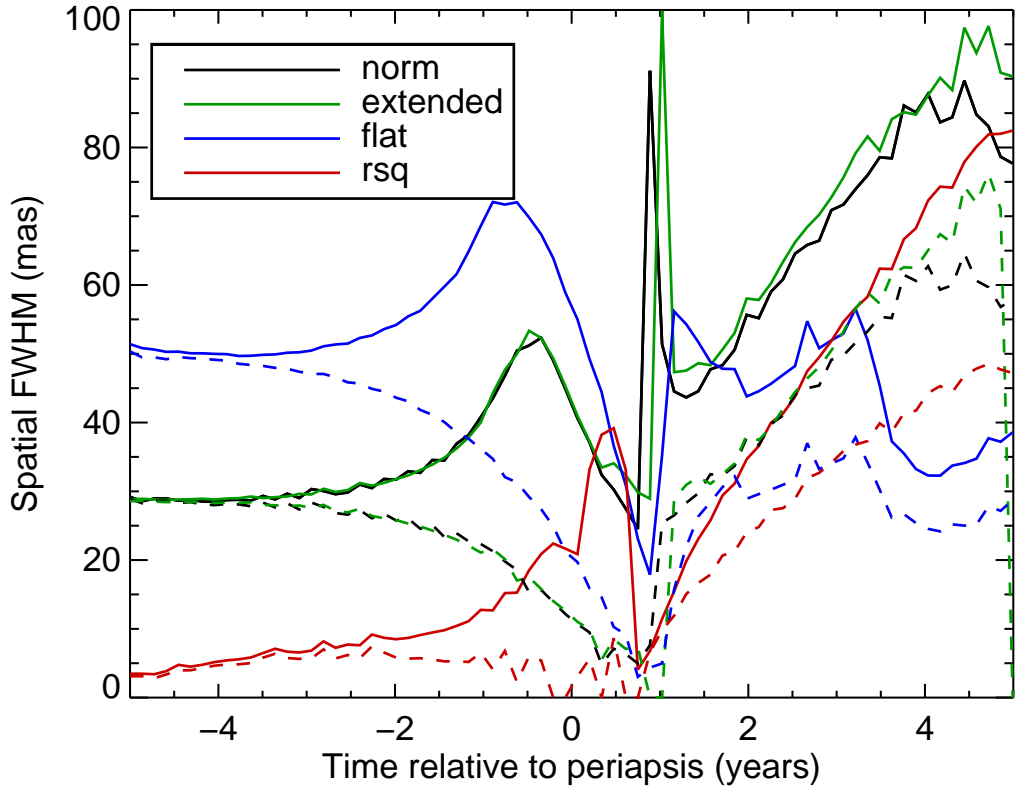}
\caption{
Comparison of the FWHM of the spatial extent of Br-$\gamma$ line emission, as seen from Earth, in models with normal Gaussian (black), extended Gaussian (green), flat (blue), and $r^{-2}$ (red) cloud density profiles. 
The long (solid) and short (dashed) axes of a 2D elliptical Gaussian fitted to the Br-$\gamma$ image at each time are shown.
All simulations are carried out with a wind background model. 
The ``norm'' and ``extended'' models have an initial FWHM of about 30 mas, and ``flat'' model 50 mas, and the ``rsq'' model 2 mas. 
The ``norm'', ``extended'' and ``flat'' models show a gradual increase of the spatial extent in one direction and a drop in the other, as the cloud is stretched along its orbit and compressed perpendicular to its orbit approaching periapsis. 
As the total Br-$\gamma$ emission begins to decline, due to gas at periapsis being shock heated, the spatial extent decreases.
This continues until about 1 year after cloud-center periapsis, when the peak of the Gaussian fit jumps from the remainder of the unshocked cloud to the shocked, post-periapsis fan of expanding material.   
The fan shows an increase size (both long an short axis) as it expands. 
The ``rsq'' model has a very small initial FWHM, but expands initial due to its high central pressure to about 10 mas by 2 years before periapsis. 
It then continues to expand along the cloud orbit due to tidal stretching until a few months after periapsis. 
The expanding fan then become the dominant emission component, and grows in size over time. 
}
\label{fig:profiles_spatial_FWHM}
\end{figure*}

The fitted spatial extent of the cloud is plotted in fig.~\ref{fig:profiles_spatial_FWHM} for our 4 cloud models.
Initially, the spatial extent is about 30~mas for the ``norm'' and ``extended'' clouds, and 50~mas for the ``flat'' cloud.
The observed spatial extent of 42~mas reported by \citet{gillessen13b} falls between these models, but the ``norm'' and ``extended'' models reach this size by 1 year before periapsis.

The ``norm'', ``extended'' and ``flat'' models show a gradual increase of the spatial extent in one direction and a drop in the other, as the cloud is stretched along its orbit and compressed perpendicular to its orbit approaching periapsis. 
As the total Br-$\gamma$ emission begins to decline, due to gas at periapsis being shock heated, the spatial extent along the orbit (long axis) decreases.
This continues until about 1 year after cloud-center periapsis, when the peak of the Gaussian fit jumps from the remainder of the unshocked cloud to the shocked, post-periapsis fan of expanding material.   
The fan shows an increase size (both long an short axis) as it expands. 

The ``rsq'' model has a very small initial FWHM (2~mas), but expands initial due to its high central pressure to about 10~mas by 2 years before periapsis. 
It then continues to expand along the cloud orbit due to tidal stretching until a few months after periapsis. 
The expanding fan then become the dominant emission component, and grows in size over time, similar to the other models.

\begin{figure*}
\includegraphics[scale=.45]{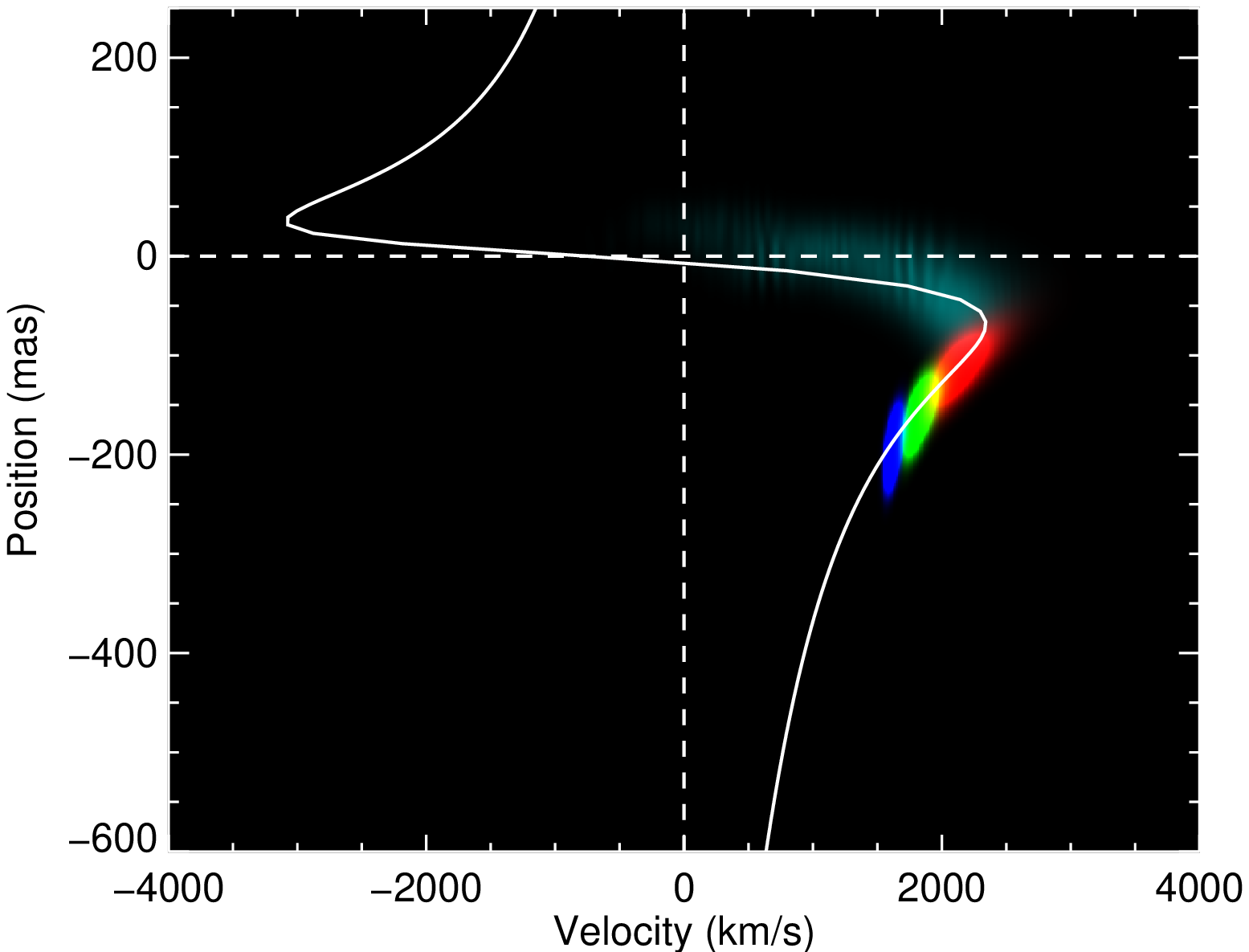}
\includegraphics[scale=.45]{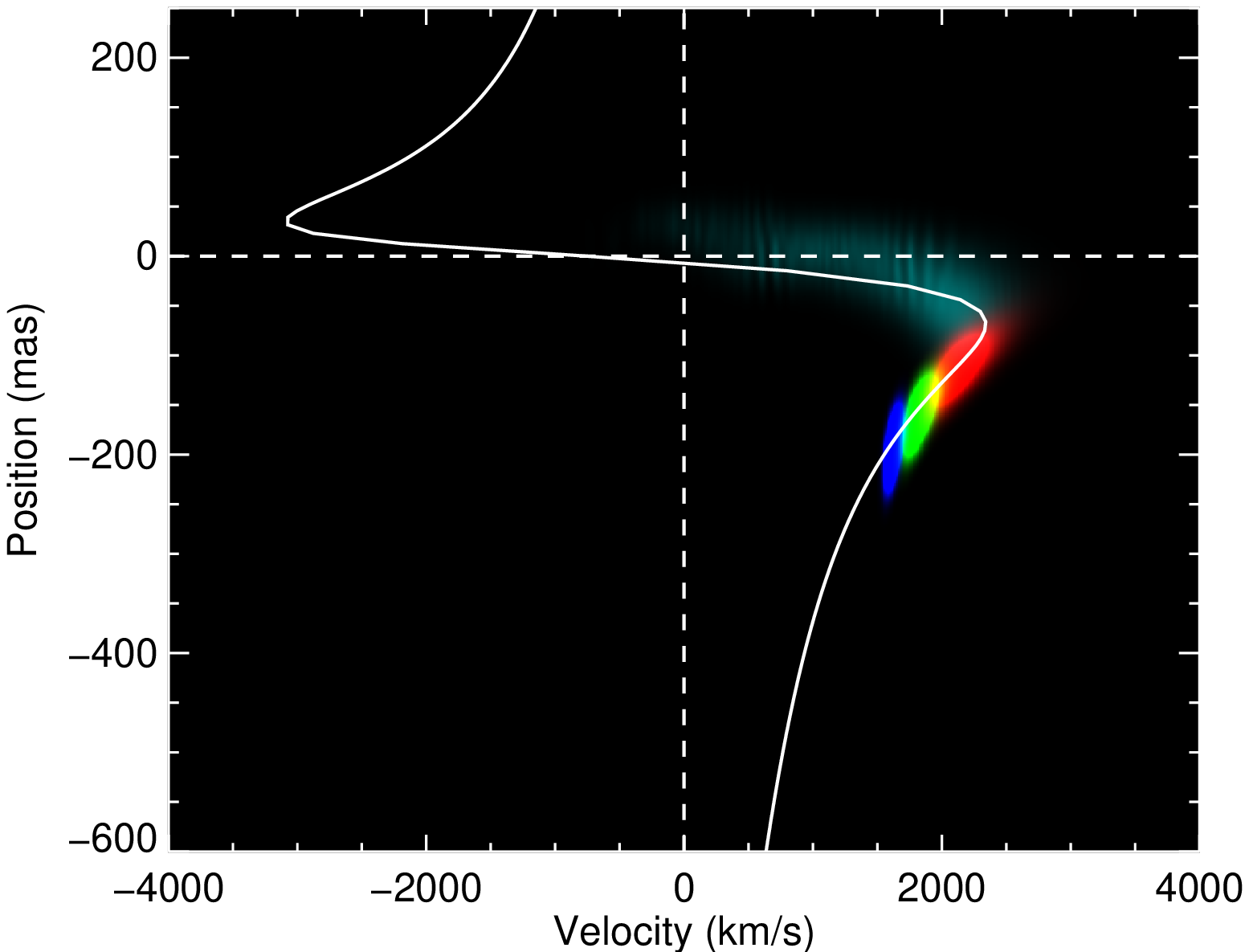}

\includegraphics[scale=.45]{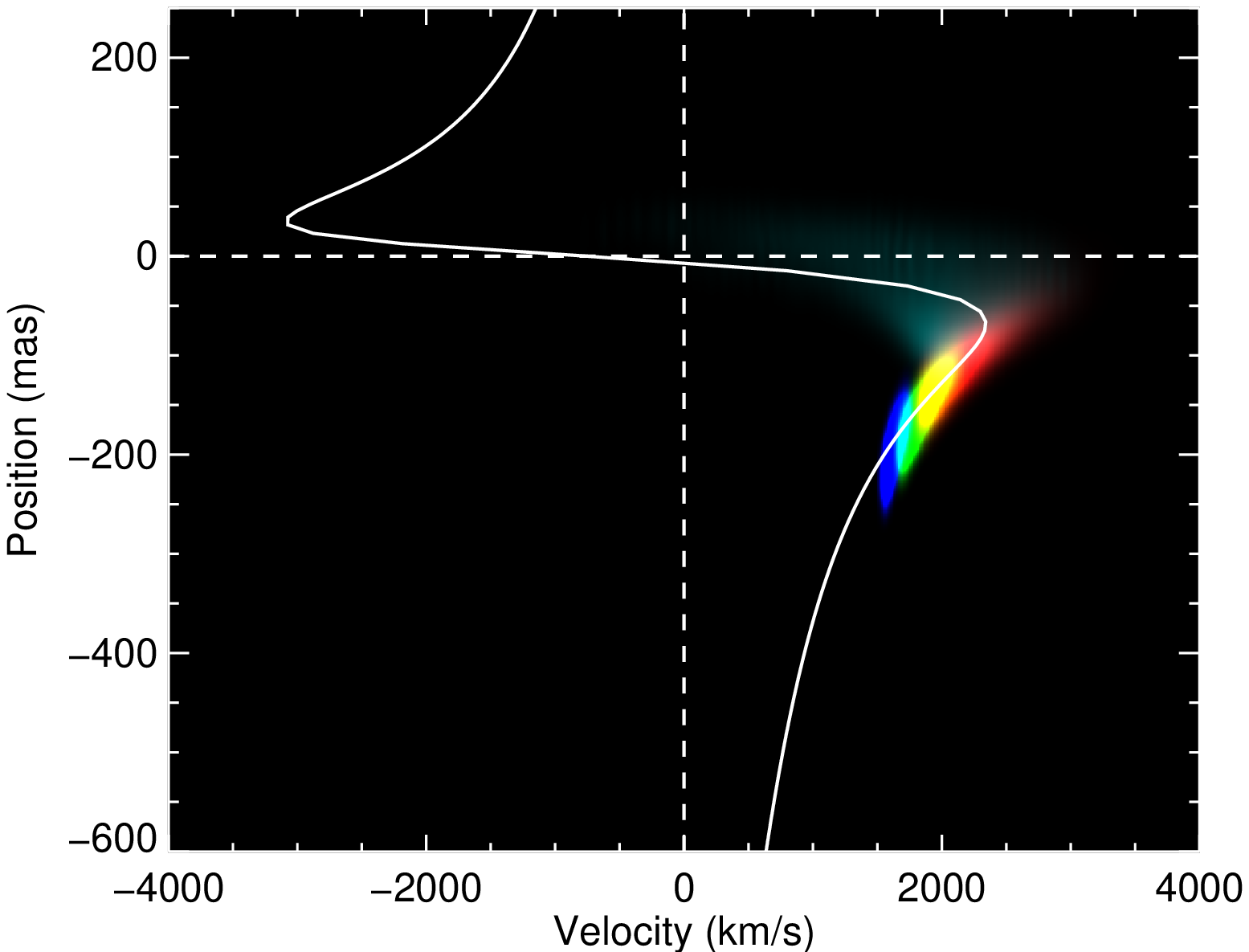}
\includegraphics[scale=.45]{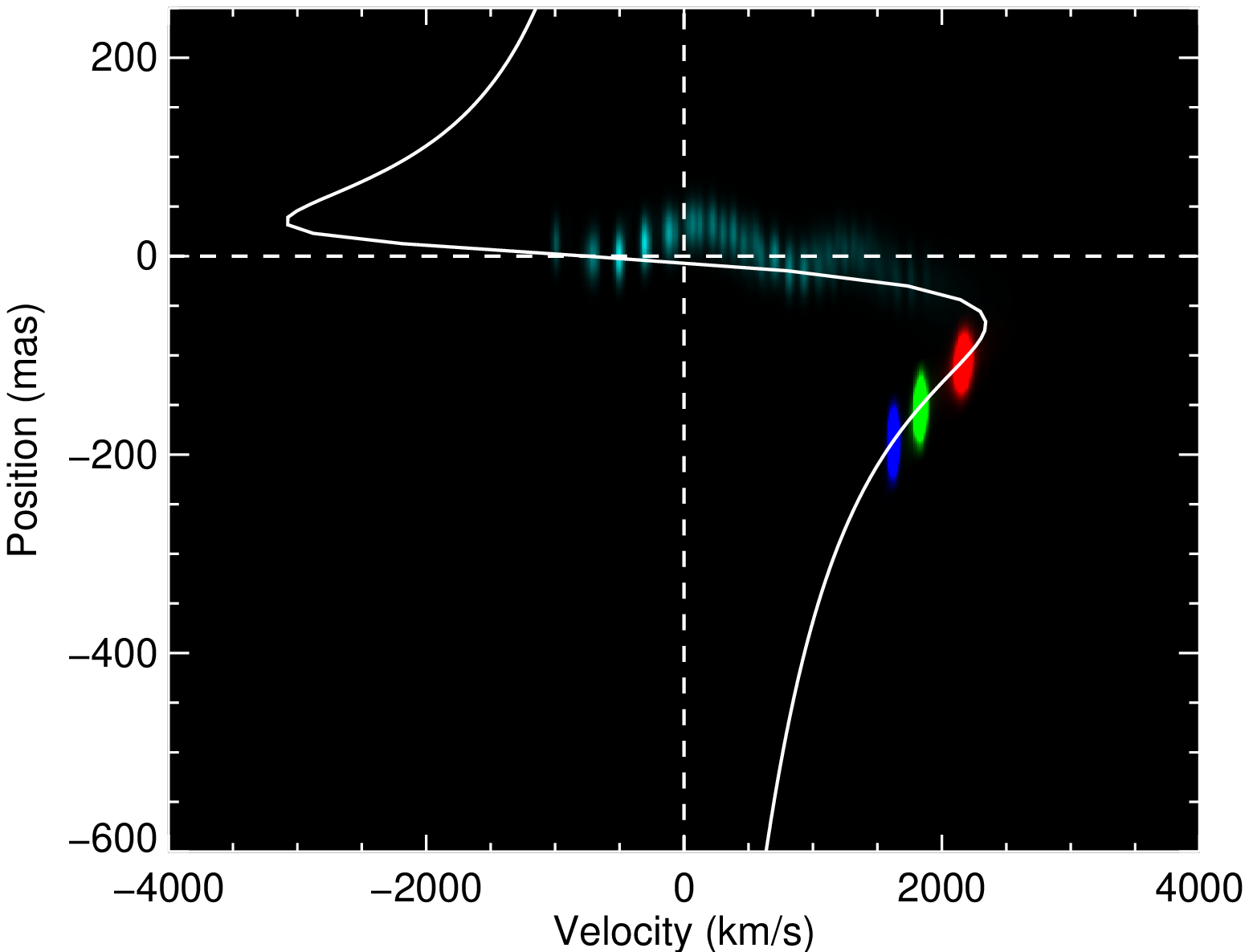}
\caption{
Position-velocity diagram of Br-$\gamma$ emission for the normal Gaussian (upper left), extended Gaussian (upper right), flat (lower left), and $r^{-2}$ (lower right) cloud density profiles. 
Position is along the cloud orbit, and velocity is along the observer's line of sight.
All simulations are carried out with a wind background model. 
Colors are cloud emission (linear scale) at 3 years (blue), 2 years (green), and 1 year (red) before periapsis and at periapsis (cyan).
The white line is the track of the cloud-center orbit.
The clouds follow the initial orbit very well, and the increasing velocity dispersion of the cloud is evident.
At periapsis, the main part of the cloud emission is still far from periapsis in velocity, at $\approx +2000$~km/s rather than $\approx-1000$~km/s.
There is a fading stream of material from the main emission source to periapsis, where is disappears due to shock heating.
The banding visible in the stream is due to limited resolution across a large change in velocity.
The spatial and velocity extent of the ``norm'' and ``extended'' clouds are similar, while the ``flat'' cloud is more extended, and the ``rsq'' cloud is is much smaller, and breaks up into a set of discrete points at periapsis.
}
\label{fig:profiles_posvel_all}
\end{figure*}

We also construct position-velocity diagrams for our synthetic Br-$\gamma$ data, comparable to the observed position-velocity of, e.g., \citet{gillessen13a}.
Every point in our simulation is mapped to the nearest point along cloud-center orbit.
The Br-$\gamma$ emission is then binned by this position and line-of-sight velocity to create a 2D image.
Each image is smoothed by 37.5~mas in position and 50~km/s in velocity.
Position-velocity diagrams for our 4 cloud models are shown in fig.~\ref{fig:profiles_posvel_all}.
Images for 4 different times are displayed on each plot, each in a different color.
The blue, green, red, and cyan colors correspond to simulation times of 3, 2, 1, and 0 years before periapsis, respectively.
Each image is on a linear scale, normalized to $1/10$th the maximum brightness at any time.
The white line in each plot is the track of the cloud-center orbit in position-velocity space.

The clouds follow the initial orbit very well, and the increasing velocity dispersion of the cloud is evident.
At periapsis (faint cyan image), the main part of the cloud emission is still far from periapsis in velocity, at $\approx +2000$~km/s rather than $\approx-1000$~km/s.
There is a fading stream of material from the main emission source to periapsis, where is disappears due to shock heating.
The banding visible in the stream is due to limited resolution across a large change in velocity.
The spatial and velocity extent of the ``norm'' and ``extended'' clouds are similar, while the ``flat'' cloud appears larger.
The ``rsq'' cloud is is much smaller, and breaks up into a set of discrete points at periapsis.

Comparing with \citet{gillessen13a} (fig.~7), the ``norm'', ``extended'', and ``flat'' models all show a similar increase in velocity spread as the cloud approaches periapsis.
The ``rsq'' cloud is much more point-like, particularly in line-of-sight velocity, than G2.

\section{Conclusions}
\label{sec:conclusions}

With respect to the accretion of G2 material onto Sgr $A^*$, we are able to draw the following conclusions:



- No substantial change with respect to the always present background accretion rate of Sgr $A^*$ should be expected due to the passage of G2.
Regardless of the background density and structure (wind/no wind, hot/cold, high/low density, etc.) only between $3$\% and $21$\% of the material accreted from 0 to 5 years after periapsis is from the cloud.  
The amount of cloud material accreted ranges from $0.03$\% to $10$\% of the total cloud mass by 5 years after periapsis.
Using high density backgrounds, i.e. 100 times our default density, and the two background profiles from \citet{anninos12}, produces cloud and total accretion rates in the range of previous simulations \citep[e.g.][]{anninos12, decolle14}.

- Changing the accretion radius has a large impact on the total accretion rate, particularly if it is large enough that part of the cloud passes inside this radius at periapsis.  
Smaller accretion radii produce a smaller cloud and total accretion rate.  
Using an appropriately small accretion radius (such that the cloud is outside the accretion radius at periapsis), the cloud still only accounts for between 1\% and 21\% of the total material accreted from 0 to 5 years after periapsis.


- Cloud structure has a relatively small impact on the accretion rate.  Our most extreme model, an extended Gaussian envelope, roughly doubles the amount of cloud material accreted, but this cloud still accounts for only $26\%$ of the accreted mass from 0 to 5 years after periapsis.

Although the absolute accretion rate onto the black hole is highly dependent on the specifics of the simulation setup, in all cases the passage of a cloud does not produce a large change in the total accretion rate onto the black hole, and the relative amount of cloud material accreted is always small, less than $26\%$ of the total mass accreted after periapsis.  The only exceptions are when the background inflow is initially stable shocked, spherically symmetric configuration, which is destabilized by the cloud leading to an order of magnitude reduction is the accretion rate.  Even out to 30 years after periapsis, we do not find a significant increase in total accretion.

From comparing modeling of G2 cloud emission to observational data, we find the following:

- The Br-$\gamma$ emission can give insight into the cloud structure if G2 is, in fact, a gas cloud.  In our simulations, there is a small peak in emission a few months to a year before periapsis, followed by a linear decrease and the disappearance of hydrogen line emission as the cloud is shocked.  The speed of the decline depends on how centrally concentrated the cloud is, with a central concentration giving a later peak and a faster decline.  Regardless of cloud structure, emission is reduced a factor of 1000 by 1 year after periapsis, making it unobservable.  


- The evolution of the bolometric luminosity is less sensitive to cloud structure. It is characterized by distinct rise and fall, peaking at periapsis with most emission occurring from 1 year before to 1 year after periapsis.  The luminosity peaks at $\sim 0.5 - 5 \times10^{35}$~erg/s and after 1 year, emission falls to a few $\times 10^{33}$ erg/s.  


- The line-of-sight velocity of the Br-$\gamma$ line closely follows the cloud-center Keplerian velocity, except from just before until 8 to 14 months after periapsis.  During this time, Br-$\gamma$ emission is dominated by unshocked, pre-periapsis gas and the velocity remains redshifted well after the cloud center passes periapsis.

- The FWHM of the line velocity gradually increases and decreases over the several years before and after periapsis for the Gaussian and flat density profiles.  For the centrally concentrated $r^{-2}$ profile, the FWHM remains small, $<300$~km/s at all times, except for about 2 months before to 8 months after periapsis, when the line is very broad ($>1500$~km/s).

Fitting observations of G2 likely requires two components: an extended cloud responsible for most of the pre-pericenter line emission and increasing FWHM, and a very compact component, such as a dusty stellar object, that dominates the bolometric luminosity and is responsible for the very narrow, blue-shifted Br-$\gamma$ line detected after periapsis in August 2014.
Simultaneously accounting for the lack of an observed peak in L' and X-ray observations near periapsis, and the high luminosity of the Br-$\gamma$ emission, requires the extended gas be fairly cold, $\sim1,500$K or less.
The gaseous component likely needs to be fairly extended, e.g. closer to our ``flat'' model, and may need to contain less material than the 4 Earth masses used in our simulations to prevent the appearance of a peak in L' luminosity near periapsis.

By the present (1$+$ year after periapsis), any emission from an extended gaseous component of G2 should have faded significantly, both in Br-$\gamma$ and broadband emission.  
Any future detections will be of a compact core or dusty stellar object.  
If it is not tidally disrupted, the bolometric emission from such an object should be roughly the same as before periapsis, and the line emission should have a narrow FWHM.

In summation, our simulations lead us to conclude that G2 is unlikely to be a single component object and that, if it is purely a gas cloud, it will not be detectable in the future.

\section*{Acknowledgments}

BJM was supported by an NSF Astronomy and Astrophysics Postdoctoral Fellowship under grant AST1102796, by the Aspen Center for Physics under NSF grant 1066293, and by the NSF under grant AST1333514.
The software used in this work was in part developed by the DOE NNSA ASC- and DOE Office of Science ASCR-supported Flash Center for Computational Science at the University of Chicago.
Resources supporting this work were provided by the NASA High-End Computing (HEC) Program through the NASA Advanced Supercomputing (NAS) Division at Ames Research Center.
Some of the simulations presented in this paper were performed on the Deepthough2 cluster, maintained and supported by the Division of Information Technology at the University of Maryland College Park.

\software{FLASH \citep[v. 4.2.2][]{Fryxell00, Dubey08}}

\bibliography{references}





\begin{table*}
\centering
\caption{Model Data}

\begin{tabular}{l||r|r|r|r|r|r||r|r|r|r}
  \hline
    name & acc rad$^1$ & background$^2$ & profile$^3$ & res$^4$ & cooling$^5$ & acc$^6$ & total acc (g)$^7$ & cloud acc (g)$^8$ & cloud/total$^9$ & total/nc$^{10}$ \\ \hline
norm & 1x & wind & norm & 1x & yes & yes & 6.01e+25 & 7.50e+24 & 0.21 & 1.28 \\
extended & 1x & wind & extended & 1x & yes & yes & 8.04e+25 & 1.47e+25 & 0.26 & 1.66 \\
flat & 1x & wind & flat & 1x & yes & yes & 6.91e+25 & 7.26e+24 & 0.20 & 1.15 \\
rsq & 1x & wind & $r^{-2}$ & 1x & yes & yes & 7.95e+25 & 1.16e+25 & 0.24 & 1.54 \\
2r & 2x & wind & norm & 1x & yes & yes & 1.76e+26 & 1.44e+25 & 0.13 & 0.91 \\
4r & 4x & wind & norm & 1x & yes & yes & 3.63e+26 & 3.70e+25 & 0.14 & 0.79 \\
8r & 8x & wind & norm & 1x & yes & yes & 6.51e+26 & 1.92e+26 & 0.06 & 0.64 \\
16r & 16x & wind & norm & 1x & yes & yes & 4.49e+27 & 3.82e+27 & 0.54 & 1.34 \\
norm\_nw & 1x & no wind & norm & 1x & yes & yes & 2.05e+26 & 3.33e+24 & 0.03 & 0.70 \\
4r\_nw & 4x & no wind & norm & 1x & yes & yes & 1.28e+27 & 1.38e+25 & 0.01 & 0.73 \\
16r\_nw & 16x & no wind & norm & 1x & yes & yes & 5.38e+27 & 3.58e+27 & 0.24 & 0.86 \\
1r\_nc & 1x & wind & none & 1x & yes & yes & 6.40e+25 & -- & -- & -- \\
2r\_nc & 2x & wind & none & 1x & yes & yes & 1.63e+26 & -- & -- & -- \\
4r\_nc & 4x & wind & none & 1x & yes & yes & 4.37e+26 & -- & -- & -- \\
8r\_nc & 8x & wind & none & 1x & yes & yes & 7.11e+26 & -- & -- & -- \\
16r\_nc & 16x & wind & none & 1x & yes & yes & 9.19e+26 & -- & -- & -- \\
1r\_nc\_nw & 1x & no wind & none & 1x & yes & yes & 2.67e+26 & -- & -- & -- \\
4r\_nc\_nw & 4x & no wind & none & 1x & yes & yes & 1.56e+27 & -- & -- & -- \\
16r\_nc\_nw & 16x & no wind & none & 1x & yes & yes & 2.52e+27 & -- & -- & -- \\
100bg & 1x & 100x wind & norm & 1x & yes & yes & 9.32e+27 & 5.17e+26 & 0.06 & 1.31 \\
100bg\_nw & 1x & 100x no wind & norm & 1x & yes & yes & 2.37e+28 & 2.36e+27 & 0.18 & 1.32 \\
100bg\_nc & 1x & 100x wind & none & 1x & yes & yes & 7.19e+27 & -- & -- & -- \\
100bg\_nc\_nw & 1x & 100x no wind & none & 1x & yes & yes & 2.87e+28 & -- & -- & -- \\
hot & 1x & hot wind & norm & 1x & yes & yes & 5.43e+25 & 3.95e+24 & 0.13 & 1.29 \\
hot\_nw & 1x & hot no wind & norm & 1x & yes & yes & 4.87e+26 & 3.12e+25 & 0.16 & 0.24 \\
hot\_nc & 1x & hot wind & none & 1x & yes & yes & 5.05e+25 & -- & -- & -- \\
hot\_nc\_nw & 1x & hot no wind & none & 1x & yes & yes & 1.00e+27 & -- & -- & -- \\
bg1 & 1x & bg1 & norm & 1x & yes & yes & 2.17e+28 & 7.32e+26 & 0.15 & 0.14 \\
bg1\_nc & 1x & bg1 & none & 1x & yes & yes & 4.61e+28 & -- & -- & -- \\
bg2 & 1x & bg2 & norm & 1x & yes & yes & 2.80e+28 & 4.81e+26 & 0.07 & 0.16 \\
bg2\_nc & 1x & bg2 & none & 1x & yes & yes & 5.26e+28 & -- & -- & -- \\
no\_cool & 1x & wind & norm & 1x & no & yes & 6.45e+25 & 6.17e+24 & 0.15 & 1.10 \\
no\_cool\_nc & 1x & wind & none & 1x & no & yes & 6.07e+25 & -- & -- & -- \\
1/2\_res & 1x & wind & norm & 1/2x & yes & yes & 7.29e+25 & 4.00e+24 & 0.08 & 1.14 \\
1/2\_res\_nc & 1x & wind & none & 1/2x & yes & yes & 7.12e+25 & -- & -- & -- \\
noacc & -- & wind & norm & 1x & yes & no & -- & -- & -- & -- \\
  \hline 
\end{tabular}

\par
$^1$ Accretion radius times $1.47\times10^{14}$~cm \\
$^2$ Background structure type \\
$^3$ Cloud density profile \\
$^4$ Resolution in cloud times $1.17\times10^{14}$~cm \\
$^5$ Is cooling turned on? \\
$^6$ Is mass allowed to accrete? \\
$^7$ Total mass accreted (cloud + background) from 5 years before to 5 years after periapsis \\
$^8$ Cloud mass accreted from 5 years before to 5 years after periapsis \\
$^9$ Ratio of cloud to total mass accreted from periapsis to 5 years after periapsis \\
$^{10}$ Ratio of total mass accreted to total mass accreted in equivalent simulation with no cloud present, from periapsis to 5 years after periapsis 

\label{table:models}
\end{table*}



\clearpage

%


\clearpage


\end{document}